\documentclass[prb,twocolumn,showpacs,amsmath,amssymb,superscriptaddress]
{revtex4}

\usepackage{graphicx}

\begin{document}

\title{Kondo model in nonequilibrium: Interplay between voltage,
temperature, and crossover from weak to strong coupling}
\author{Frank Reininghaus}
\author{Mikhail Pletyukhov}
\author{Herbert Schoeller}
\affiliation{Institut f\"ur Theorie der Statistischen Physik, RWTH Aachen, 52056
Aachen, Germany and JARA -- Fundamentals of Future Information Technology}
\date{\today}
\begin{abstract}
We consider an open quantum system in contact with fermionic metallic reservoirs in a
nonequilibrium setup. For the case of spin, orbital or potential fluctuations, we
present a systematic formulation of real-time renormalization group at finite temperature,
where the complex Fourier variable of an effective Liouvillian is used as flow parameter.
We derive a universal set of differential equations free of divergencies written
as a systematic power series in terms of the frequency-independent two-point vertex only,
and solve it in different truncation orders by using a universal set of boundary conditions. We apply
the formalism to the description of the weak to strong coupling crossover of the isotropic
spin-$\frac{1}{2}$ nonequilibrium Kondo model at zero magnetic field. From the temperature and
voltage dependence of the conductance in different energy regimes we determine various
characteristic low-energy scales and compare their universal ratio to known results. For a
fixed finite bias voltage larger than the Kondo temperature, we find that
the temperature-dependence of the differential conductance exhibits non-monotonic behavior
in the form of a peak structure. We show that the peak position and peak width scale
linearly with the applied voltage over many orders of magnitude in units of the Kondo
temperature. Finally, we compare our calculations with recent experiments.
\end{abstract}


\maketitle

\section{Introduction}
\label{sec:introduction}

For many decades, the Kondo model has attracted a great amount of interest in
condensed matter physics. The Kondo effect was first
discovered\cite{ResistanceMinimum} and analyzed\cite{KondoPerturbationTheory} in
bulk metals which contain magnetic impurities, where the exchange coupling $J$
between a localized spin-$\frac12$ and the conduction electrons leads to a
screening of the spin and to an increased resistivity at low
temperatures (see Ref.~\onlinecite{hewson} for a review). More recently, it was
first predicted theoretically\cite{kondo_theo_glazman_raikh,kondo_theo_ng_lee}
and then confirmed experimentally\cite{kondo_exp_goldhaber_gordon,
kondo_exp_cronenwett} that the Kondo effect also occurs in quantum dots in the
Coulomb blockade regime, where the net spin on the dot can form a single
impurity that is exchange-coupled to the conduction electrons in two or more
reservoirs. It turns out that the Kondo effect causes an enhancement of the
conductance through the quantum dot at low temperatures, and that the
conductance can reach the unitary value $2\frac{e^2}{h}$ for very low
temperatures and zero bias voltage.\cite{kondo_exp_van_der_wiel} Quantum dots do
not only permit us to control the coupling between the impurity to the
conduction electrons, but also allow us to study the behavior of the impurity in a
nonequilibrium setup by applying a finite bias voltage.\cite{kondo_exp_simmel,
kondo_revival_kouwenhoven_glazman}

\subsection{Previous theoretical work}
From a theoretical point of view, the Kondo model can be deduced from the
single impurity Anderson model by integrating out the charge degrees of freedom
using the Schrieffer-Wolff transformation.\cite{SchriefferWolff} Various
methods have been applied to the Anderson and Kondo models in three different
regimes:

\subparagraph{Equilibrium.} Methods that have been applied
successfully to the Anderson and Kondo models in equilibrium include
Fermi-liquid theory,\cite{nozieres_74} the Bethe
Ansatz,\cite{bethe_ansatz1, bethe_ansatz2, bethe_ansatz3, konik_etal_01_02}
conformal field theory,\cite{conformal_field_theory1, conformal_field_theory2}
and the numerical renormalization group\cite{wilson,costi,weichselbaum} (NRG). An
important result is that the zero bias conductance $G_{V=0}(T)$ through a single impurity at
finite temperature is unitary at $T=0$,
\begin{align}
\label{eq:unitary_conductance}
G_0 = G(T=V=0) = \frac{2 e^2}{h},
\end{align}
and is a universal function of the ratio $\frac{T}{T_K}$, where
the Kondo temperature $T_K$ is a characteristic energy scale that governs the low-energy
behavior of the impurity. In two-loop poor man scaling methods \cite{hewson,poor_man_scaling}
it is defined by
\begin{align}
\label{eq:TK_pms}
T_K = D\sqrt{J_0} e^{-\frac{1}{2J_{0}}},
\end{align}
where $D$ is the band width of the reservoirs, and $J_0$ is the exchange coupling
between the impurity spin and the conduction electrons. The Kondo temperature
is related to the width of the peak in $G_{V=0}(T)$ at $T=0$. Therefore, a precise
definition of a characteristic low-energy scale is the temperature for which the
conductance drops to half its maximum value:
\begin{align}
\label{eq:TK*_definition}
G_{V=0}(T=T_K^*)=\frac{1}{2} G_0,
\end{align}
We denote this energy scale by $T_K^*$, in contrast to $T_K$ which is not
uniquely defined in the literature.

\subparagraph{Expansions in the strong coupling regime.} In the case that both
temperature and voltage are much smaller than the Kondo temperature, Fermi
liquid theory has been used\cite{nozieres_74, oguri_JPSJ05, sela_malecki_PRB09}
to obtain an expansion of the differential conductance up to second order in
$\frac{T}{T_K}$ and $\frac{V}{T_K}$. The result is
\begin{align}
\label{eq:Expansion_cT_cV}
G(T,V)/G_0 = 1-c_T\left(\frac{T}{T_K}\right)^2
-c_V\left(\frac{V}{T_K}\right)^2,
\end{align}
where the ratio of the coefficients $c_V$ and $c_T$ is
\begin{align}
\label{eq:ratio_cV_cT}
\frac{c_V}{c_T}=\frac{3}{2\pi^2}.
\end{align}
For the ratio of $c_V$ and $c_T$ it is not important which definition of $T_K$
is chosen. If one uses the energy scale $T_K^*$ instead of $T_K$ in the expansion
\eqref{eq:Expansion_cT_cV} we write
\begin{align}
\label{eq:Expansion_cT_cV_*}
G(T,V)/G_0 = 1-c_T^*\left(\frac{T}{T_K^*}\right)^2
-c_V^*\left(\frac{V}{T_K^*}\right)^2.
\end{align}
This defines uniquely the coefficients $c_T^*$ and $c_V^*$, which are universal
numbers (i.e. independent of the details of the high-energy cutoff function) of
$O(1)$. Recently, the coefficient $c_T^*$ has been determined from very precise
numerical renormalization group calculations with the result \cite{merker_PRB13}
\begin{equation}
\label{eq:c_TV*}
c_T^*\approx 6.58,\quad c_V^*=\frac{3}{2\pi^2}c_T^*\approx 1.00,
\end{equation}
which serves as a quality benchmark for the reliabilty of other many-body methods
in the regime of very low energies.
A delicate issue for the precise calculation of these coefficients is the fact
that the band width $D$ has to be many orders of magnitude larger than the Kondo
temperature $T_K$ in order to obtain universal results in the scaling limit
\begin{align}
\label{eq:scaling_limit}
D\rightarrow\infty
\quad,\quad
J_0\rightarrow 0
\quad,\quad
T_K = \text{const}.
\end{align}
Numerically, as explained in Ref.~\onlinecite{hanl_PRB14}, it is very difficult
to achieve this for the Kondo model whereas for the underlying Anderson impurity
model it has only recently become possible to extrapolate the universal value of
$c_T^*$.\cite{merker_PRB13} In this paper, we will show that
our analytical method allows for a different way to achieve universality directly
for the Kondo model.

\subparagraph{Weak coupling regime.} If there is an energy scale in the system,
such as the temperature, the voltage, or the magnetic field, which is much
larger than the Kondo temperature, perturbative renormalization group
(RG) methods can be used. They perform an expansion of the physical quantities
in terms of a renormalized, but still small, coupling, provided that the RG flow
of the coupling does not cause divergencies. These methods were pioneered by
poor man's scaling\cite{poor_man_scaling} and include the following:
\begin{itemize}
\item Scaling methods that include a phenomenological decay rate $\Gamma$ as a
cutoff for the RG flow.\cite{rosch_kroha_woelfle_PRL01,
rosch_paaske_kroha_woelfle_PRL03, glazman_pustilnik_05, doyon_andrei_PRB06}
\item The flow equations method, where the competition between terms of
different orders in the coupling constant prevents divergencies during the RG
flow.\cite{kehrein_PRL05}
\item The real time renormalization group (RTRG), which, unlike the previous
methods, can explain the emergence of a decay rate $\Gamma$ even in the lowest
order truncation of the RG equations. The RTRG has been used with either the
reservoir bandwidth\cite{korb_reininghaus_hs_koenig_PRB07} or an imaginary
frequency cutoff, which cuts off the Matsubara poles of the Fermi distribution
function,\cite{RTRG_FS,hs_reininghaus_PRB09} as the flow parameter.
\end{itemize}

\subsection{Recent developments}

Recently, attempts have been made to fill the gap between the well-understood
strong coupling and weak coupling regimes of the nonequilibrium Kondo model.
The RTRG, which had been applied earlier to the Kondo model in the weak coupling
regime, has recently been used with the Fourier variable $E$ as the flow
paramter to consider the Kondo model at arbitrary voltage and zero temperature,
and vice versa.\cite{pletyukhov_hs_PRL12} This $E$-flow scheme of the RTRG
is able to reproduce the NRG results for the equilibrium conductance up to
deviations of a few percent, and yields results for the nonequilibrium
conductance which are consistent both with Fermi-liquid theory in the strong
coupling regime, and with weak coupling expansions. The results were also in agreement
with measurements of the differential conductance in an InAs nanowire quantum dot.\cite{kretinin_PRB84}
Moreover, Ref.~\onlinecite{pletyukhov_hs_PRL12} predicted that
the nonequilibrium differential conductance $G_{T=0}(V)$ at zero temperature
and a voltage that matches the energy scale $T_K^*$ as defined by
Eq.~\eqref{eq:TK*_definition} is approximately two thirds of the unitary conductance:
\begin{align}
\label{eq:GV_2_3}
G_{T=0}(V=T_K^*)&=\frac{2}{3}G_0.
\end{align}
This prediction, which has been confirmed by recent
experiments,\cite{kretinin_PRB85} provides an alternative way to determine the
Kondo temperature $T_K^*$ of a system in an experiment, which is usually much simpler to
perform because all measurements can be done at constant temperature.
Using the $E$-flow scheme of RTRG, similiar results were recently obtained for
the $S=1$ Kondo model,\cite{hoerig_mora_schuricht_PRB14} where Eq.~\eqref{eq:GV_2_3} changes
to $G_{T=0}(V=T_K^*)=0.605 G_0$.

Moreover, from
the voltage dependence of the conductance at zero temperature one can define an
energy scale $T_K^{**}$ as the voltage for which the conductance drops to half its
maximum value
\begin{align}
\label{eq:TK**_definition}
G_{T=0}(V=T_K^{**})=\frac{1}{2} G_0.
\end{align}
Correspondingly one can define Fermi liquid coefficients $c_T^{**}$ and $c_V^{**}$
by taking the scale $T_K^{**}$ as reference scale
\begin{align}
\label{eq:Expansion_cT_cV_**}
G(T,V)/G_0 = 1-c_T^{**}\left(\frac{T}{T_K^{**}}\right)^2
-c_V^{**}\left(\frac{V}{T_K^{**}}\right)^2.
\end{align}
An interesting issue is the determination of the universal ratio of the two
energy scales $T_K^*$ and $T_K^{**}$, from which one can determine the absolute values
of the universal Fermi liquid coefficients $c_T^{**}$ and $c_V^{**}$ via
\begin{align}
\label{eq:formula_cT_cV_**}
c_T^{**} = c_T^* \left(\frac{T_K^{**}}{T_K^*}\right)^2,\quad
c_V^{**} = c_V^* \left(\frac{T_K^{**}}{T_K^*}\right)^2.
\end{align}

A Keldysh effective action theory has recently been applied to a highly
unsymmetrical Anderson model, which prohibits double occupancy of the
impurity.~\cite{smirnov_grifoni_PRB03} The method supports the
prediction~\eqref{eq:GV_2_3} and has recently been extended to the situation
where the temperature, the voltage, and the magnetic field are all
non-zero.\cite{smirnov_grifoni_NJP13}

Another perturbative method to describe universal
scaling at low and intermediate energy scales has been proposed in Ref.~\onlinecite{spataru_PRB10},
where the $GW$ approximation within the $\sigma G \sigma W$ formalism has been used for
the symmetric Anderson model. They predicted a value for the ratio $T_K^*/T_K^{**}\sim 0.66$,
which is quite close to the result $T_K^*/T_K^{**}\sim 0.62$ obtained by our method.

Other studies of the nonequilibrium Anderson impurity model exist (see, e.g.,
Refs.~\onlinecite{eckel_etal_NJP10} and \onlinecite{andergassen_etal_review10}
for reviews). However, these suffer from the problem that the Coulomb
interaction cannot be chosen arbitrarily large, or, equivalently, the
corresponding Kondo exchange coupling $J_0$ cannot be made arbitrarily small. This
hinders achievement of universality in the scaling limit.

\subsection{Scope of this paper}
\label{sec:scope}
The purpose of this paper is twofold. In the first part, we will describe
in all detail the idea of the $E$-flow scheme of the RTRG, as proposed in
Ref.~\onlinecite{pletyukhov_hs_PRL12}.
We note that this scheme is essentially different from the one
developed in Ref.~\onlinecite{RTRG_FS}, where a cutoff of the Matsubara poles of
the Fermi functions was used, and the RG equations were derived by the principle
of invariance when reducing the cutoff. In contrast, the $E$-flow scheme uses the
Fourier variable $E$ itself as flow parameter, yielding a physical result for all
quantities at each stage of the RG flow. The technical derivation of the RG equations is very
different compared to Ref.~\onlinecite{RTRG_FS} since one does not make use of the
principle of invariance. Instead, one can set up directly a systematic and well-defined perturbative
expansion of the derviatives of all physical quantities w.r.t. $E$ in terms of the
effective two-point vertex. Since $E$ can be considered in
the whole complex plane, the RG equations can be solved along arbitrary paths in
the complex plane. This provides a natural scheme to define analytic continuations
of all retarded quantities into the lower half of the complex plane, even on a
pure numerical level. For these reasons, the $E$-flow scheme is a natural RG scheme
capable of addressing the physics of nonequilibrium stationary states, together with
the full time evolution starting from an initially uncoupled system from the reservoirs (for more
general initial conditions for quantum quenches and time-dependent Hamiltonians, see
Refs.~\onlinecite{kashuba_etal_13} and~\onlinecite{kashuba_etal_12}). Technically,
the $E$-flow scheme allows for a systematic resummation of
all logarithmic divergencies at high and low energies (i.e., short and long times)
simultaneously and provides the possibility to solve the RG flow also starting from
the infrared regime. As we will explain below, the latter turns out to be important to
determine the universal part of the solution. The supplementary part of
Ref.~\onlinecite{pletyukhov_hs_PRL12} contains a short description of the ideas
of the $E$-flow scheme, whereas the present paper will reveal all technical details.
Moreover, we will also go beyond Ref.~\onlinecite{pletyukhov_hs_PRL12} and develop a scheme
which can be generalized to all orders and we will show that it is sufficient
to set up a systematic power series in terms of the frequency-independent two-point
vertex only. We will focus on fermions and consider
the case of a generic quantum dot in the Coulomb blockade regime (i.e., charge
fluctuations are suppressed) which is coupled to noninteracting reservoirs with a
flat density of states (DOS). Other extensions for charge fluctuations or
frequency-dependent DOS are also possible and have recently been started in
connection with the interacting resonant level model \cite{kashuba_etal_13}
and the Ohmic spin-boson model.\cite{kashuba_schoeller_PRB13}

An important issue of this paper concerns universality, i.e., the way how one
can set up the scaling limit \eqref{eq:scaling_limit}, which determines that part of the
solution which is independent of the specific choice of the high-energy cutoff function.
Whereas the limit $D\rightarrow\infty$ can be performed directly for the RG equations
(since all frequency integrals are convergent), it is necessary to find appropriate
universal initial or boundary conditions to solve the differential equations. This is
achieved by using a perturbative calculation for various quantities at high energies,
together with the boundary condition of unitary conductance for $E=V=T=0$. In this way,
no specific form for the high-energy cutoff function is needed. In comparison to
Ref.~\onlinecite{pletyukhov_hs_PRL12}, we propose an improved scheme to set up
the initial conditions which, for the Kondo model, guarantees universality already for
exchange couplings of the order of $J_0\sim 0.04$, i.e., by using Eq.~\eqref{eq:TK_pms},
for $\frac{D}{T_K}\sim 10^6$.

Furthermore, we will discuss critically the crucial issue why the $E$-flow scheme can
sometimes even provide quantitatively reliable information for the strong coupling
regime although the RG equations are truncated in a perturbative manner. We will
explain why this issue is related to the complex nature of the flow parameter $E$
such that the stationary case is not related to any fixed point of the RG but
corresponds to some intermediate point in the RG flow where the solution is still
analytic in $E$. In contrast, the fixed points correspond to a flow parameter
$E^*=\pm\Omega-i\Gamma^*$, where $\Omega>0$ are the oscillation frequencies
and $\Gamma^*>0$ the relaxation/decoherence rates of the time evolution.

In the second part of the paper, we will apply the $E$-flow scheme
to the special case of the isotropic
spin-$\frac{1}{2}$ and $1$-channel Kondo model in nonequilibrium at zero
magnetic field. In contrast to Ref.~\onlinecite{pletyukhov_hs_PRL12},
we will consider the general case that both temperature and voltage are non-zero
(and not only one of these scales) and analyze the interplay between temperature
and voltage. We discuss situations where this interplay leads to a nonmonotonic
temperature-dependence of the conductance at fixed finite voltage, and compare
our results to recent experiments.
Furthermore, due to our improved scheme for the inital conditions, we will present
a new result for the universal coefficient $c_V^*$ and compare it to
the known result \eqref{eq:c_TV*}. Surprisingly, we find that the deviation in third order truncation
is only $\sim 1\%$, providing evidence that our solution for the nonlinear conductance
is reliable in the whole range of voltages.

This paper is organized as follows. In Sec.~\ref{sec:model}, we present the
generic model of a quantum dot in the Coulomb blockade regime and the special
case which is considered in more detail here, namely, the isotropic Kondo
model. In Sec.~\ref{sec:liouville_space}, we introduce the description of
the dynamics of the system in terms of superoperators in Liouville space, which
forms the basis of the RTRG. Section~\ref{sec:rg_formalism} describes the
$E$-flow scheme of the RTRG for the generic model.
Section~\ref{sec:E_flow_idea} explains the general idea of the
method, whereas readers who are interested in the technical details can find
a step-by-step derivation of the RG equations in
Secs.~\ref{sec:E_flow}--\ref{subsec:E_flow_current}.
Section~\ref{sec:rg_kondo} demonstrates how the $E$-flow scheme of the RTRG
can be applied to the isotropic Kondo model.
Section~\ref{sec:results} presents the results of our calculations and a
comparison with recent experiments. Finally, we summarize the most important
ideas and results of this paper in Sec.~\ref{sec:summary}. We use units
$e=k_B=\hbar=1$ throughout this paper.

\section{Model}
\label{sec:model}

We consider a system which consists of a quantum dot with fixed charge (Coulomb
blockade regime) and external non-interacting reservoirs. The quantum dot and
the reservoirs are coupled in such a way that spin and/or orbital fluctuations
can be induced on the dot. The total Hamiltonian of the system is
\begin{align}
H_{\text{tot}}=H+H_{\text{res}}+V.
\end{align}
The term
\begin{align}
H=\sum_s E_s\left|s\right\rangle\left\langle s\right|
\end{align}
is the part that corresponds to the isolated quantum dot with eigenstates
$\left|s\right\rangle$ and eigenvalues $E_s$.
\begin{align}
H_{\text{res}}&=\sum_{\alpha}H_\alpha,\\
H_\alpha&=\sum_{\sigma}\int
d\omega\left(\omega+\mu_\alpha\right)a_{+\alpha\sigma}(\omega)
a_{-\alpha\sigma}(\omega)
\end{align}
describes the reservoirs in continuum representation, where the operators
$a_{\eta\alpha\sigma}(\omega)$ are creators and annihilators (for $\eta=+$ and
$-$, respectively) for electrons with spin $\sigma$ in reservoir $\alpha$,
and $\omega$ is the energy relative to the chemical potential $\mu_\alpha$. We
will often use multiindices
\begin{align}
1&\equiv\eta_1\alpha_1\sigma_1\omega_1
\end{align}
to simplify the notation, and sum or integrate implicitly over indices which
appear twice in a term. If no ambiguities can occur, the index 1 will be left
out, e.g.,
\begin{align}
1&\equiv\eta\alpha\sigma\omega, & 1'&\equiv\eta'\alpha'\sigma'\omega'.
\end{align}
The reservoir operators fulfill the anticommutator relation
\begin{align}
\label{eq:anticomm_relation}
\left\{a_1,a_{1'}\right\}&=D(\omega)\delta_{1\bar1'},
\end{align}
where
\begin{align}
\label{eq:high_energy_cutoff}
D(\omega)&=\frac{D^2}{\omega^2+D^2}
\end{align}
is a dimensionless high-energy cutoff function for the leads with band width
$2D$, $\bar1$ is a shorthand notation for switching the index $\eta$, i.e.,
$\bar1\equiv-\eta,\alpha\sigma\omega$, and
\begin{align}
\delta_{11'}&=\delta_{\eta\eta'}\delta_{\alpha\alpha'}\delta_{\sigma\sigma'}
\delta(\omega-\omega').
\end{align}
We note that the DOS of lead $\alpha$ with spin $\sigma$ is given by
\begin{align}
\label{eq:d.o.s.}
\rho_{\alpha\sigma}(\omega)&=\rho_{\alpha\sigma}^{(0)}D(\omega),
\end{align}
where the constant $\rho_{\alpha\sigma}^{(0)}$ is absorbed in the field operators
such that the anticommutation relation \eqref{eq:anticomm_relation} is fulfilled.
Finally, the term
\begin{align}
V&=\frac12\sum_{\eta\eta'\alpha\alpha'\sigma\sigma'}\int d\omega\int d\omega'
\nonumber\\
&\quad
g_{\eta\alpha\sigma,\eta'\alpha'\sigma'}(\omega,\omega')
:a_{\eta\alpha\sigma}(\omega)a_{\eta'\alpha'\sigma'}(\omega'):\\
&=\frac12 g_{11'}:a_1 a_{1'}: \nonumber
\end{align}
describes the coupling between quantum dot and reservoirs, where
$g_{11'}=g_{\eta\alpha\sigma,\eta'\alpha'\sigma'}(\omega,\omega')$ is an
operator that induces spin and/or orbital fluctuations on the quantum dot, and
$:\ldots:$ denotes normal ordering of the reservoir operators. Note that
$g_{11'}$ can be non-zero only if $\eta=-\eta'$ because $V$ should not change
the charge on the quantum dot.

A special case of this generic model that will be examined more closely in this
paper is the isotropic spin-$\frac12$ and $1$-channel Kondo model with
spin-unpolarized leads. In this case, the coupling operator $g_{11'}$ takes the form
\begin{align}
\label{eq:g_Kondo}
g_{11'}=
\frac12
\begin{cases}
J^{(0)}_{\alpha\alpha'}\underline{S}\cdot\underline{\sigma}_{\sigma\sigma'}
& \text{for $\eta=-\eta'=+$,} \\
-J^{(0)}_{\alpha'\alpha}\underline{S}\cdot\underline{\sigma}_{\sigma'\sigma}
& \text{for $\eta=-\eta'=-$,}\end{cases}
\end{align}
where $\underline{S}$ is the spin-$\frac12$ operator on the quantum dot, and
$\underline{\sigma}$ is the vector of Pauli matrices.

The operator that corresponds to the electron current from reservoir $\gamma$ to
the quantum dot is
\begin{align}
I^\gamma&=\frac{d}{dt}N^\gamma=-i\left[H_{\text{tot}},N^\gamma\right]=-i\left[V,
N^\gamma\right],
\end{align}
where $N^\gamma$ is the number of electrons in reservoir $\gamma$. The current
operator can be written in the form
\begin{align}
I^\gamma&=\frac12 i^\gamma_{11'}:a_1a_{1'}:,
\end{align}
where
\begin{align}
i^\gamma_{11'}&=-2ic^\gamma_{11'}g_{11'}, \\
c^\gamma_{11'}&=-\frac12\left(\eta\delta_{\alpha\gamma}+\eta'\delta_{
\alpha'\gamma'}\right).
\end{align}
The current at time $t$ is given by
\begin{align}
\left\langle I^\gamma\right\rangle(t) =\text{Tr}\left[I^\gamma\rho_{\text{tot}}(t)\right],
\end{align}
where $\rho_{\text{tot}}(t)$ is the total density matrix of the system at time $t$.

\section{Representation in Liouville space}
\label{sec:liouville_space}
\subsection{Superoperators and Fourier transform}
Following the procedure described in Ref.~\onlinecite{RTRG_FS}, we introduce the
concept of superoperators in Liouville space, which act on ordinary operators
in Hilbert space. In particular, the Liouvillian $L_{\text{tot}}$ is the superoperator which,
when applied to an arbitrary operator $b$, yields the commutator of that
operator with the Hamiltonian of the system:
\begin{align}
L_{\text{tot}} b=[H_{\text{tot}},b].
\end{align}
It can be used to write a simple expression for the reduced density matrix
$\rho(t)$ of the quantum dot at time $t$, provided that the density matrix at
time $t_0$ can be factorized into an arbitrary dot part $\rho(t_0)$ and a
product $\rho_{\text{res}}=\prod_\alpha \rho_\alpha$ of grandcanonical density matrices for the
reservoirs:
\begin{align}
\rho(t)=\text{Tr}_{\text{res}}e^{-iL_{\text{tot}}(t-t_0)}\rho(t_0)\rho_{\text{res}}.
\end{align}
In this equation, $\text{Tr}_{\text{res}}$ denotes the trace over the reservoir
degrees of freedom only. Together with the trace over the quantum dot degrees
of freedom, denoted by $\text{Tr}$, it yields the total trace
\begin{align}
\text{Tr}_{\text{tot}}=\text{Tr}\text{Tr}_{\text{res}}.
\end{align}
In the same way as $L_{\text{tot}}$, a current superoperator $L_\gamma$ can be defined by
\begin{align}
L_\gamma b=\frac{i}{2}\left\{I^\gamma,b\right\}.
\end{align}
The current at time $t$ is then given by
\begin{align}
\left\langle I^\gamma\right\rangle(t)&=-i\text{Tr}\left[L_\gamma\rho_{\text{tot}}(t)\right]
\\
&=-i\text{Tr}\text{Tr}_{\text{res}}\left[L_\gamma
e^{-iL_{\text{tot}}(t-t_0)}\rho(t_0)\rho_{\text{res}}\right].
\end{align}
In the following, it will be convenient to use the Fourier transforms
(note that all functions are only defined for $t>t_0$ such that the Fourier transform
is identical to the Laplace transform, where $-iE$ denotes the Laplace variable. Our
definition is similar to the definition of the Fourier transform of retarded
response functions such that all nonanalytic features occur in the lower half of the
complex plane.)
\begin{align}
\rho(E)&=\int_{t_0}^\infty dt\,e^{iE(t-t_0)}\rho(t) \nonumber \\
&=\text{Tr}_{\text{res}}\frac{i}{E-L_{\text{tot}}}\rho(t_0)\rho_{\text{res}},
\label{eq:rho_S_Fourier}
\\
\left\langle I^\gamma\right\rangle(E)&=\int_{t_0}^\infty
dt\,e^{iE(t-t_0)}\left\langle I^\gamma\right\rangle(t) \nonumber \\
&=\text{Tr}\text{Tr}_{\text{res}}L_\gamma\frac{1}{E-L_{\text{tot}}}\rho(t_0)\rho_{\text{res}}.
\label{eq:IL_Fourier}
\end{align}
These can be used to calculate the stationary density matrix $\rho^{\text{st}}$ and
the stationary current $\left\langle I^\gamma\right\rangle^{\text{st}}$:
\begin{align}
\label{eq:stationary_density_matrix}
\rho^{\text{st}}&=-i\lim_{E\rightarrow i0^+}E\,\rho(E),\\
\left\langle I^\gamma\right\rangle^{\text{st}}&=-i\lim_{E\rightarrow i0^+}
E\,\left\langle I^\gamma\right\rangle(E).
\label{eq:stationary_current_1}
\end{align}

\subsection{Description in terms of the effective Liouvillian of the system}
Following the procedure described in Ref.~\onlinecite{RTRG_FS}, the Liouvillian
$L_{\text{tot}}=\left[H_{\text{tot}},\cdot\right]$ can be split into three parts,
\begin{align}
L_{\text{tot}}=L^{(0)}+L_{\text{res}}+L_V,
\end{align}
where each of these corresponds to one of the terms in the Hamiltonian
$H_{\text{tot}}=H+H_{\text{res}}+V$:
\begin{align}
\label{eq:L_definition}
L^{(0)}b&=\left[H,b\right], &
L_{\text{res}}b&=\left[H_{\text{res}},b\right], &
L_Vb&=\left[V,b\right].
\end{align}
Using the bare quantum dot superoperator $G^{(0)pp'}_{11'}$, called vertex, and the lead
superoperator $J^p_1$, which are defined by their action on an arbitrary
operator $b$,
\begin{align}
\label{eq:G_definition}
G^{(0)pp'}_{11'}b&=\delta_{pp'}
\begin{cases}
g_{11'}b & \text{for $p=+$,} \\
-bg_{11'} & \text{for $p=-$,}
\end{cases} \\
J^p_1b&=
\begin{cases}
a_1b & \text{for $p=+$,} \\
ba_1 & \text{for $p=-$,}
\end{cases}
\end{align}
the coupling superoperator can be written as
\begin{align}
\label{eq:L_V}
L_V=\frac{1}{2}p' G^{(0)pp'}_{11'}:J^p_1 J^{p'}_{1'}:.
\end{align}
In principle, it would be possible to also include the Keldysh indices $p$ and
$p'$ in the multiindices $1$ and $1'$. However, it will be shown later that
only the sum of the vertex over the Keldysh indices, i.e.,
\begin{align}
\label{eq:vertex_averaged}
G^{(0)}_{11'}=\sum_{pp'}G^{(0)pp'}_{11'},
\end{align}
remains in the final RG equations (in renormalized form). Therefore, it is more convenient to treat the
Keldysh indices separately for the time being.

In analogy to the vertex $G^{(0)pp'}_{11'}$, we define a bare current vertex
$I^{\gamma(0) pp'}_{11'}$ by
\begin{align}
\label{eq:current_vertex}
I^{\gamma(0) pp'}_{11'}&=\delta_{pp'}p\,c^\gamma_{11'}G^{(0)pp'}_{11'}.
\end{align}
It enables us to find the representation
\begin{align}
L_\gamma=\frac12 p' I^{\gamma(0) pp'}_{11'} :J^p_1 J^{p'}_{1'}:
\end{align}
of $L_\gamma$ [cf. Eq.~\eqref{eq:L_V}].

We expand Eqs.~\eqref{eq:rho_S_Fourier} and~\eqref{eq:IL_Fourier} in $L_V$,
perform the trace $\text{Tr}_{\text{res}}$ over the reservoir degrees of freedom, apply
Wick's theorem w.r.t. the reservoir degrees of freedom,
and define the irreducible kernel $\Sigma(E)$, which is the sum
of all diagrams that are connected by reservoir contractions (see
Ref.~\onlinecite{RTRG_FS} for details). A contraction between two vertices
corresponds to the term
\begin{align}
\label{eq:contraction}
\gamma_{11'}^{pp'}(\omega,\omega')=\delta_{1\bar{1}'}\delta(\bar{\omega}+\bar{
\omega}') \gamma^{p'}(\bar{\omega}),
\end{align}
where $\bar\omega$ is a shorthand notation for $\bar\omega=\eta\omega$,
\begin{align}
\label{eq:gamma_p}
\gamma^{p'}(\bar{\omega})=p'f(p'\bar{\omega})D(\omega),
\end{align}
and
\begin{align}
f(\omega)=\frac{1}{e^{\omega/T}+1}
\end{align}
denotes the Fermi function at temperature $T$. Analogously, the irreducible
current kernel $\Sigma_\gamma(E)$ is the sum of all connected diagrams where the
first vertex $G^{(0)}$ is replaced by a current vertex $I^{\gamma(0)}$.

We define an effective Liouvillian $L(E)$ of the system, which contains all
effects that are due to the coupling to the reservoirs, by
\begin{align}
L(E)=L^{(0)}+\Sigma(E).
\end{align}
This permits us to rewrite the reduced density matrix~\eqref{eq:rho_S_Fourier} and
the current~\eqref{eq:IL_Fourier} in a form where the reservoirs do not
appear explicitly:
\begin{align}
\rho(E)&=\frac{i}{E-L(E)}\rho(t_0),
\label{eq:density_matrix_Fourier}
\\
\left\langle I^\gamma\right\rangle(E)&=
\text{Tr}\Sigma_\gamma(E) \frac{1}{E-L(E)}\rho(t_0).
\label{eq:current_Fourier}
\end{align}
Defining the Liouvillian and the current kernel in time space by the inverse
of the Fourier transform, $L(E)=\int_0^\infty dt e^{iEt}L(t)$ and
$\Sigma_\gamma(E)=\int_0^\infty dt e^{iEt}\Sigma_\gamma(t)$, we can write
\eqref{eq:density_matrix_Fourier} and \eqref{eq:current_Fourier} in
time space as
\begin{align}
i\dot\rho(t)&=\int_{t_0}^t dt' L(t-t')\rho(t'),
\label{eq:density_matrix_time}
\\
\left\langle I^\gamma\right\rangle(t)&=
-i\text{Tr}\int_{t_0}^t \Sigma_\gamma(t-t') \rho(t').
\label{eq:current_time}
\end{align}
The formal analogy of Eq.~\eqref{eq:density_matrix_time} to the von Neumann
equation demonstrates most clearly that $L(t)$ is an effective Liouvillian
containing memory effects.
Since $L(t),\Sigma_\gamma(t)\sim \theta(t)$ are retarded response functions,
i.e., only defined for positive times $t>0$, $L(E)$ and $\Sigma_\gamma(E)$
are analytic functions in the upper half of the complex plane and the
usual Kramers-Kronig relations hold.

Applying the inverse Fourier transform to \eqref{eq:density_matrix_Fourier}
and \eqref{eq:current_Fourier} we obtain an explicit formula for
the time evolution for $t>t_0$
\begin{align}
\label{eq:time_rho}
\rho(t)&=\frac{i}{2\pi}\int_{-\infty+i0^+}^{\infty+i0^+}\!\!\!\!\!\!\!\!dE\,
\frac{e^{-iE(t-t_0)}}{E-L(E)}\,\rho(t_0),\\
\label{eq:time_current}
\left\langle I^\gamma\right\rangle(t)&=
\frac{1}{2\pi}\text{Tr}\int_{-\infty+i0^+}^{\infty+i0^+}\!\!\!\!\!\!\!\!dE\,
\Sigma_\gamma(E) \frac{e^{-iE(t-t_0)}}{E-L(E)}\rho(t_0),
\end{align}
where the contour of integration is slightly above the real axis to ensure
convergence. Closing the integration contour in the lower half of the complex plane
we see that the individual terms of the time evolution follow from enclosing
the poles and branch cuts of the resolvent $1/[E-L(E)]$ and the current kernel
$\Sigma_\gamma(E)$. The stationary solution follows from
Eqs.~\eqref{eq:stationary_density_matrix} and~\eqref{eq:stationary_current_1}:
\begin{align}
\label{eq:stationary_rho}
L(i0^+)\rho^{\text{st}}&=0,\\
\label{eq:stationary_current}
\left\langle I^\gamma\right\rangle^{\text{st}}&=
-i\text{Tr}\Sigma_\gamma(i0^+)\rho^{\text{st}}.
\end{align}

The remaining challenge is to find a way to calculate the irreducible kernels
$\Sigma(E)$ [or, equivalently, $L(E)$] and $\Sigma_\gamma(E)$.

\section{RG formalism}
\label{sec:rg_formalism}
In this section, we will discuss how the effective Liouvillian $L(E)$ of the
system can be evaluated using a real-time renormalization group (RTRG) approach.
In contrast to the approach presented in Ref.~\onlinecite{RTRG_FS}, where a
cutoff was defined by cutting off the Matsubara poles of the Fermi functions, we
use an alternative flow scheme which uses the Fourier variable $E$ as the
flow parameter, which was proposed in Ref.~\onlinecite{pletyukhov_hs_PRL12}.
The new approach is called the $E$-flow scheme in the following.

Here, we derive the $E$-flow RG equations for the case where only
fermionic two-point vertices $G^{p_1p_2}_{12}$ are present in the bare perturbation
theory and  describe either spin, orbital or potential fluctuations.
Moreover, we assume that the bare vertices do not depend on the frequency
variables $\bar{\omega}_i=\eta_i\omega_i$.

As already summarized in Sec.~\ref{sec:scope}, the $E$-flow scheme
is technically very different compared to the Matsubara RG scheme described in
Ref.~\onlinecite{RTRG_FS}. Therefore, the detailed description of the $E$-flow scheme
in this Section does not rely on the Matsubara scheme and we will only use the
diagrammatic rules of the perturbative expansion in terms of the bare vertices as
starting point, as described in Ref.~\onlinecite{RTRG_FS}.

Before entering the technical details on how to determine RG equations within
the $E$-flow scheme from the specific diagrammatic rules, we will first motivate
what the idea of the $E$-flow scheme of RTRG is and why it is the most appropriate
choice for the determination of the time evolution.

\subsection{The idea of the $E$-flow scheme of RTRG}
\label{sec:E_flow_idea}
\begin{figure}[htbp!]
  \includegraphics[width=\linewidth]{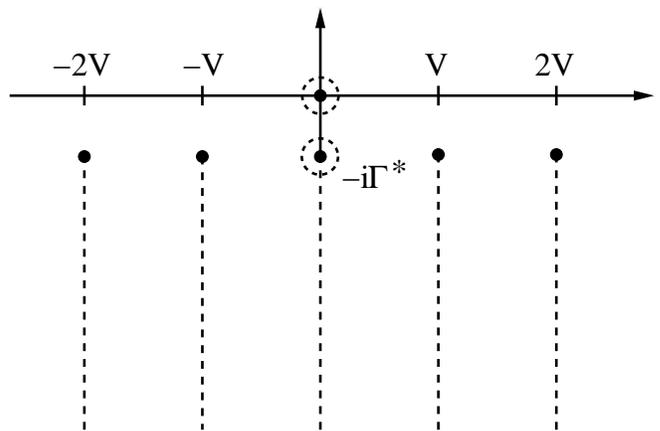}
  \caption{The analytic structure of the resolvent $1/[E-L(E)]$ for the isotropic
Kondo model at zero temperature and finite voltage $V$. There are two pole positions
at $E=0$ and $E=-i\Gamma^*$ together with branch cuts of $L(E)$ starting at
$E=-i\Gamma^*+nV$ with some integer $n$.}
\label{fig:branchcuts_kondo}
\end{figure}

For small couplings between the quantum dot and the reservoirs, the most obvious
choice to calculate the effective Liouvillian $L(E)$ is to use a perturbative
expansion in terms of the bare vertices $G_{11'}^{(0)pp'}$. These vertices
are dimensionless and we denote their order of magnitude by $O(G)$. The
expansion of $L(E)$ can then formally be written as
\begin{align}
\label{eq:perturbative_expansion}
L(E)&=L^{(0)}+L^{(2)}(E)+L^{(3)}(E)+\dots\,
\end{align}
where $L^{(n)}(E)\sim O(G^n)$ denotes the Liouvillian in order $n$, $L^{(0)}=L$ is the
bare Liouvillian, and the term with $n=1$ is missing due to normal-ordering. For the
current kernel $\Sigma_\gamma(E)$ an analogous expansion holds but also the lowest order
term $n=0$ is absent.
The problem with the series \eqref{eq:perturbative_expansion} is that, for $n\geq 2$,
the internal frequency integrations can be logarithmically divergent at large
energies for $D\rightarrow\infty$. In order $n$, the divergencies occur in the form
$\ln^k\left(\frac{D}{\text{max}\{|E|,\Delta\}}\right)$, where $\Delta=T,V,\dots$
is some physical energy scale (except $E$), and $k\leq n-1$. From the perturbative
series it can be seen that the Fourier variable $E$ occurs always in linear
combination with the internal frequencies in the form
$E+\sum_i\bar{\omega}_i$, i.e., the imaginary part of the Fourier variable always acts as a high-energy
cutoff. Thus, for $|E|$ larger than any other physical energy scale, {\it all}
logarithmic divergencies occur in the form $\ln^k\left(\frac{D}{-iE}\right)$.
By convention, we have chosen $-iE$ in the argument of the logarithm such that,
for the natural choice of the logarithm, all branch cuts point into the direction
of the negative imaginary axis. This leads to exponentially decaying integrands
for the integrals around the branch cuts to obtain the time evolution from
Eq.~\eqref{eq:time_rho}. Concerning the precise position of the branching points
it can be shown \cite{Pletyukhov_Schuricht_Schoeller_PRL10,kashuba_schoeller_PRB13}
that they are given by the non-zero poles $z_i^\pm=\pm\Omega_i-i\Gamma_i$ of the
resolvent $1/[E-L(E)]$ in the lower half of the complex plane ($\Omega_i,\Gamma_i>0$),
shifted by multiples of the voltage, i.e., generically the branching points appear
at $z_i^\pm+mV$ with some integer $m$. In Fig.~\ref{fig:branchcuts_kondo}, we show
the position of the poles and the branch cuts of the resolvent $1/[E-L(E)]$ for
the specific example of the isotropic Kondo model considered in this paper, where
the non-zero pole of the resolvent is given by $-i\Gamma^*$ with
the spin relaxation rate $\Gamma^*$. At finite temperature, it turns out that all branch
cuts are replaced by an infinite number of poles separated by the Matsubara frequencies.

In order to get rid of the divergencies, the idea of the $E$-flow scheme of RTRG
is not to consider an expansion of the effective Liouvillian $L(E)$
in terms of the bare vertices $G_{11'}^{(0)pp'}$ but to consider an expansion of the
{\it second derivative} $\frac{\partial^2}{\partial E^2}L(E)$ of the effective Liouvillian
in terms of {\it effective} vertices $G_{11'}^{pp'}(E_X)$. The latter quantities are
defined as the sum over all connected diagrams with two outgoing reservoir lines.
The quantities $X\equiv 12\dots n$, containing all possible sets of indices,
determine a shift of the Fourier variable by linear combinations of the chemical
potentials of the leads via $E_{12\dots n}=E+\sum_{i=1}^n \eta_i\mu_{\alpha_i}$.
As shown below this expansion can be achieved by a unique resummation of certain
subclasses of diagrams which has also the effect that only the full effective
Liouvillian $L(E_X)$ occurs in this series. Most importantly, we will show that if
the second derivative is taken for the effective Liouvillian, the resulting
series does no longer contain any logarithmic divergence at high energies,
such that we can take the limit $D\rightarrow\infty$ to calculate the frequency
integrals of any diagram in any order of the effective vertices. The same can be
shown to hold for the {\it first derivative} of the effective vertices such that the
RG equations within the $E$-flow scheme can be symbolically written as
\begin{align}
\label{eq:Liouvillian_E_flow_generic}
\frac{\partial^2}{\partial E^2}L(E)&={\cal{F}}_L\left\{L(E_X),G(E_{X'})\right\},\\
\label{eq:vertex_E_flow_generic}
\frac{\partial}{\partial E}G(E)&={\cal{F}}_G\left\{L(E_X),G(E_{X'})\right\},
\end{align}
where ${\cal{F}}_{L/G}$ denote some functionals which have to be determined from the
diagrammatic rules, see the next section. We note that the RG equations involve only
the two-point vertex $G_{11'}^{pp'}(E_X)$, whereas in Ref.~\onlinecite{pletyukhov_hs_PRL12}
a set of coupled RG equations for all $n$-point vertices occur. Moreover, as we will see
in the next section, it can be shown that the right-hand side of the RG equations can
be rewritten as a well-defined power series in terms of the {\it frequency-independent}
two-point vertex only (i.e. the index $1$ no longer involves the frequency). This simplifies
the analysis of the RG equations in higher order truncation schemes beyond third order.

Similar RG equations can be set up to
calculate the effective current kernel and vertex. Since the limit
$D\rightarrow\infty$ can be taken, these RG equations are universal, i.e., are
independent of the specific choice of the high-energy cutoff function
\eqref{eq:high_energy_cutoff}. If this limit is taken, the RG equations are only
valid in the regime $|E|\ll D$ and a corresponding initial condition has to be set
up in this regime.
In principle, it is also possible to include the high-energy cutoff function on
the right-hand side of the RG equations \eqref{eq:Liouvillian_E_flow_generic} and
\eqref{eq:vertex_E_flow_generic},
such that the RG equations are valid for all values of $E$ in the complex plane and
can include a specific microscopic choice to describe the physics at high energies.
In this case, the initial conditions of the RG equations at $E=i\Lambda_0$ with
$\Lambda_0\gg D$ are just given by the bare values
of the Liouvillian and the vertices. However, the advantage of the $E$-flow scheme
is that the scaling limit can be built in directly such that the limit
$D\rightarrow\infty$ can be performed from the very beginning before solving the
RG equations, and only the universal part of the solution is obtained. To achieve
this, we need to find an appropriate initial condition for $\Lambda_0\ll D$. In
this regime and neglecting terms of $O(1/D)$, the bare perturbation series will
contain the band width $D$ only within the logarithmic terms
$\ln^k\left(\frac{D}{-iE}\right)$. All these logarithmic terms are generated
by the universal RG equations, if $E=iD$ is used as initial value.
Therefore, in order to set up universal initial conditions, we set $E=iD$ in the bare
perturbation series {\it after} having neglected all terms of $O\left(\frac{|E|
}{D}\right)$, in order
to remove all logarithmic and nonuniversal terms in the initial condition. Furthermore,
$D$ is taken much larger than all other physical energy scales to avoid nonuniversal
terms of $O(\Delta/D)$. In addition, we consider only those lowest order terms of
the perturbative expansion which are universal, i.e., independent of the choice of
the high-energy cutoff function. If the lowest-order term is non-universal we take zero
for the initial condition. As a consequence of
this procedure, only the universal part of the solution is picked out and
the band width $D$ enters only as initial value $E=iD$ of the Fourier variabe but does
not appear explicitly in the initial value of the Liouvillian or the vertices.
Together with the initial values of the vertices, the band width $D$ will finally
enter into some characteristic non-universal low-energy scale $T_K$ of the
problem, like the Kondo temperature for the Kondo problem. Once this scale is defined,
the scaling limit \eqref{eq:scaling_limit} is defined such that this scale stays
constant in the limit where the
band width $D\rightarrow\infty$ and the bare couplings are sent to zero (such that only
the lowest order terms of the perturbative expansion dominate the initial condition).

The prescribed way to determine the initial condition at $E=iD$ works very well for the
initial condition of all dimensionless quantities, like the vertices and the first
derivative $\frac{\partial}{\partial E}L(E)$ of the effective Liouvillian. However,
for the effective Liouvillian $L(E)$ itself a problem occurs since it contains
terms which are proportional to $E$. As proposed in
Ref.~\onlinecite{kashuba_schoeller_PRB13}, the effective Liouvillian can be
decomposed in two terms
\begin{align}
\label{eq:L_decomposition}
L(E)&=L_\Delta(E) + E L'(E),
\end{align}
where $L_\Delta(E)\sim\Delta$ contains all terms proportional to some physical
scale except $E$, and $E L'(E)$ contains all terms proportional to $E$. The quantities
$L_\Delta(E)$ and $L'(E)$ are slowly varying logarithmic functions in $E$, and the
above procedure to determine the initial condition at $E=iD$ can be applied to them.
However, setting $E=iD$ in \eqref{eq:L_decomposition} leads to a term $iDL'(iD)$
proportional to $D$ itself. Furthermore, it turns out that the coefficient in front
of this term is non-universal for the Kondo problem. Neglecting this term in the
initial condition would lead to a large error since the term diverges linearly in $D$.
Therefore, we have to find a different way to set up the initial condition for $L(E)$.
One way is to set up directly RG equations for the quantities $L_\Delta(E)$ and
$L'(E)$ as proposed in Ref.~\onlinecite{kashuba_schoeller_PRB13}, which can be used
very effectively for a generic weak-coupling solution of the RG equations.
\cite{goettel_etal_13}
However, the decomposition \eqref{eq:L_decomposition} is not unique and some
ambiguity is left to describe problems in strong coupling. Therefore, for the
Kondo problem, we choose here a different strategy by first solving the RG equations,
when all other physical scales $\Delta=T,V,\dots$ are set to zero, and
starting the RG flow at $E=0$. This point corresponds to the stationary case, and
it is known exactly that the conductance is unitary at this point. This boundary
condition is used as an input to fix the unknown initial condition of the Liouvillian.
The RG flow is then first solved for $\Delta=0$ starting from $E=0$ up to $E=iD$
and the result is used as initial condition for the RG flow at finite $\Delta\ll D$.

Since the RG equations involve the Liouvillian and the two-point vertex at the shifted
variables $E_X=E+\bar{\mu}_{1\dots n}$, an initial condition is needed for all these values.
In Ref.~\onlinecite{pletyukhov_hs_PRL12}, the same initial condition has
been taken at all these points but, for the Kondo model, it turns out that for this choice
the solution of the RG equations in the low energy regime $|E|\ll T_K$ is not
independent of the initial value $E=iD$ even if $D$ differs by many orders of magnitude
from the physical scales $\Delta\sim T_K,T,V$. The problem is that there is an
instability of the low-energy solution against exponentially small changes (of the order
of $T_K$) of the Liouvillian at high energies. Therefore, the relative difference between $L(E)$
and $L(E+nV)$, with $n\ne 0$, is important and cannot be neglected for large $E$ at fixed
voltage $V$. In this paper, we will solve this problem by solving the RG equations at
$T=V=0$ from $E=0$ up to $E=iD$ and, subsequently, from $E=iD$ to $E=iD+nV$, providing
different initial conditions for all quantities at the shifted variables. Using this
procedure, one finds that the scaling limit is achieved already for values of the
exchange coupling of the order of $J_0\sim 0.04$, i.e., by using Eq.~\eqref{eq:TK_pms},
for $\frac{D}{T_K}\sim 10^6$.

The $E$-flow scheme is a new concept in RG methods, since
it uses a complex flow parameter. This allows the solution of the RG equations
along an arbitrary path in the complex plane and all effective quantities can
be analytically continued from the upper to the lower half of the complex plane.
Only if a branching point is encircled the solution does not return to the
same value. Thus, even numerically one can determine the precise position of all
branching points and can fix the shape of the branch cuts in a convenient
way. To calculate the time evolution it is not necessary to calculate
the integrals in Eqs.~\eqref{eq:time_rho} and \eqref{eq:time_current} along the
real axis which is numerically not very convenient due to strongly oscillating
integrands. Choosing the shape of the branch cuts along the negative imaginary
axis starting from a branching point/pole of the resolvent $1/[E-L(E)]$ at position
$z_B=z_i^\sigma + mV$, one can close the integration contour of \eqref{eq:time_rho} and
\eqref{eq:time_current} in the lower half of the
complex plane and can address each individual term of the time evolution separately
by calculating the integration around each individual branch cut. This requires
the knowledge of the effective Liouvillian for $z=z_B -ix\pm 0^+$, with $x>0$,
which can be determined by solving the RG equations along the path
\begin{align}
\label{eq:rg_path}
E=z_B+i\Lambda\pm 0^+,
\end{align}
starting at $\Lambda\sim D$ down to $\Lambda=-\infty$. Using \eqref{eq:time_rho} for
$\rho(t)$, the branch cut integral leads to a term
\begin{align}
\label{eq:form_time}
F(t)\,e^{-iz_B t}
\end{align}
for the time evolution, where the position of the branching point/pole determines
the exponential and $F(t)$ is a pre-exponential function given by
\begin{align}
\nonumber
F(t)&=\frac{1}{2\pi}\int_0^\infty dx\,e^{-x(t-t_0)}
\left(\frac{1}{z_B-ix-L(z_B-ix+0^+)}\right.\\
\label{eq:preexponential}
&\left.-\frac{1}{z_B-ix-L(z_B-ix-0^+)}\right)\rho(t_0).
\end{align}
A similar equation holds for the time evolution for the current by using
Eq.~\eqref{eq:time_current}. Due to the exponentially
decaying integrand, the long time behavior of $F(t)$ can be determined by analyzing the
scaling behavior of the Liouvillian close to the branching point $z_B$.
\cite{kashuba_schoeller_PRB13}
Since each term of the time evolution has a different
oscillation frequency and a different decay rate due to different positions of the
branching points, it is very hard to distinguish the different terms if a method
is used which can only calculate the sum of all terms. Thus, the $E$-flow scheme
is a very natural and effective scheme for a systematic determination of the time
dynamics for problems in dissipative quantum mechanics.

Within the $E$-flow scheme, also the notion of fixed points of the RG flow has to be
generalized. In conventional RG methods, the flow parameter is a real cutoff $\Lambda$
and the fixed points are defined as those points where the RG flow of all quantities
stops for $\Lambda\rightarrow 0$. Within the $E$-flow scheme, there is no unique path
for the flow parameter. For each given set of initial conditions, there is a certain
set of branching points $z_B$ in the lower half of the complex plane where the RG flow
stops. Thus, if the RG flow is solved along the path $z_B+i\Lambda$, the fixed point
is defined as the value of all quantities which is obtained for $\Lambda\rightarrow 0$.
This means that the fixed point itself is associated with $z_B$ such that $z_B$ can
be equivalently called a fixed point. If the initial conditions are changed then
also the position of the branching points can change, i.e., it makes no sense in
general to associate several fixed points with a single branching point. Therefore,
in the following, we will denote the branching points as fixed points of the RG.
As already mentioned above, the scaling behavior around these fixed points determines
the long-time behavior of pre-exponential functions for the time evolution.

The stationary solution requires only the
knowledge of the effective Liouvillian (or the effective current kernel)
close to $E=0$, see Eqs.~\eqref{eq:stationary_rho} and \eqref{eq:stationary_current}.
Around this point the effective Liouvillian is analytic for the isotropic $1$-channel
Kondo model, where the branching points are located at $-i\Gamma^*+nV$. Therefore, in
contrast to other RG methods, the RG flow is still sufficiently away from the
fixed points, and the expansion in $E$, $T$, or $V$ is analytic
around this point. This is the reason why even for the strong coupling case
$T,V\lesssim T_K$, there is some hope that the stationary conductance
$G(T,V)$ can be quite close to the exact value even if the RG equations are
truncated perturbatively in the effective vertices. Although this truncation
is not controlled in a strict mathematical sense since the effective vertices
are still of order $1/3$ at $E=0$ and $T,V\lesssim T_K$, one can check the reliability
of the method by comparing the results in second
and third order truncation. Despite the fact that it cannot be anticipated whether
the result will converge by increasing the truncation order, a nearly identical
result in second and third order gives some hint that an asymptotic series may be
present leading to a very good result already in a low-order truncation. As we will
see, this is indeed the case for the isotropic $1$-channel Kondo model, where we can
additionally check the quality of our results by comparing with the temperature
dependence of the conductance at zero voltage obtained from numerically exact
NRG calculations.

Close to the branching points $z_B$, the situation is very different. Here, the
Liouvillian is non-analytic and no finite-order truncation scheme will lead to
a reliable result in the strong coupling case, where the vertices are of $O(1)$
close to the branching point. This can indeed be checked for the Kondo model at
$T=V=0$, where completely different results are obtained in second and
third order truncation for the time evolution.\cite{pletyukhov_hs_PRL12}
The same holds for quantum-critical models, like the $2$-channel Kondo model, where
the branching point starts at $E=0$, such that even stationary quantities cannot
be calculated by a perturbative truncation scheme.
Only for weak-coupling problems, where the effective vertices stay small close
to the branching points, a truncation in finite order is controlled.
For such models, the $E$-flow scheme is a useful method to resum systematically
all powers of logarithmic terms $g\ln(T_K t)$ in leading or sub-leading order to
calculate the long-time behavior, where $g$ is some small dimensionless coupling
constant. This has been demonstrated recently for the Ohmic spin-boson model,
\cite{kashuba_schoeller_PRB13}, where different power-law exponents have been
obtained for the time evolution compared to previous results. To the best of our knowledge,
this is not possible within any other RG method at the moment.

As already noted in Ref.~\onlinecite{kashuba_schoeller_PRB13}, the values of the
effective vertices close to the branching points can be of order $O(1)$ even if
they are small at the stationary point $E=0$. This is the case, e.g., for the $1$-channel
Kondo model at large bias voltages compared to the Kondo temperature.
Thus, weak-coupling problems for stationary quantities can turn into strong-coupling
ones concerning the long-time evolution. The physical reason is that the long-time
dynamics is not cut off by any decay rate $\Gamma_i$ since the cutoff parameter of
the RG flow determining that term of the time evolution associated with the
branching point $z_B$ is given by the linear combination $|E-z_B|\sim 1/t$ which
tends to zero for $t\rightarrow\infty$.

\subsection{RG equations for the effective Liouvillian and the effective vertex}
\label{sec:E_flow}

We will now derive the basic RG equations \eqref{eq:Liouvillian_E_flow_generic} and
\eqref{eq:vertex_E_flow_generic} for the effective Liouvillian and the
effective vertices. We start from the diagrammatic representation of the effective
Liouvillian in terms of the bare vertices as derived in Ref.~\onlinecite{RTRG_FS}.
Each diagram consists of a sequence of vertices connected by dot propagators
denoted by the resolvents
\begin{align}
\label{eq:resolvent}
R(E)&=\frac{1}{E-L(E)},
\end{align}
where we have already resummed all self-energy insertions to obtain the full
effective dot propagator. The reservoir field operators of the vertices are
connected by reservoir contractions $\gamma\equiv\gamma^{pp'}_{11'}(\omega,\omega')$
as defined in Eq.~\eqref{eq:contraction}. The diagrammatic series can be written as
\begin{align}
\nonumber
L(E)\,&=\,L^{(0)}\,+\,\sum_{m=2}^\infty\,\,\sum_{\text{diagrams}}\,
\frac{(-1)^{N_p}}{S}\,\left(\prod \gamma\right)_{\text{irr}}\\
\label{eq:diagrammatic_series}
&\hspace{-1cm}\times G^{(0)}\,R_{X_1}(E)\dots G^{(0)}\,R_{X_{m-1}}(E)\,G^{(0)}\,.
\end{align}
Here, $L^{(0)}$ and $G^{(0)}\equiv G^{(0)pp'}_{11'}$ are the bare Liouvillian
and the bare vertices as defined in Eqs.~\eqref{eq:L_definition} and
\eqref{eq:G_definition}. The resolvents
\begin{align}
\label{eq:resolvent_shift}
R_X(E)&=R(E_X+\bar{\omega}_X),
\end{align}
with
\begin{align}
\label{eq:shift}
E_X&=E+\bar{\mu}_X,\\
\bar{\mu}_{X}&=\sum_{j\in X}\eta_j\mu_{\alpha_j},&
\bar{\omega}_{X}&=\sum_{j\in X}\eta_j\omega_j,
\end{align}
are determined by the set of indices $X$ which are associated with the reservoir lines
crossing over the corresponding resolvent, where each index is taken from the
vertex connected to this line and standing left to the resolvent. $N_p$ is the
number of crossings of reservoir lines and $S=\prod_k m_k!$ is a symmetry factor
arising if two vertices are connected by $m_k$ reservoir lines.
$\left(\prod \gamma\right)_{\text{irr}}$ denotes the product over all
reservoir contractions, where the subindex $\lq\lq \text{irr}\lq\lq$ means that
only connected diagrams without any self-energy insertions are allowed. Using
Eq.~\eqref{eq:contraction}, we see that only pairs of indices $(1,\bar{1})$ can be
connected by reservoir lines. Thus, the lowest order diagram for the effective
Liouvillian reads
\begin{multline}
\label{eq:L_lowest}
\raisebox{-1.5em}{
\includegraphics[scale=0.45]{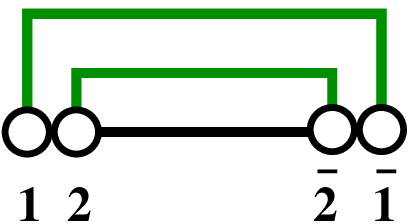}
}
=\frac{1}{2!}\gamma^{\bar{p}_1}(\bar{\omega}_1)\gamma^{\bar{p}_2}(\bar{\omega}_2)
\\
\times
G^{(0)}_{12} R_{12}(E) G^{(0)\bar{p}_2 \bar{p}_1}_{\bar{2}\bar{1}}\quad,
\end{multline}
where $G^{(0)}_{12}=\sum_{p_1p_2}G^{(0)p_1p_2}_{12}$ denotes the bare vertex averaged
over the Keldysh indices. Implicitly we sum always over all indices and integrate over
all frequencies $\bar{\omega}_i$.

Similar to the effective Liouvillian, one can also set up a diagrammatic series
for the effective vertex $G^{p_1\dots p_n}_{1\dots n}(E)$, which is defined by
the sum of all connected diagrams with $n$ free reservoir lines with indices
$1\dots n$ and Keldysh indices $p_1\dots p_n$. In the following, we will call
these objects $n$-point vertices. Since we consider here only bare
vertices with $n=2$, the effective vertices must have an even number of external lines.
The diagrammatic series for $G^{p_1\dots p_n}_{1\dots n}(E)$ is exactly the same as
for the effective Liouvillian $L(E)$ (which can be considered as a zero-point
vertex) with the following additional rules:

(i) By convention, all reservoir lines are directed to the right, and the corresponding
frequencies and chemical potentials of the external lines have to be included in the
resolvents.

(ii) If the sequence of the external indices from left to right is given
by $P_1\dots P_n$, where $P$ is any permuation of $1\dots n$, the diagram
gets a factor $(-1)^P$, i.e., a minus sign for an odd permutation. This minus
sign accounts correctly for the minus sign from crossings of external lines if
an effective vertex is used in a certain diagram instead of a bare vertex.

(iii) If the external lines are associated with different vertices, one has to sum
over all permutations of the external lines. If two external lines are associated with
the same vertex, only one sequence of the indices has to be considered.

(iv) The external vertices are normal-ordered, i.e., if an effective vertex is used
instead of a bare one in a certain diagram it is not allowed to connect the
effective vertex with itself.

These rules give, e.g., for the second order diagrams for the effective two-point vertex
\begin{multline}
\label{eq:G_lowest}
G^{p_1p_2}_{12}(E)-G^{(0)p_1p_2}_{12}=
\raisebox{-1.5em}{
\includegraphics[scale=0.45]{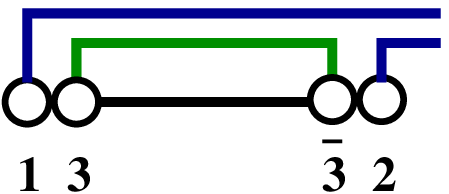}
}
-(1\leftrightarrow 2)
\\
=\gamma^{\bar{p}_3}(\bar{\omega}_3)
G^{(0)p_1p_3}_{13} R_{13}(E) G^{(0)\bar{p}_3 p_2}_{\bar{3}2}
-(1\rightarrow 2).
\end{multline}
Note that we integrate only over the frequency variable $\bar{\omega}_3$ but
not over the external ones $\bar{\omega}_{1/2}$. If we want to exhibit the
frequency dependence of the effective vertices explicitly we will also use
the representation
\begin{align}
\label{eq:omega_explicit}
&G^{p_1\dots p_n}_{1\dots n}(E)\rightarrow
G^{p_1\dots p_n}_{1\dots n}(E;\bar{\omega}_1,\dots,\bar{\omega}_n),
\end{align}
where on the right-hand side the indices $i\equiv \eta_i\alpha_i\sigma_i$ do no longer
contain the frequency variable. Furthermore, when omitting the Keldysh indices,
we define by $G_{1\dots n}(E)$ the $n$-point vertex averaged over the Keldysh indices.

Once the effective vertices are defined, one can use them instead of bare ones
in the diagrammatic series by resumming subclasses of connected diagrams.
According to rule (i), within a certain diagram the energy argument of the
effective vertex has to be chosen identical to the one of the resolvent
standing left to this vertex, i.e., the combination
\begin{align}
\label{eq:G_rule}
R(E_X+\bar{\omega}_X)G^{p_1\dots p_n}_{1\dots n}(E_X+\bar{\omega}_X)
\end{align}
will occur. If the first vertex from the left in a diagram is replaced
by an effective one it has the energy argument $E$. For example, replacing both vertices
in Eq.~\eqref{eq:L_lowest} by effective ones, we obtain the expression
\begin{multline}
\label{eq:L_lowest_effective}
\frac{1}{2!}\gamma^{\bar{p}_1}(\bar{\omega}_1)\gamma^{\bar{p}_2}(\bar{\omega}_2)
\\
\times
G_{12}(E) R_{12}(E)
G^{\bar{p}_2 \bar{p}_1}_{\bar{2}\bar{1}}(E_{12}+\bar{\omega}_{12}) \quad.
\end{multline}
However, we note that because of double-counting, it is not possible to replace all bare vertices
by effective ones in the diagrammatic series and omitting certain diagrams. For example, when inserting
Eq.~\eqref{eq:G_lowest} for the two two-point vertices into Eq.~\eqref{eq:L_lowest_effective}, we find a double
counting of third order diagrams for the effective Liouvillian. The same happens for the
diagram \eqref{eq:G_lowest} of the two-point vertex when we replace the two bare vertices by
effective ones. Only for all $n$-point vertices with $n>2$ a straightforward inspection shows
that all diagrams can be resummed in a unique way such that only two-point vertices remain.
Furthermore, as will be explained below, after this resummation all internal frequency integrations
are well-defined in the limit $D\rightarrow\infty$. This is the reason why we need a
reformulation of the diagrammatic series in terms of the RG equations \eqref{eq:Liouvillian_E_flow_generic}
and \eqref{eq:vertex_E_flow_generic} only for the effective Liouvillian and
the two-point vertices.

Using similar proofs as in Ref.~\onlinecite{RTRG_FS}, one can show that the
effective vertices have the following properties for fermions and $n$ even (the case
which we consider here)
\begin{eqnarray}
\label{eq:G_permutation}
G^{p_1\dots p_i\dots p_j\dots p_n}_{1\dots i \dots j \dots n}(E) &=&
- G^{p_1\dots p_j\dots p_i\dots p_n}_{1\dots j \dots i \dots n}(E),\\
\label{eq:G_c_trafo}
G^{p_1\dots p_n}_{1\dots n}(E)^c &=&
- G^{-p_n\dots -p_1}_{\bar{n}\dots \bar{1}}(-E^*),\\
\label{eq:G_trace}
\text{Tr}G_{1\dots n}(E)&=&0,
\end{eqnarray}
where the $c$-transformation is defined in matrix notation by
$(A^c)_{s_1s_2,s_1^\prime s_2^\prime}=(A_{s_2 s_1,s_2^\prime s_1^\prime})^*$.
In particular, $L(E)^c=-L(-E^*)$ guarantees the important property that the
reduced density matrix $\rho(t)$ given by Eq.~\eqref{eq:time_rho} is Hermitian,
which is related to the Hermiticity of the original Hamiltonian.\cite{RTRG_FS}
The property $\text{Tr}L(E)=0$ guarantees conservation of probability,
$\text{Tr}\rho(t)=\text{Tr}\rho(t_0)$.

Furthermore, we note that all $n$-point vertices are analytic functions in the upper
half of the complex plane w.r.t. the Fourier variable $E$ and
the external frequencies $\bar{\omega}_i$. This follows from the fact that these
variables occur only in the argument of the resolvents standing between the bare vertices
in the form $R(E+\bar{\omega}_i+\dots)$ together with the property that the resolvent is
an analytic function in the upper half of the complex plane. Here we have assumed that
the bare vertices are frequency independent. If an
effective vertex is used instead of a bare one in a diagram it appears in the form
(after integrating out all $\delta$-functions between the internal frequencies of
connected vertices)
\begin{align}
\nonumber
&G^{p_1\dots p_n}_{1\dots n}(E_{1\dots m}+\bar{\omega}_{1\dots m};\\
\label{eq:vertex_form}
&\hspace{1.5cm}-\bar{\omega}_1,\dots,-\bar{\omega}_m,\bar{\omega}_{m+1},\dots,\bar{\omega}_n),
\end{align}
where the indices $1,\dots, m$ and $m+1,\dots, n$ belong to the contractions which
point to the left or to the right direction, respectively. Using the diagrammatic series
for Eq.~\eqref{eq:vertex_form}, we again see that the quantity is analytic w.r.t. $E$ and all $\bar{\omega}_i$.
Therefore, even if effective vertices are taken instead of bare ones, the internal frequency
integrations can be closed in the upper half of the complex plane and only the nonanalytic
features of the functions $\gamma^{p'}(\bar{\omega})$
defined in Eq.~\eqref{eq:gamma_p} have to be considered, which are the Matsubara poles
of the Fermi functions and the pole $iD$ of the high-energy cutoff function $D(\bar{\omega})$.
This is very helpful for practical calculations. In particular, as we will show in the following by using
a proper reformulation of the diagrammatic series in terms of RG equations and effective
vertices, it will turn out that the limit $D\rightarrow\infty$ can be performed and
only the Matsubara poles of the Fermi functions remain. In this case, it is useful to
split the Fermi function into symmetric and antisymmetic parts by
\begin{align}
\label{eq:splitting_Fermi}
f(\omega)&=1/2 + f^a(\omega),& f^a(\omega)=f(\omega)-1/2.
\end{align}
When inserted in Eq.~\eqref{eq:gamma_p}, this leads to the decomposition
\begin{align}
\label{eq:gamma_p_splitting}
\gamma^{p'}(\bar{\omega})&=p'\gamma^s(\bar{\omega})+\gamma^a(\bar{\omega}),
\end{align}
with
\begin{align}
\label{eq:gamma_sa}
\gamma^s(\bar{\omega})&=\frac{1}{2}D(\bar{\omega}), &
\gamma^a(\bar{\omega})&=f^a(\bar{\omega})D(\bar{\omega}).
\end{align}
Thus, for $D\rightarrow\infty$, the internal frequency integration can be written as
a sum over the Matsubara poles of the antisymmetric part of the Fermi functions
on the positive imaginary axis,
\begin{align}
\label{eq:frequency_integration}
\int d\bar{\omega} \left[f(\bar{\omega})-\frac{1}{2}\right] F(\bar{\omega}) &=
-2\pi i T \sum_{\omega_n>0} F(i\omega_n)\\
&\stackrel{T\rightarrow 0}{\longrightarrow}
-i\int_0^\infty d\omega F(i\omega),
\end{align}
where $\omega_n=(2n+1)\pi T$ denote the fermionic Matsubara frequencies. We will return
to this point at the end of this section when analyzing the analytic structure of
the RG equations.

We next turn to the central question as to how the limit $D\rightarrow\infty$ can be performed.
The convergence of the frequency integrals at high energies can easily be checked by
counting the number of integrations and resolvents. For example, for the diagram \eqref{eq:L_lowest}
of the effective Liouvillian we have two frequency integrations and thus we need three
resolvents for convergence. However, since there is only one resolvent present, we see that
we need at least two derivatives w.r.t. $E$ to obtain a convergent integral. For the diagram
\eqref{eq:G_lowest} of the two-point vertex we have one frequency integral, so we
need two resolvents for convergence. Since there is only one resolvent present, we need
one derivative w.r.t. $E$ for convergence. This is the reason why we consider a
perturbative expansion for $\frac{\partial^2}{\partial E^2}L(E)$ and
$\frac{\partial}{\partial E}G^{p_1p_2}_{12}(E)$ to perform the limit $D\rightarrow\infty$.
Using the diagrammatic representation \eqref{eq:diagrammatic_series}, the $E$-derivative
can only act on the resolvents $R_X(E)$ occurring between the bare vertices. If a
specific resolvent of a certain diagram is chosen for the derivative, one can classify
the diagrams by the number of contractions running over this resolvent, i.e., contractions
which connect vertices standing left to the resolvent with vertices standing right to it.
Cutting all these contractions virtually, the diagram splits into two parts and one can
subsequently resum all connected diagrams to the left and to the right of the resolvent.
This resummation leads to effective vertices such that no contraction is left which
connects effective vertices both standing either left or right to the resolvent. As a result,
we obtain diagrams which contain only the effective vertices instead of bare ones.
Moreover, only diagrams are allowed where all internal contractions connect effective vertices
which are on different sides of the resolvent where the derivative is taken. This
leads to the following equations up to $O(G^3)$:
\begin{multline}
\label{eq:L_rg_1}
\frac{1}{2}\frac{\partial^2}{\partial E^2}L(E)=
\frac{1}{2}\frac{1}{2}
\raisebox{-0.5em}{
\includegraphics[scale=0.45]{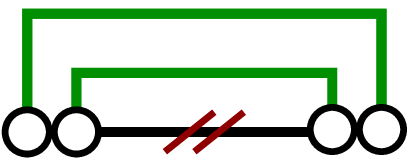}
}
\\
+
\raisebox{-0.5em}{
\includegraphics[scale=0.45]{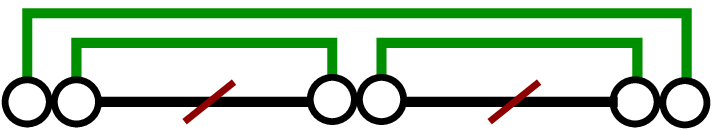}
}
+\mathcal{O}\left(G^4\right)
\end{multline}
for the second derivative of the effective Liouvillian, and
\begin{align}
&\frac{\partial}{\partial E}G^{p_1p_2}_{12}(E)=
\Bigg[
\raisebox{-0.75em}{
\includegraphics[scale=0.45]{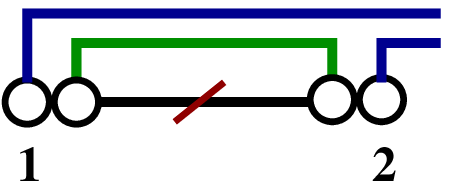}
}
-(1\leftrightarrow 2)\Bigg]
\nonumber\\
&\ \
+\frac{1}{2}
\raisebox{-1em}{
\includegraphics[scale=0.45]{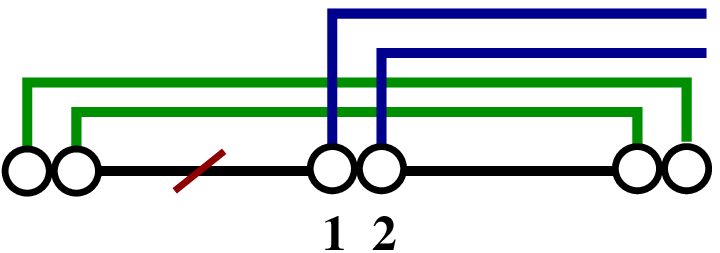}
}
+\frac{1}{2}
\raisebox{-1em}{
\includegraphics[scale=0.45]{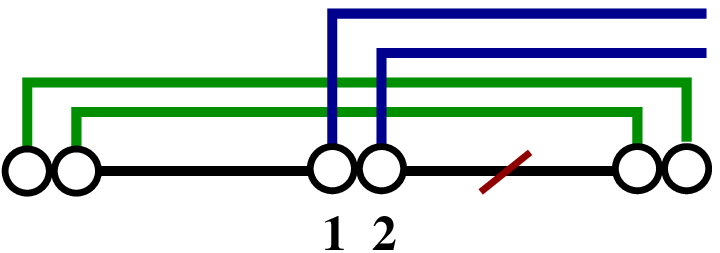}
}
\nonumber\\
&\ \
+\Bigg[
\raisebox{-1em}{
\includegraphics[scale=0.45]{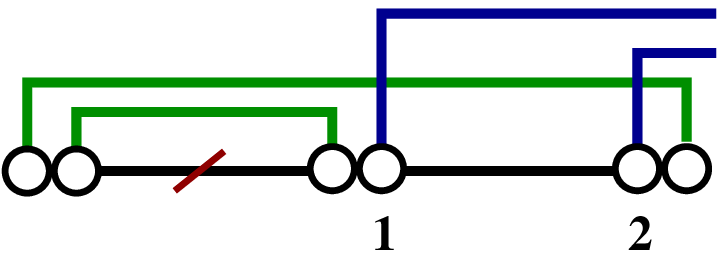}
}
+
\raisebox{-1em}{
\includegraphics[scale=0.45]{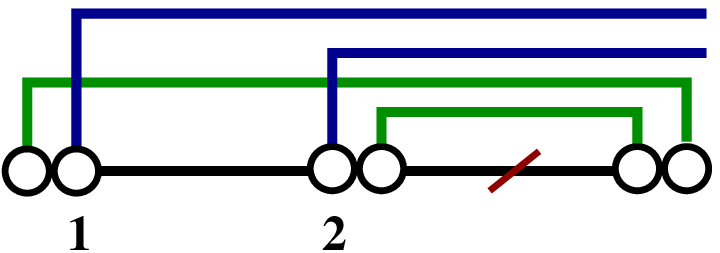}
}
\nonumber\\
&\qquad\qquad
-(1\leftrightarrow 2)\Bigg]
+
O(G^4)
\label{eq:G_rg_1}
\end{align}
for the first derivative of the two-point vertex. In these diagrams, the slash indicates the
$E$-derivative $\frac{\partial}{\partial E}$ and a double-circle represents the full effective
two-point vertex (a convention which we use always in all following diagrams).
Symmetry factors $\frac{1}{n!}$, arising either from the factor $S$
in Eq.~\eqref{eq:diagrammatic_series}, or from the $E$-derivatives $\frac{1}{n!}\partial^n_E$,
are explicitly quoted for convenience. Most importantly, even if one neglects the
frequency dependence of the effective vertices, all frequency integrations converge in
these equations in the infinite band width limit $D\rightarrow\infty$. This is the
reason why the frequency dependence of the vertices can be systematically treated perturbatively
as will be shown in the following sections.

In $O(G^4)$, diagrams involving the four-point vertex can occur for the derivative
of $G^{p_1 p_2}_{12}(E)$ like, e.g.,
\begin{align}
\label{eq:G_rg_higher_order}
\frac{1}{2}\,\raisebox{-1.em}{
\includegraphics[scale=0.45]{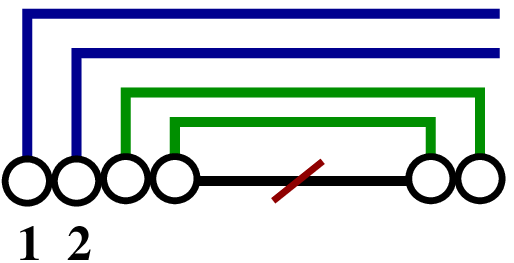}
}
\end{align}
Neglecting the frequency dependence of the two effective vertices, the frequency integrations
do not converge since two resolvents and two integrations are present. However, all $n$-point
vertices with $n>2$ can be expressed in terms of two-point vertices where all frequency
integrations are convergent. For this reason these vertices are called irrelevant, which means that
they can be treated perturbatively. For example, the lowest order terms of $O(G^3)$ for the
four-point vertex are given by
\begin{align}
\nonumber
&G^{p_1p_2p_3p_4}_{1234}(E)
\\
\label{eq:G_n=4}
&=\sum_{P,P_2<P_3}(-1)^P\raisebox{-1.5em}{
\includegraphics[scale=0.45]{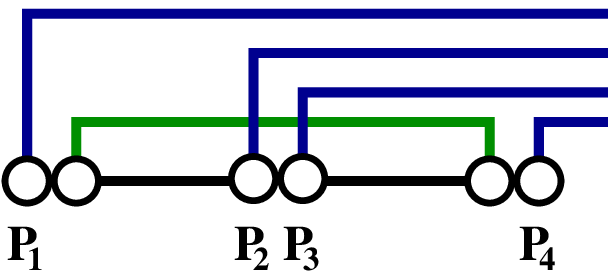}
}
+O(G^4),
\end{align}
where $(P_1,P_2,P_3,P_4)$ denotes a permutation of $(1,2,3,4)$. Obviously the frequency integration
converges for $D\rightarrow\infty$ and we can insert this expression for the four-point vertex
into Eq.~\eqref{eq:G_rg_higher_order}. This leads to the terms
\begin{align}
&\frac{1}{2}\,\raisebox{-1.em}{
\includegraphics[scale=0.45]{G_rg4.eps}
}=
\raisebox{-0.75em}{
\includegraphics[scale=0.45]{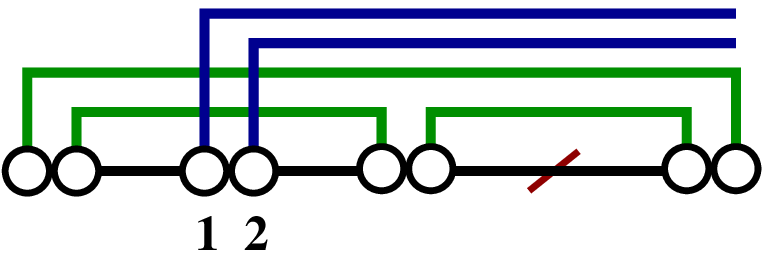}
}
\nonumber\\
&+\Bigg[\frac{1}{2}
\raisebox{-1em}{
\includegraphics[scale=0.45]{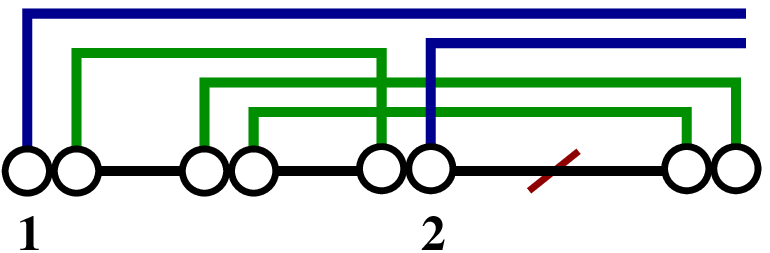}
}
+
\raisebox{-1em}{
\includegraphics[scale=0.45]{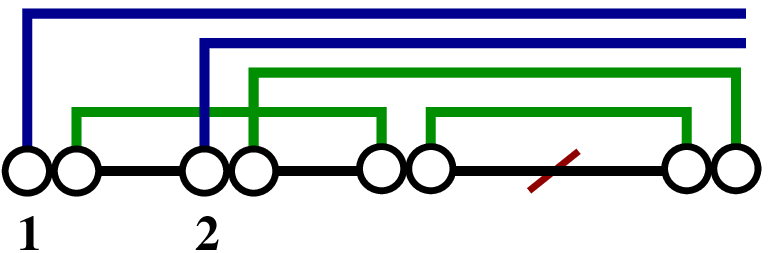}
}
\nonumber\\
&+
\raisebox{-1em}{
\includegraphics[scale=0.45]{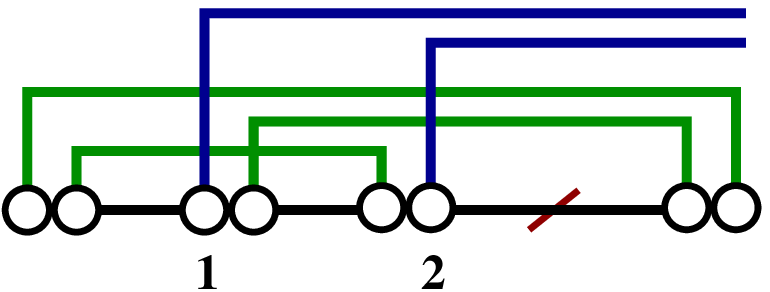}
}
-(1\leftrightarrow 2)\Bigg]
+O(G^5)
\label{eq:G_rg_4th_order}
\end{align}
for the derivative of $G^{p_1 p_2}_{12}(E)$ in $O(G^4)$, where the frequency integrations
are all convergent for $D\rightarrow\infty$. Thus, we see that the frequency dependence
of the four-point vertex is very important for the convergence of the frequency integrations
in Eq.~\eqref{eq:G_rg_higher_order}. The reason is that the frequency dependence
of the four-point vertex is not logarithmic but, as can be seen from Eq.~\eqref{eq:G_n=4}, behaves as
$1/\bar{\omega}_i$ for large frequencies. Therefore, instead of writing complicated coupled RG equations
for all higher-order effective vertices, it is more convenient to directly resum the
diagrams for $\frac{\partial}{\partial E}G^{p_1 p_2}_{12}(E)$, such that only two-point
vertices occur left and right to the resolvent where the derivative is taken. A straightforward
inspection shows that this leads directly to the diagrams of Eq.~\eqref{eq:G_rg_4th_order} in $O(G^4)$,
and in all orders there is no problem with double counting and all frequency
integrations are convergent for $D\rightarrow\infty$. A similar procedure
can be used for $\frac{\partial^2}{\partial E^2}L(E)$ to obtain a series for the second derivative
of the Liouvillian in terms of the two-point vertex with convergent frequency integrations
in all orders. As a result, we obtain the RG equations \eqref{eq:Liouvillian_E_flow_generic} and
\eqref{eq:vertex_E_flow_generic} for the effective Liouvillian and the two-point vertex.

We note that the limit $D\rightarrow\infty$ can only be performed if all bare quantities are
replaced by effective ones and the derivative w.r.t. the Fourier variable $E$ is taken. The
effective Liouvillian and the two-point vertices contain the band width $D$ implicitly via the
initial value $E=iD$ (see Section~\ref{subsec:initial_conditions} for the determination of the
initial conditions). Therefore, concerning these quantities, the infinite band width limit has
to be taken in the sense of the scaling limit, i.e., the bare coupling constants are
simultaneously sent to zero, such that a characteristic low energy scale $T_K^*$ remains constant.

In this paper, we will restrict ourselves to a truncation scheme where all terms in $O(G^2)$ or
$O(G^3)$ are considered on the right-hand side of the RG equations, which, in the following, will be called
truncation in second and third order, respectively. Therefore, we will restrict ourselves to the
RG equations \eqref{eq:L_rg_1} and \eqref{eq:G_rg_1} in the following. Nevertheless, if desired,
the systematic construction of the RG equations allows straightforwardly to go beyond third
order truncation schemes.

Since the limit $D\rightarrow\infty$ has been performed, all internal frequency integrations
for any diagram of the RG equations of the effective Liouvillian or the two-point vertex can
be replaced by the sum over the Matsubara poles of the antisymmetric part of the Fermi function,
see Eq.~\eqref{eq:frequency_integration}. In particular, this means that the symmetric part
$p'\gamma^s(\bar{\omega})=p'/2$ of the contraction $\gamma^{p'}(\bar{\omega})$ does not
contribute in the limit $D\rightarrow\infty$, and only the antisymmetric part
$\gamma^a(\bar{\omega})=f(\bar{\omega})-1/2$ remains. Since the latter is independent of
the Keldysh indices, we find that we need to consider only the vertices $G_{1\dots n}(E)$
averaged over the Keldysh indices. This is very helpful for practical calculations and
an important advantage over other nonequilibrium formalisms, such as the Keldysh formalism, where
the whole matrix structure in Keldysh space has to be taken into account. Although the
symmetric part of the Fermi function does not enter the RG equations, we note that it is
important for the determination of the initial condition (see Sec.~\ref{subsec:initial_conditions}).

Following Ref.~\onlinecite{RTRG_FS}, another important consequence of the fact that only
the vertices averaged over the Keldysh indices occur in the RG equations is that the effective
Liouvillian occurring in the resolvents between the effective vertices cannot contain
the zero eigenvalue. This follows since the projector $P_0(E)$, which projects any operator
on the operator $b(E)$ with $L(E)b(E)=0$, fulfills $P_0(E)G_{12}(E)=0$, see Ref.~\onlinecite{RTRG_FS}.
This allows for a general analysis of the analytical properties of the RG equations
even before calculating the sum over the Matsubara frequencies. Replacing the real
frequencies by positive Matsubara frequencies via $\bar{\omega}_X\rightarrow i\sum_{j\in X}\omega_{n_j}$,
each resolvent $R(E_X+i\sum_{j\in X}\omega_{n_j})$ occurring between the two-point
vertices has a pole for $E_X+i\sum_{j\in X}\omega_{n_j}=z_i^\sigma$,
where $z_i^\pm$ are the non-zero poles of the resolvent $R(E)$. Since all $\omega_n$ are
positive, there is an infinite set of poles at
\begin{align}
\label{eq:poles}
E &= z_i^\pm + nV - i m 2\pi T,
\end{align}
for any two integers $n,m$ with $m>0$. For $T\rightarrow 0$, the infinite set of poles
turns into a branch cut with branching point $z_i^\pm + nV$ pointing in the direction
of the negative imaginary axis. Thus, with our choice of how to calculate the internal
frequency integration, we have determined the shape of the nonanalyticities in the lower
half of the complex plane or, equivalently, have chosen a specific way of how to analytically
continue the effective vertices into the lower half of the complex plane. In the original
form \eqref{eq:diagrammatic_series} of the diagrammatic series with integrations
over the real axis, all branch cuts would appear on the real axis which would be very
inconvenient for an evaluation of the time evolution via inverse Fourier transform.

Although all internal frequency integrations can be written as sums over the Matsubara
frequencies, such a procedure is not very convenient for an explicit solution of the
RG equations since a set of differential equations has to be solved numerically
where the frequency dependence of the vertices is parametrized by an infinite
set of Matsubara frequencies. Because of the huge number of variables, this is numerically very
complicated. Therefore, in the following sections, we will show how we can set up systematic
RG equations only for the {\it frequency independent} vertices
\begin{align}
\label{eq:G_omega=0}
G_{12}(E)\equiv G_{\eta_1\alpha_1\sigma_1,\eta_2\alpha_2\sigma_2}(E;\bar{\omega}_1=0,\bar{\omega}_2=0),
\end{align}
where the indices $i=\eta_i\alpha_i\sigma_i$ will no longer contain the frequency variable
from now on. Furthermore, we will show how the frequency dependence in the argument of the
effective Liouvillian $L(E_X+\bar{\omega}_X)$ occurring in the resolvents between
the effective vertices can be systematically eliminated.
The procedure consists of two steps. First we will transform the $E$-derivative on the right-hand side
of the RG equations \eqref{eq:Liouvillian_E_flow_generic} and \eqref{eq:vertex_E_flow_generic}
to frequency derivatives and use integration by parts to shift them to the derivative
of the Fermi functions. Secondly, we will use a perturbative expansion for the frequency
dependence of the two-point vertices and the effective Liouvillian.

\subsection{Transforming the $E$-derivatives to frequency-derivatives}

Before using a parametrization of the frequency dependence of the vertices and the
resolvents to calculate analytically the integrations over the internal frequencies
for the various diagrams of the RG equation, it is useful to first replace the
derivative w.r.t. $E$ by a frequency derivative. This is possible since the
resolvents $R_{1\dots n}=R(E_{1\dots n}+\bar{\omega}_{1\dots n})$
depend on the sum of the Fourier variable and the frequencies $\bar{\omega}_i$.
Therefore, the $E$-derivative can be written as frequency-derivative
$\frac{\partial}{\partial\bar{\omega}_i}$ and we can apply integration by parts
to calculate the frequency integrals. For example, for the first term on the right-hand side
of Eq.~\eqref{eq:G_rg_1} we obtain [we have permuted the two indices of the
vertices by using antisymmetry,
$G^{p_1p_2}_{12}(E,\bar{\omega}_1,\bar{\omega}_2)=
-G^{p_2p_1}_{21}(E,\bar{\omega }_2, \bar{\omega}
_1)$]
\begin{widetext}
\begin{align}
\nonumber
\raisebox{-1em}{
\includegraphics[scale=0.45]{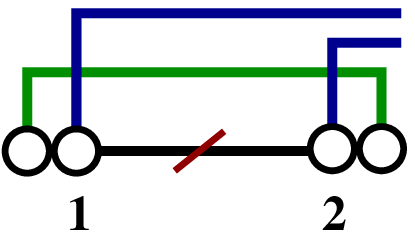}
}
&=-
\raisebox{-1em}{
\includegraphics[scale=0.45]{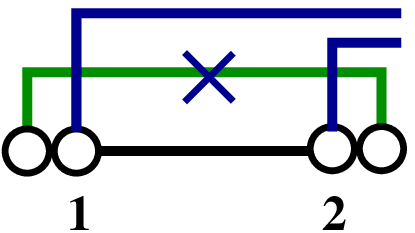}
}
-
\raisebox{-1em}{
\includegraphics[scale=0.45]{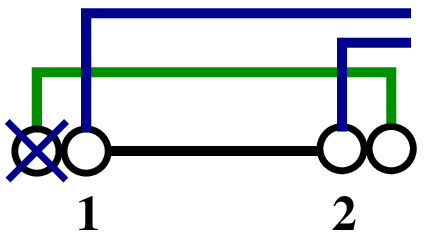}
}
-
\raisebox{-1em}{
\includegraphics[scale=0.45]{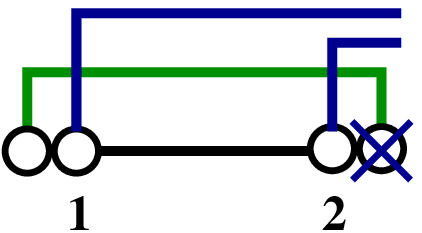}
}
\\ \nonumber\\
\label{eq:G_partial}
&=-
\raisebox{-1em}{
\includegraphics[scale=0.45]{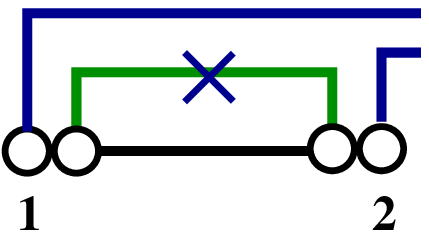}
}
-
\raisebox{-1em}{
\includegraphics[scale=0.45]{G_Eder_4.eps}
}
-
\raisebox{-1em}{
\includegraphics[scale=0.45]{G_Eder_5.eps}
}
+\mathcal{O}\left(G^4\right).
\end{align}
\end{widetext}
The cross indicates the frequency derivative with respect to the frequency
$\bar{\omega}$ of either the corresponding reservoir contraction or the
frequency argument of one of the vertices. The first transformation is exact and
follows from integration by parts. For the derivation of the second line, we have
used
\begin{align}
\label{eq:G_cross}
\raisebox{-0.5em}{
\includegraphics[scale=0.45]{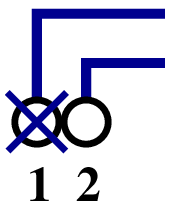}
}
&=
\raisebox{-0.5em}{
\includegraphics[scale=0.45]{G_Eder_1.eps}
}
+\mathcal{O}\left(G^3\right),
\\
\label{eq:G_cross2}
\raisebox{-0.5em}{
\includegraphics[scale=0.45]{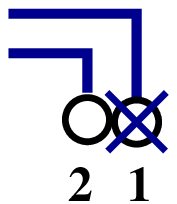}
}
&=
\raisebox{-0.5em}{
\includegraphics[scale=0.45]{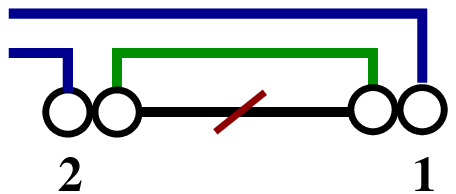}
}
+\mathcal{O}\left(G^3\right),
\end{align}
which follows analogously to Eq.~\eqref{eq:G_rg_1} by using the fact that, in the
original diagrammatic series, a frequency associated with an external line
occurs only in those resolvents which lie below the external line but not in
other resolvents (here we assume that the bare vertices are frequency independent).
Note that these relations can only be applied if it is specified whether the external
line involving the frequency derivative is directed either towards the left or
towards the right.

Analogously, we can treat the first term on the right-hand side of Eq.~\eqref{eq:L_rg_1} by
two integrations by parts and using the fact that
\begin{align}
\frac{\partial^2}{\partial\bar{\omega}_1\partial\bar{\omega}_2}
G^{p_1p_2}_{12}(E,\bar{\omega}_1,\bar{\omega}_2)\sim O(G^3).
\end{align}
We obtain
\begin{align}
\nonumber
&\raisebox{-0.5em}{
\includegraphics[scale=0.45]{L_Eder_1.eps}
}
=\raisebox{-0.5em}{
\includegraphics[scale=0.45]{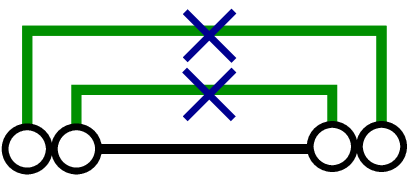}
}
\\
&
+2
\raisebox{-0.5em}{
\includegraphics[scale=0.45]{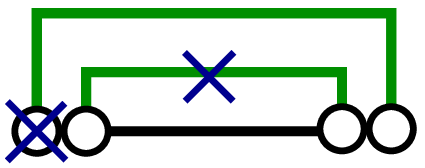}
}
+2
\raisebox{-0.5em}{
\includegraphics[scale=0.45]{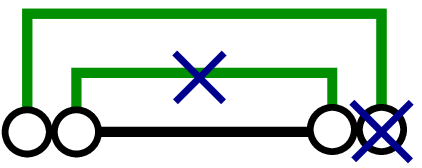}
}\nonumber
+\mathcal{O}\left(G^4\right)
\nonumber\\
=
&\raisebox{-0.5em}{
\includegraphics[scale=0.45]{L_cross_1.eps}
}
-
4
\raisebox{-0.5em}{
\includegraphics[scale=0.45]{L_Eder_2.eps}
}
+\mathcal{O}\left(G^4\right),
\nonumber\\
\label{eq:L_partial}
\end{align}
where, in the second step, we have again used Eqs.~\eqref{eq:G_cross}
and~\eqref{eq:G_cross2}.

Inserting Eqs.~\eqref{eq:L_partial} and~\eqref{eq:G_partial}
in~\eqref{eq:L_rg_1} and~\eqref{eq:G_rg_1}, respectively, we see that many
diagrams in $\mathcal{O}\left(G^3\right)$ cancel each other, which enables us to
write the final third order RG-equations in a very compact and generic
form:
\begin{align}
\label{eq:L_rg_2}
\frac{\partial^2}{\partial E^2}L(E)=
&\frac{1}{2}
\raisebox{-0.5em}{
\includegraphics[scale=0.45]{L_cross_1.eps}
}
+\mathcal{O}\left(G^4\right),
\\
\nonumber
\frac{\partial}{\partial E}G^{p_1p_2}_{12}(E,\bar{\omega}_1,\bar{\omega}_2)=
&-\Bigg[
\raisebox{-1em}{
\includegraphics[scale=0.45]{G_cross_1v.eps}
}
-(1\leftrightarrow 2)\Bigg]
\\
&\hspace{-2.5cm}
-
\frac{1}{2}
\raisebox{-1em}{
\includegraphics[scale=0.45]{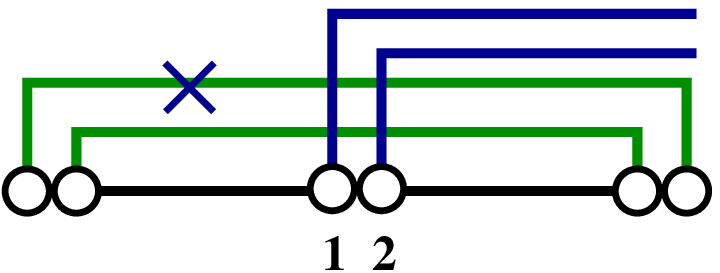}
}
+\mathcal{O}\left(G^4\right).
\label{eq:G_rg_2}
\end{align}

\subsection{Frequency dependence of the vertices}
\label{sec:G_freq}

To evaluate the frequency integrals on the right-hand side of the RG
equations~\eqref{eq:L_rg_2} and~\eqref{eq:G_rg_2} explicitly, we need a
consistent approximation for the  frequency dependence of the vertices and the
Liouvillian. In contrast to $n$-point vertices with $n>2$, this can
be achieved for the two-point vertex since the frequency dependence
of $G_{12}(E;\bar{\omega}_1,\bar{\omega}_2)$ is logarithmic. Therefore,
we can use a perturbative expansion in terms of the frequency
independent vertex $G_{12}(E)$. Although such an expansion will
contain arbitrary powers of logarithmic terms in the frequencies,
it does not lead to any divergence of the integrals when inserted
into the RG diagrams. To expand $G_{12}(E;\bar{\omega}_1,\bar{\omega}_2)$
in terms of the frequency-independent two-point vertex $G_{12}(E)$, we start from the
diagrammatic series in terms of the bare vertices and split each resolvent
$R_X(E)$ as
\begin{align}
\label{eq:resolvent_splitting}
R_X(E)=R_X(E)|_{\bar{\omega}_{X_{ex}}=0} + \Delta_{X_{ex}} R_X(E),
\end{align}
where $X=X_{in}\cup X_{ex}$ consists of internal indices $X_{in}$ and external
ones $X_{ex}$. The first term is the resolvent where all external frequencies
are set to zero, and
\begin{align}
\label{eq:delta_R}
\Delta_{X_{ex}} R_X(E)=R(E_X+\bar{\omega}_X)-R(E_X+\bar{\omega}_{X_{in}})
\end{align}
falls off like $(1/\bar{\omega}_{X_{in}})^2$ w.r.t. all internal frequency variables
$\bar{\omega}_{X_{in}}$. Inserting Eq.~\eqref{eq:resolvent_splitting} for each resolvent, we obtain a
sequence of resolvents $R_X(E)|_{\bar{\omega}_{X_{ex}}=0}$ and $\Delta_{X_{ex}} R_X(E)$ between the bare
vertices. Since $R_X(E)|_{\bar{\omega}_{X_{ex}}=0}$ are the resolvents without the external
frequencies, we can resum all diagrams between two subsequent $\Delta_{X_{ex}}R_X(E)$ in terms of the
two-point vertices at zero external frequency, similar to the procedure described
in the previous section. Up to $O(G^3)$, this gives the equation
\begin{align}
\nonumber
\raisebox{-1.em}{
\includegraphics[scale=0.45]{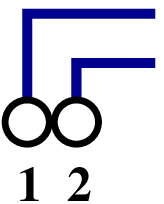}
}
&=
\raisebox{-1.em}{
\includegraphics[scale=0.45]{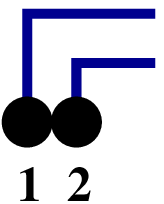}
}
+
\raisebox{-1.em}{
\includegraphics[scale=0.45]{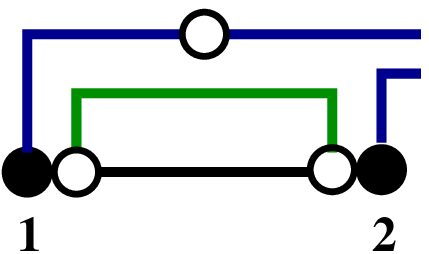}
}
-
\raisebox{-1.em}{
\includegraphics[scale=0.45]{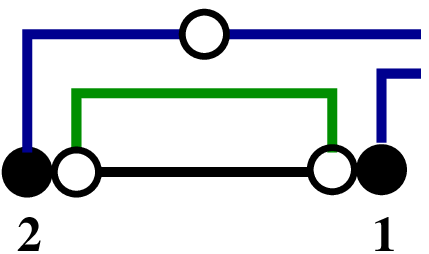}
}
\\
\nonumber
&\hspace{-1.1cm}
+\,\frac{1}{2}\raisebox{-1.em}{
\includegraphics[scale=0.45]{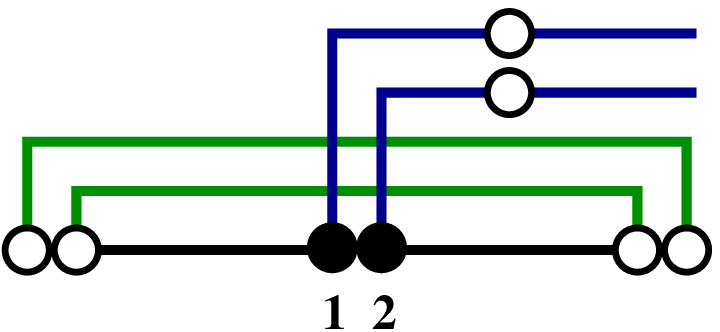}
}
+
\Bigg[
\raisebox{-1.em}{
\includegraphics[scale=0.45]{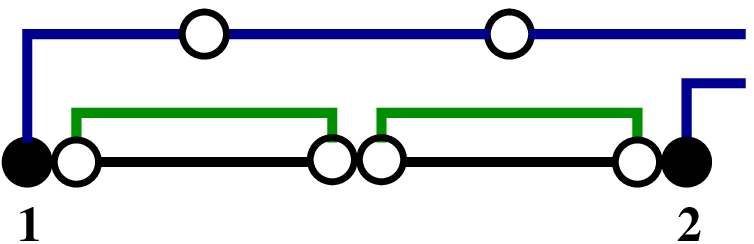}
}
\\
\label{eq:G_omega_mix}
&\hspace{-1cm}
+\raisebox{-1.em}{
\includegraphics[scale=0.45]{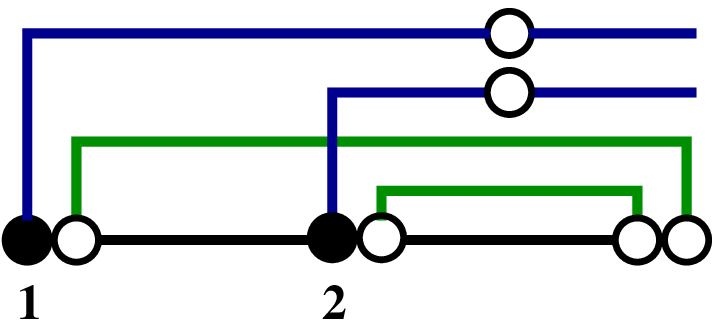}
}
-(1\leftrightarrow 2)\Bigg]
+\mathcal{O}\left(G^4\right).
\end{align}
Here the filled dots indicate that the corresponding frequency of the vertex is set to zero.
A contraction with an open circle and index $X'$ indicates that the
resolvent $R_X(E)$ corresponding to the vertical cut at the position of that circle has
to be replaced by $\Delta_{X'}R_X(E)$. If several contractions with
open circles at the same position of a certain resolvent appear, $X'$ contains the set
of all indices of these contractions. Since $\Delta_{X_{ex}}R_X(E)$ falls off like
$(1/\bar{\omega}_{X_{in}})^2$, all frequency integrals are convergent in the limit $D\rightarrow\infty$.
We note that the left resolvent in the last diagram on the right-hand side of Eq.~\eqref{eq:G_omega_mix}
involves the external frequency $\bar{\omega}_1$
since one can sum the two diagrams where the external frequency does not occur and where
$\Delta_1 R$ appears. This is a generic feature for all resolvents $R_X(E)|_{\bar{\omega}_{X_{ex}}=0}$
which stand below a set of external lines since the replacement
$R_X(E)|_{\bar{\omega}_{X_{ex}}=0}\rightarrow \Delta_{X_{ex}} R_X(E)$
produces also a valid diagram such that the two terms can be added to $R_X(E)$.

To get rid of the remaining frequency dependence of the vertices w.r.t. the internal
frequency variables in Eq.~\eqref{eq:G_omega_mix}, one can iterate this equation and obtains
up to $O(G^3)$
\begin{align}
\nonumber
\raisebox{-1.em}{
\includegraphics[scale=0.45]{G_full.eps}
}
&=
\raisebox{-1.em}{
\includegraphics[scale=0.45]{G_free.eps}
}
+
\raisebox{-1.em}{
\includegraphics[scale=0.45]{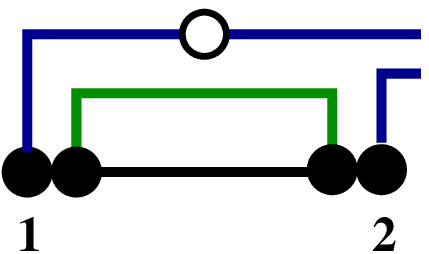}
}
-
\raisebox{-1.em}{
\includegraphics[scale=0.45]{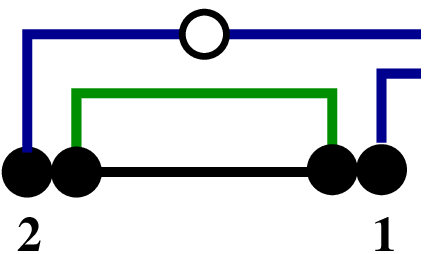}
}
\\
\nonumber
&\hspace{-1.1cm}
+\,\frac{1}{2}\raisebox{-1.em}{
\includegraphics[scale=0.45]{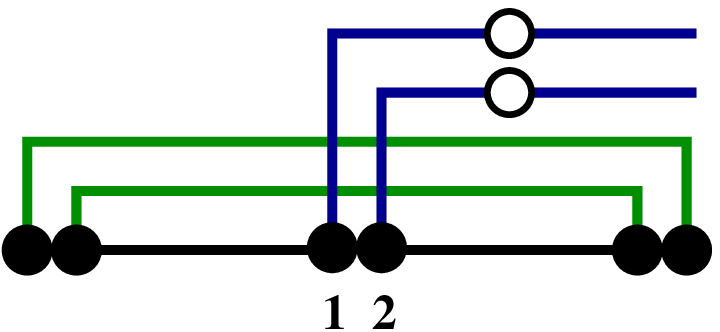}
}
+
\Bigg[
\raisebox{-1.em}{
\includegraphics[scale=0.45]{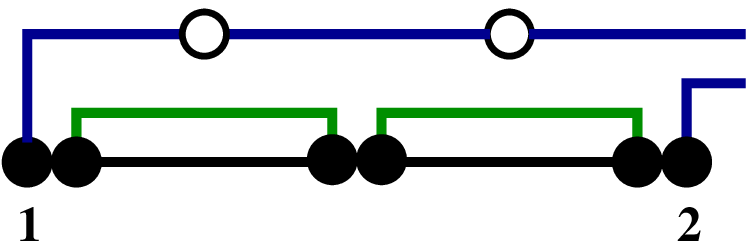}
}
\\
\nonumber
&\hspace{-1.1cm}
+\raisebox{-1.em}{
\includegraphics[scale=0.45]{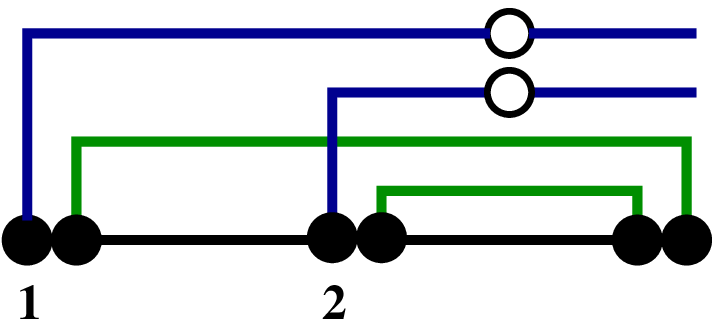}
}
+\raisebox{-1.em}{
\includegraphics[scale=0.45]{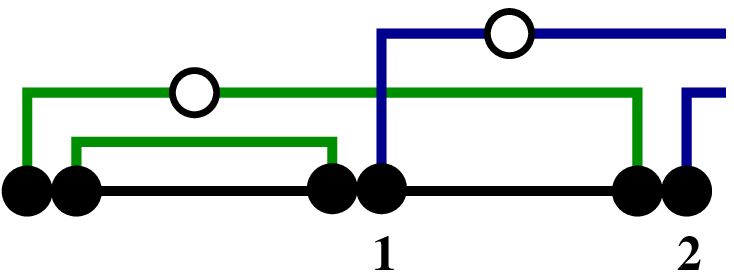}
}
\\
\label{eq:G_omega}
&\hspace{-1.1cm}
+\raisebox{-2.em}{
\includegraphics[scale=0.45]{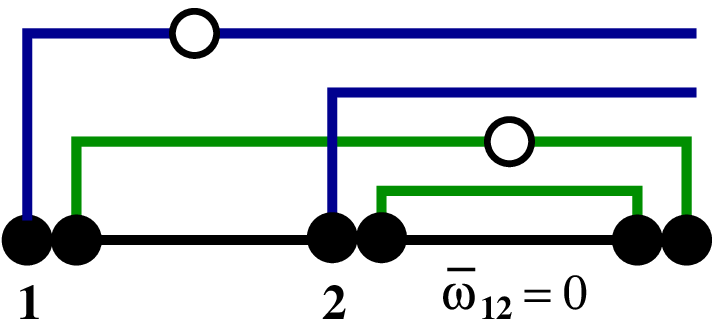}
}
-(1\leftrightarrow 2)\Bigg]
+\mathcal{O}\left(G^4\right).
\end{align}
Here, the last two diagrams occur due to the internal frequency dependence of the two vertices
of the second diagram on the right-hand side of Eq.~\eqref{eq:G_omega_mix}. Note that in the last diagram
we have to set $\bar{\omega}_{12}=0$ for the right resolvent since the right vertex of the
second diagram on the right-hand side of Eq.~\eqref{eq:G_omega_mix} does not depend on the external frequencies.
Proceeding in this way in all orders we see that we
obtain a systematic perturbative expansion of the frequency dependent two-point vertex in terms
of the frequency-independent ones which is free of any divergence for $D\rightarrow\infty$.
We note again that, for large external frequencies $|\bar{\omega}_i|\gg |E|$, this expansion
involves arbitrary powers in $\ln|\frac{\bar{\omega}}{E}|$, i.e., it is not a meaningful
expansion to determine the high-frequency behavior of the vertex. However, setting
$\bar{\omega}_{1/2}=0$ in Eq.~\eqref{eq:G_rg_2} and inserting Eq.~\eqref{eq:G_omega} for the
dependence of the two-point vertices on the internal frequencies in Eqs.~\eqref{eq:L_rg_2}
and \eqref{eq:G_rg_2}, there is no divergence at high frequencies since, due to the
presence of the resolvents and the derivatives of the Fermi functions, the integrand falls off
either like $(1/\bar{\omega})^2$ or exponentially w.r.t. all internal frequency variables,
such that additional logarithmic powers do not change the convergence.

Equation~\eqref{eq:G_omega}
refers to the case where the two external lines are directed to the right,
and similar equations can be written for the other cases. We note that the sign
factors for the terms on the right-hand side where the indices $1$ and $2$ are interchanged
accounts explicitly for the crossing of the two external lines. Therefore, when
using these relations in a certain diagram, the sign factor must not be
written explicitly since it is automatically accounted for in the diagrammatic
rules.

Setting $\bar{\omega}_{1/2}=0$ in Eq.~\eqref{eq:G_rg_2} and inserting Eq.~\eqref{eq:G_omega}
in~\eqref{eq:L_rg_2} and~\eqref{eq:G_rg_2}, we obtain
\begin{widetext}
\begin{align}
\label{eq:L_rg_3}
\frac{\partial^2}{\partial E^2}L(E)&=
\frac{1}{2}
\raisebox{-0.5em}{
\includegraphics[scale=0.45]{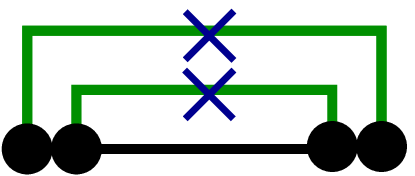}
}
+
\raisebox{-0.5em}{
\includegraphics[scale=0.45]{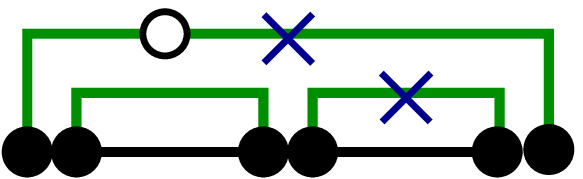}
}
+
\raisebox{-0.5em}{
\includegraphics[scale=0.45]{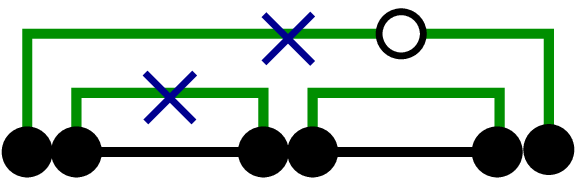}
}
+\mathcal{O}\left(G^4\right),
\\ \nonumber \\ \nonumber
\frac{\partial}{\partial E}G_{12}(E)&=
-\left[
\raisebox{-1em}{
\includegraphics[scale=0.45]{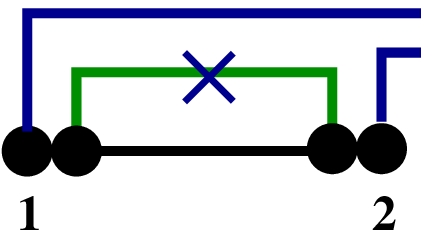}
}
+
\raisebox{-1em}{
\includegraphics[scale=0.45]{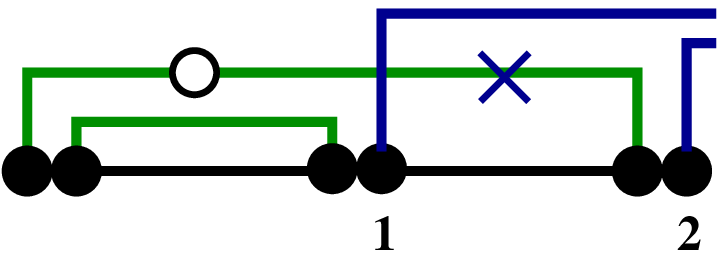}
}
+
\raisebox{-1em}{
\includegraphics[scale=0.45]{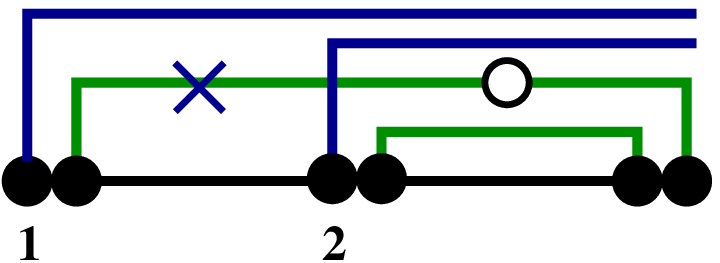}
}
-(1\leftrightarrow 2)\right]
\\ \nonumber \\
\label{eq:G_rg_3}
&\hspace{1cm}
-\frac{1}{2}
\raisebox{-1em}{
\includegraphics[scale=0.45]{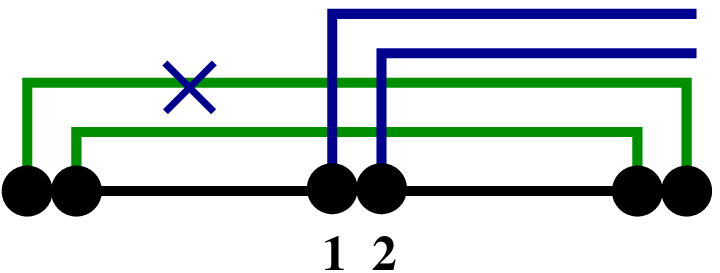}
}%
+\mathcal{O}\left(G^4\right).
\end{align}
\end{widetext}
At zero temperature, the evaluation of the RG equations is simplified
considerably because all diagrams in Eqs.~\eqref{eq:L_rg_3} and~\eqref{eq:G_rg_3}
which contain a contraction with a circle and a cross vanish. The reason is that
the cross indicates a contraction which is differentiated with respect to the
frequency, and
\begin{align}
\frac{\partial}{\partial\bar\omega}\gamma^{p'}\left(\bar\omega\right)
=f'\left(\bar\omega\right)=-\delta\left(\bar\omega\right)
\end{align}
at zero temperature and $D\rightarrow\infty$, such that the difference $\Delta R$
yields zero.

\subsection{Frequency dependence of the propagator}

Finally, to calculate the integrals over the internal frequency variables on the right-hand side
of the RG equations \eqref{eq:L_rg_3} and \eqref{eq:G_rg_3}, one needs a
consistent approximation for the frequency dependence of the resolvent
\begin{align}
\label{eq:resolvent_freq}
R_X(E)=\frac{1}{E_X+\bar{\omega}_X-L(E_X+\bar{\omega}_X)},
\end{align}
where $\bar{\omega}_X=\sum_{i\in X}\bar{\omega}_i$ contains the integration variables
$\bar{\omega}_i$ together with the frequencies of external lines. This requires a perturbative
expansion of the difference
\begin{align}
\Delta_{\bar{\omega}}L(E)=L(E+\bar{\omega})-L(E)
\end{align}
in terms of the frequency-independent vertices.
Treating this difference similar to the two-point vertex as described in the
previous section, we find that the frequency integrals do not converge in the
limit $D\rightarrow\infty$, similar to the fact that, for infinitesimal differences,
two derivatives w.r.t. $E$ are needed to guarantee convergence (see Sec.~\ref{sec:E_flow}).
Therefore, we need a convenient discrete version for the second derivative.
We use the following definition for the second variation for a finite shift
$\bar{\omega}$
\begin{align}
\label{eq:2nd_derivative_discrete}
\Delta_{\bar{\omega}}^2 L(E)&\equiv L(E+\bar{\omega})-L(E)-
\bar{\omega}\frac{\partial}{\partial E}L(E)\sim O(G^2),
\end{align}
which, for $\bar{\omega}=\delta E \rightarrow 0$, gives the second variation
$\delta^2 L(E)=\frac{1}{2}\delta E^2\frac{\partial^2}{\partial E^2}L(E)$.
$\Delta^2_{\bar{\omega}}L(E)$ is of second order in the two-point vertex, since
a Taylor expansion produces the terms $\frac{1}{n!}\bar{\omega}^n\frac{\partial^n}{\partial E^n}L(E)$
with $n\geq 2$. For all these terms we can use the procedure described in
Section~\ref{sec:E_flow} by starting from the diagrammatic series in terms of the
bare vertices, taking the derivatives of the resolvents, and resumming the diagrams
in between to the full two-point vertices. This gives convergent terms in the
limit $D\rightarrow\infty$ and, for all $n\geq 2$, the diagrams are of $O(G^2)$ since
at least one resolvent is needed for the derivative. However, this procedure is not
very practical since all terms with $n\geq 2$ contribute to the lowest order $G^2$.
To resum all terms we apply the same procedure separately for the difference
$L(E+\bar{\omega})-L(E)$ and for $\bar{\omega}\frac{\partial}{\partial E}L(E)$,
following Section~\ref{sec:G_freq} and \ref{sec:E_flow}, respectively. All terms
for $L(E+\bar{\omega})-L(E)$, which contain more than one difference
\begin{align}
\label{eq:om_diff_resolvent}
&\Delta_{\bar{\omega}} R_X(E)\equiv R_X(E+\bar{\omega})-R_X(E)
\end{align}
are at least of $O(G^3)$ and contain already convergent frequency integrals
for $D\rightarrow\infty$. For the other terms with only one $\Delta_{\bar{\omega}}R_X(E)$,
this is not the case and here we need the difference to the correponding term of
$\bar{\omega}\frac{\partial}{\partial E}L(E)$, where the derivative of the resolvent is
taken. This means that Eq.~\eqref{eq:om_diff_resolvent} is changed to
\begin{align}
\label{eq:resolvent_2nd_discrete_deriv}
\Delta^2_{\bar{\omega}}R_X(E)=\Delta_{\bar{\omega}}R_X(E)-\bar{\omega}\frac{\partial}{\partial E}R_X(E).
\end{align}
This is a form for the discrete version of the second derivative of the resolvent which
can be seen after some straightforward manipulations
\begin{align}
\nonumber
\Delta^2_{\bar{\omega}}R_X(E)&=-\frac{1}{2}\left\{
\Delta_{\bar{\omega}}R_X [\bar{\omega}-\Delta_{\bar{\omega}}L_X]R_X\right.\\
\nonumber
&\hspace{-1cm}
\left.+R_X [\bar{\omega}-\Delta_{\bar{\omega}}L_X]\Delta_{\bar{\omega}}R_X\right\}
+ R_X (\Delta^2_{\bar{\omega}} L_X) R_X\\
\label{eq:resolvent_2nd_discrete_deriv_2}
&\hspace{-1.5cm}
= -\Delta_{\bar{\omega}}R_X [\bar{\omega}-\Delta_{\bar{\omega}}L_X]R_X
+ R_X (\Delta^2_{\bar{\omega}} L_X) R_X
\end{align}
where $L_X(E)=L(E_X+\bar{\omega}_X)$. According to Eq.~\eqref{eq:2nd_derivative_discrete}, the
last term is of $O(G^2)$, which gives at least $O(G^4)$ for $\Delta^2_{\bar{\omega}}L(E)$.
Therefore, in order $O(G^2)$ we obtain for $\Delta^2_{\bar{\omega}}L(E)$ the expression
\eqref{eq:L_lowest_effective}, where $R_{12}(E)$ has to be replaced by the first
term of \eqref{eq:resolvent_2nd_discrete_deriv_2}. All frequency integrations exist in
the limit $D\rightarrow\infty$, such that the symmetric part of the Fermi function does
not contribute, and only the two-point vertex averaged over the Keldysh indices is needed.
This gives the following result for the lowest order contribution to the second variation
\begin{align}
\nonumber
&\Delta^2_{\bar{\omega}}L(E)=
-\frac{1}{2}f^a(\bar{\omega}_1)f^a(\bar{\omega}_2)G_{12}(E)\Delta_{\bar{\omega}}R_{12}(E)\cdot\\
\label{eq:Delta_L_lowest}
& \cdot [\bar{\omega}-\Delta_{\bar{\omega}}L_{12}(E)]
R_{12}(E) G_{\bar{2}\bar{1}}(E_{12}) + O(G^3),
\end{align}
where, in lowest order, only the frequency-independent vertices enter.

Using Eq.~\eqref{eq:2nd_derivative_discrete}, we can now approximate the frequency dependence of the
resolvent $R_X(E)$, which is given by Eq.~\eqref{eq:resolvent_freq}. We use
$L(E_X+\bar{\omega}_X)=L(E_X)+\Delta_{\bar{\omega}_X}L(E_X)$ together with
\begin{align}
\label{eq:L_frequency}
\Delta_{\bar{\omega}_X}L(E_X)=\bar{\omega}_X\frac{\partial}{\partial E}L(E_X)+\Delta^2_{\bar{\omega}_X}L(E_X),
\end{align}
and expand the resolvent in
$\Delta^2_{\bar{\omega}_X}L(E_X)\sim O(G^2)$. This gives
\begin{align}
\nonumber
&R_X(E) = \frac{1}{\bar{\omega}_X + \chi(E_X)} Z(E_X)\\
\label{eq:freq_exp_resolvent}
&+ \frac{1}{\bar{\omega}_X + \chi(E_X)} Z(E_X)\Delta^2_{\bar{\omega}_X}L(E_X)
\frac{1}{\bar{\omega}_X + \chi(E_X)} + O(G^4),
\end{align}
where
\begin{align}
\label{eq:chi_Z}
\chi(E)&=Z(E)\left[E-L(E)\right], & Z(E)&=\frac{1}{1-\frac{\partial}{\partial E}L(E)}.
\end{align}
The first term on the right-hand side of Eq.~\eqref{eq:freq_exp_resolvent} is sufficient for
the RG equations \eqref{eq:L_rg_3} and \eqref{eq:G_rg_3} since we neglect $O(G^4)$.
This gives explicitly
\begin{widetext}
\begin{align}
\nonumber
\frac{\partial^2}{\partial E^2}L(E)&=
\frac{1}{2}G_{12}(E)
\frac{f'(\bar{\omega}_1)f'(\bar{\omega}_2)}{\bar{\omega}_{12}+\chi(E_{12})}Z(E_{12})
G_{\bar{2}\bar{1}}(E_{12}) \\
\nonumber
&\hspace{1cm}
+G_{12}(E){\mathcal{F}}(E_{12},\bar{\omega}_1)G_{\bar{2}3}(E_{12})
\frac{f'(\bar{\omega}_1)f'(\bar{\omega}_3)}{\bar{\omega}_{13}+\chi(E_{13})}
Z(E_{13})G_{\bar{3}\bar{1}}(E_{13})\\
\label{eq:L_rg_final}
&\hspace{1cm}
+G_{12}(E)
\frac{f'(\bar{\omega}_1)f'(\bar{\omega}_2)}{\bar{\omega}_{12}+\chi(E_{12})} Z(E_{12})
G_{\bar{2}3}(E_{12}){\mathcal{F}}(E_{13},\bar{\omega}_1)
G_{\bar{3}\bar{1}}(E_{13}),\\
\nonumber
\frac{\partial}{\partial E}G_{12}(E)&=
-\left[G_{13}(E)\frac{f'(\bar{\omega}_3)}{\bar{\omega}_{3}+\chi(E_{13})}Z(E_{13})
G_{\bar{3}2}(E_{13})\right.\\
\nonumber
&\hspace{1cm}
-G_{34}(E){\mathcal{F}}(E_{34},\bar{\omega}_3)G_{\bar{4}1}(E_{34})
\frac{f'(\bar{\omega}_3)}{\bar{\omega}_{3}+\chi(E_{13})}
Z(E_{13})G_{\bar{3}2}(E_{13})\\
\nonumber
&\hspace{1cm}
\left.
-G_{13}(E)
\frac{f'(\bar{\omega}_3)}{\bar{\omega}_{3}+\chi(E_{13})} Z(E_{13})
G_{24}(E_{13}){\mathcal{F}}(E_{1234},\bar{\omega}_3)
G_{\bar{4}\bar{3}}(E_{1234})-(1\leftrightarrow 2)\right]\\
\label{eq:G_rg_final}
&\hspace{1cm}
-\frac{1}{2}G_{34}(E)
\frac{f'(\bar{\omega}_3)}{\bar{\omega}_{34}+\chi(E_{34})} Z(E_{34})
G_{12}(E_{34})\frac{f^a(\bar{\omega}_4)}{\bar{\omega}_{34}+\chi(E_{1234})} Z(E_{1234})
G_{\bar{4}\bar{3}}(E_{1234}),
\end{align}
\end{widetext}
where we have introduced the definition
\begin{align}
\nonumber
\mathcal{F}(E,\bar\omega)&=
\int d\bar\omega' f^a\left(\bar\omega'\right)\cdot\\
\label{eq:integrated_prop_difference_frequency}
&\cdot
\left[\frac{1}{\bar\omega+\bar\omega'+\chi(E)}
-\frac{1}{\bar\omega'+\chi(E)}\right] Z(E).
\end{align}
After inserting the spectral decomposition of the effective Liouvillian, all frequency integrals
can be calculated analytically which will be done later for the specific example of the Kondo model.

For completeness, we note that the RG equations can be systematically improved by going beyond $O(G^3)$.
In this case, one needs also the second term on the right-hand side of Eq.~\eqref{eq:freq_exp_resolvent}, i.e., the
second variation $\Delta^2_{\bar{\omega}}L(E)$ is needed up to $O(G^2)$ by using Eq.~\eqref{eq:Delta_L_lowest}.
To evaluate Eq.~\eqref{eq:Delta_L_lowest} up to $O(G^2)$, the first term on the right-hand side of
Eq.~\eqref{eq:freq_exp_resolvent} is sufficient to approximate the frequency dependence of the resolvents.
Using
\begin{align}
\nonumber
R_X(E+\bar{\omega})&=R(E_X+\bar{\omega}_X+\bar{\omega})\\
\label{eq:freq_resolv_1}
&=\frac{1}{\bar{\omega}_X+\bar{\omega}+\chi(E_X)}Z(E_X) + O(G^2)\\
\nonumber
R_X(E)&=R(E_X+\bar{\omega}_X)\\
\label{eq:freq_resolv_2}
&=\frac{1}{\bar{\omega}_X+\chi(E_X)}Z(E_X) + O(G^2),\\
\nonumber
L_X(E+\bar{\omega})&=L(E_X+\bar{\omega}_X+\bar{\omega})\\
\label{eq:freq_Liouvillian_1}
&= L(E_X) + (\bar{\omega}_X+\bar{\omega})\frac{\partial}{\partial E}L(E_X)+ O(G^2),\\
\nonumber
L_X(E)&=L(E_X+\bar{\omega}_X)\\
\label{eq:freq_Liouvillian_2}
&= L(E_X) + \bar{\omega}_X\frac{\partial}{\partial E}L(E_X)+ O(G^2),
\end{align}
we find
\begin{align}
\nonumber
&\Delta_{\bar{\omega}}R_{12}(E)
 [\bar{\omega}-\Delta_{\bar{\omega}}L_{12}(E)] R_{12}(E)\\
\nonumber
&=\bar{\omega}\Delta_{\bar{\omega}}R_{12}(E)
 [1-\frac{\partial}{\partial E}L(E_{12})] R_{12}(E)+O(G^2)\\
\nonumber
&=\bar{\omega}\left[\frac{1}{\bar{\omega}_{12}+\bar{\omega}+\chi(E_{12})}-
\frac{1}{\bar{\omega}_{12}+\chi(E_{12})}\right]\cdot
\\\nonumber
&\hspace{2cm}
\cdot\frac{1}{\bar{\omega}_{12}+\chi(E_{12})}Z(E_{12})+O(G^2)\\
\nonumber
&=-\bar{\omega}^2\frac{1}{\bar{\omega}_{12}+\bar{\omega}+\chi(E_{12})}
\left(\frac{1}{\bar{\omega}_{12}+\chi(E_{12})}\right)^2 Z(E_{12})\\
\label{eq:appr_resolv_2nd_deriv}
&\hspace{1cm} + O(G^2).
\end{align}
Using this equation in \eqref{eq:Delta_L_lowest} gives the final result for the
second variation
\begin{align}
\nonumber
\Delta^2_{\bar{\omega}}L(E)&=
\frac{1}{2}\bar{\omega}^2 G_{12}(E)\cdot\\
\nonumber
&\hspace{-1cm}
\cdot\frac{f^a(\bar{\omega}_1)f^a(\bar{\omega}_2)}{(\bar{\omega}_{12}+\bar{\omega}+\chi(E_{12}))
(\bar{\omega}_{12}+\chi(E_{12}))^2}Z(E_{12})G_{\bar{2}\bar{1}}(E_{12})\\
\label{eq:2nd_variation_final}
&\hspace{4cm} + O(G^3),
\end{align}
which has to be used in Eq.~\eqref{eq:freq_exp_resolvent} in order to calculate the frequency
dependence of the resolvent up to $O(G^2)$ needed for the evaluation of the RG equations up
to $O(G^4)$.

\subsection{Initial conditions}
\label{subsec:initial_conditions}
To determine the initial condition as described in Sec.~\ref{sec:E_flow_idea}, we consider
the lowest order diagrams for the effective Liouvillian and the two-point vertex, as given
by Eqs.~\eqref{eq:L_lowest} and \eqref{eq:G_lowest}. Taking the unrenormalized values for the
Liouvillian and the vertices on the right-hand side of these equations, denoted by $L^{(0)}$ and
$G_{12}^{(0)pp'}=\delta_{pp'}G_{12}^{(0)pp}$, we obtain
\begin{align}
\nonumber
L(E)&=L^{(0)}\\
\label{eq:L_initial_1}
&+\frac{1}{2}G_{12}^{(0)}\iint d\omega d\omega'
\frac{\gamma^p(\omega)\gamma^p(\omega')}{E_{12}-L^{(0)}+\omega+\omega'}G_{\bar{2}\bar{1}}^{(0)pp},\\
\nonumber
G_{12}(E)&=G_{12}^{(0)}+\\
\label{eq:G_initial_1}
&+\left\{G_{13}^{(0)}\int d\omega \frac{\gamma^p(\omega)}{E_{13}-L^{(0)}+\omega}G_{\bar{3}2}^{(0)pp}
-(1\leftrightarrow 2)\right\},
\end{align}
where $G_{12}^{(0)}=\sum_p G_{12}^{(0)pp}$. Inserting Eqs.~\eqref{eq:gamma_p_splitting} and
\eqref{eq:gamma_sa} for the contraction and closing the integration contours in the upper half of
the complex plane, we obtain for $D\gg |E_X-L^{(0)}|\gg T$ and $\text{Im}E>0$
\begin{widetext}
\begin{align}
\nonumber
L(E)&=L^{(0)}+\frac{1}{2}G_{12}^{(0)}
\left[-ip\frac{\pi}{2}(E_{12}-L^{(0)})+(E_{12}-L^{(0)})\ln\frac{-i(E_{12}-L^{(0)})}{D} \right.\\
\label{eq:L_initial_2}
&\hspace{12em}
\left.+\frac{1}{4}\left(\frac{\pi^2}{4}-3\right)(E_{12}-L^{(0)})-p\frac{\pi}{2}D\right]
G_{\bar{2}\bar{1}}^{(0)pp},\\
\label{eq:G_initial_2}
G_{12}(E)&=G_{12}^{(0)}+
\left\{G_{13}^{(0)}\left[-i\frac{\pi}{2}p + \ln\frac{-i(E_{13}-L^{(0)})}{D}\right]
G_{\bar{3}2}^{(0)pp}
-(1\leftrightarrow 2)\right\}.
\end{align}
\end{widetext}
The last two terms of Eq.~\eqref{eq:L_initial_2} are non-universal. We assume that the last term
linear in $D$ vanishes, otherwise the frequency-dependence
of the unrenormalized vertices has to be taken into account and the precise form of the
high-energy cutoff function as defined by the original model
containing charge fluctuations becomes important. This gives the condition
\begin{align}
\label{eq:condition_D}
p G_{12}^{(0)}G_{\bar{2}\bar{1}}^{(0)pp}=0
\end{align}
which, as we will show, is fulfilled for the Kondo problem considered in this work
(this proof can be generalized to generic $2$-level models, see Ref.~\onlinecite{goettel_etal_13}).
As a consequence, also in the first term on the right-hand side of Eq.~\eqref{eq:L_initial_2}, we
can replace $E_{12}\rightarrow\bar{\mu}_{12}$. In contrast, all
other terms for the effective Liouvillian and the two-point vertex have universal
coefficients independent of the specific choice of the high-energy cutoff function.
Note that the first term for the Liouvillian has a different algebra compared to
the third one due to the additional factor of the Keldysh index $p$. Therefore, it
can happen that the third term gives zero whereas the first one is finite, as it is,
e.g., the case for the current kernel for the Kondo model (see later). Omitting all
non-universal terms together with the logarithmic ones (which are anyhow generated
by the RG equations), and using the property \eqref{eq:condition_D}, we take the
following universal form for the initial condition at $E=iD$
\begin{align}
\label{eq:L_initial_final}
L(E=iD)&=L^{(0)}-ip\frac{\pi}{4}G_{12}^{(0)}
(\bar{\mu}_{12}-L^{(0)})G_{\bar{2}\bar{1}}^{(0)pp},\\
\label{eq:G_initial_final}
G_{12}(E=iD)&=G_{12}^{(0)}-i\frac{\pi}{2}p \left\{G_{13}^{(0)}G_{\bar{3}2}^{(0)pp}
-(1\leftrightarrow 2)\right\}.
\end{align}
As already explained in Sec.~\ref{sec:E_flow_idea}, we cannot use this result for the
initial condition of the effective Liouvillian since we have neglected non-universal terms
proportional to $E$, which are very large. Therefore, when applying the formalism to the
Kondo model, we will set up another universal boundary condition to fix the effective
Liouvillian. For the two-point vertex, the term of $O(G^2)$ in the initial condition is
only used for those matrix elements where the first term of $O(G)$ vanishes.

Leaving out the nonuniversal and logarithmic terms, the initial Liouvillian is
independent of $E$. Therefore, we take $\frac{\partial}{\partial E}L(E)=0$ at $E=iD$ or
\begin{align}
\label{eq:initial_Z}
Z(E=iD)=1.
\end{align}

\subsection{Current and differential conductance}
\label{subsec:E_flow_current}

To find the average of the current flowing into reservoir $\gamma$
[Eq.~\eqref{eq:current_Fourier}], one needs the current kernel
$\Sigma_\gamma(E)$ in Fourier space. As shown in Ref.~\onlinecite{RTRG_FS}, it
can be determined analogously to $L(E)$ by replacing the first vertex from the
left in all diagrams of Eqs.~\eqref{eq:L_rg_3} and~\eqref{eq:G_rg_3} by the
current vertex $I^\gamma_{12}(E)$.

To calculate the differential conductance, one needs the variation $\delta
I_\alpha$ for an infinitesimal variation $\delta\mu_\alpha$ of the chemical
potentials of the reservoirs. Within the $E$-flow scheme, the RG equation for
$\frac{\partial}{\partial E}\delta L(E)$ [or equivalently, for
$\frac{\partial}{\partial E}\delta \Sigma_\alpha(E)$ by replacing the first
vertex by the current vertex] can be  straightforwardly established by applying
the variation to the original diagrammatic series, where the chemical potentials
occur in the denominator of the resolvents explicitly via the energy argument
$E_X=E+\bar{\mu}_X$ and implicitly in the effective Liouvillian. Therefore, the
variation of each resolvent can be split in two terms
\begin{align}
\delta R_X(E)=\delta\bar{\mu}_X\frac{\partial}{\partial
E}R_X(E)+R_X(E) \delta L_X(E)R_X(E),
\end{align}
where the first term contains the variation from the change of the argument and
the second one the variation $\delta L_X(E)=(\delta L)(E_X+\bar{\omega}_X)$ of the
effective Liouvillian. Fixing the position of the resolvents where the variation
$\delta R_X(E)$ and where the differentiation $\frac{\partial}{\partial E}R_X(E)$ is
taken, we can resum the rest of the diagrams for $\frac{\partial}{\partial E}\delta L(E)$
in terms of two-point vertices, analogously to the procedure described in the previous
sections. Analogous to $\frac{\partial^2}{\partial E^2}L(E)$, we obtain convergent
frequency integrals in the limit $D\rightarrow\infty$ in all orders. Up to
$\mathcal{O}\left(G^3\right)$, we obtain
\begin{widetext}
\begin{align}
\nonumber
\frac{\partial}{\partial E}\delta L(E)=
&\frac{1}{2}\delta\bar{\mu}_{12}
\raisebox{-0.75em}{
\includegraphics[scale=0.45]{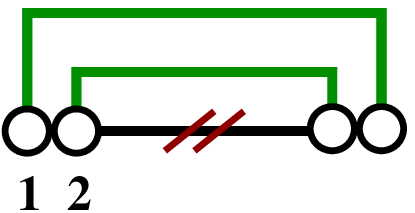}
}
+
\frac{1}{2}
\raisebox{-0.75em}{
\includegraphics[scale=0.45]{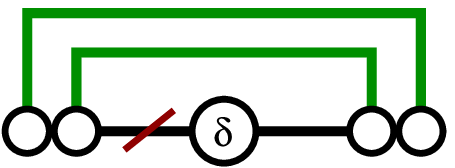}
}
+
\frac{1}{2}
\raisebox{-0.75em}{
\includegraphics[scale=0.45]{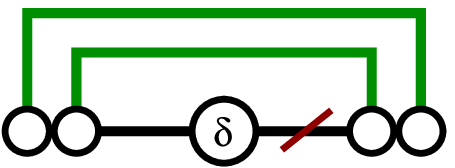}
}
\\ \nonumber \\
\label{eq:L_rg_V1}
&+
(\delta\bar{\mu}_{12}+\delta\bar{\mu}_{13})
\raisebox{-0.75em}{
\includegraphics[scale=0.45]{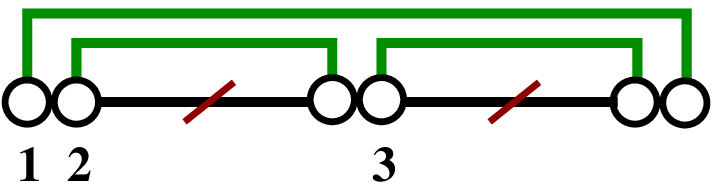}
}
+\mathcal{O}\left(G^4\right),
\end{align}
where $\delta L\sim\mathcal{O}(\delta\mu G)$ is represented by
$\includegraphics[scale=0.45]{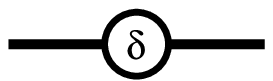}$  and can be
calculated from the first (lowest order) term on the right-hand side
of Eq.~\eqref{eq:L_rg_V1} and subsequently inserted in the second and third term on
the right-hand side of Eq.~\eqref{eq:L_rg_V1}. Applying integration by parts twice to the
first term on the right-hand side of Eq.~\eqref{eq:L_rg_V1} [analogous
to Eq.~\eqref{eq:L_partial}], we obtain
\begin{equation}
\label{eq:L_rg_V2}
\frac{\partial}{\partial E}\delta L(E)=
\frac{1}{2}\delta\bar{\mu}_{12}
\raisebox{-0.75em}{
\includegraphics[scale=0.45]{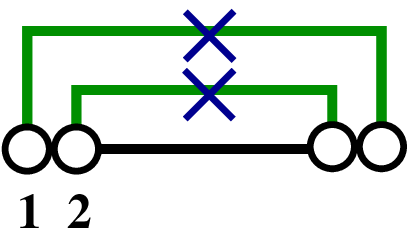}
}
-
\frac{1}{2}
\raisebox{-0.75em}{
\includegraphics[scale=0.45]{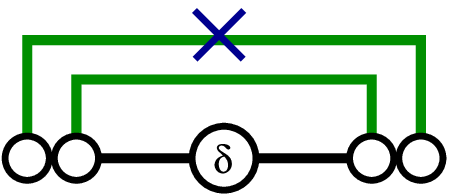}
}
+\mathcal{O}\left(G^4\right).
\end{equation}
Using Eq.~\eqref{eq:G_omega}, we get the final RG equation for the variation of the
kernel,
\begin{align}
\nonumber
\frac{\partial}{\partial E}\delta L(E)=
&\frac{1}{2}\,\delta\bar{\mu}_{12}
\raisebox{-0.75em}{
\includegraphics[height=1cm]{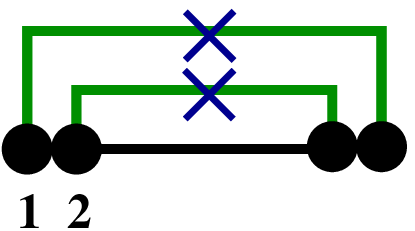}
}
-
\frac{1}{2}
\raisebox{-0.75em}{
\includegraphics[height=1cm]{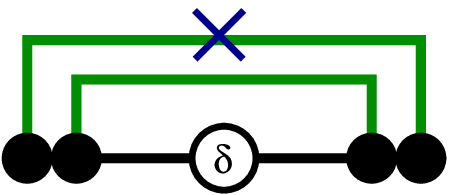}
}
\\ \nonumber \\
\label{eq:L_rg_V3}
&+\delta\bar{\mu}_{13}
\raisebox{-0.75em}{
\includegraphics[height=1cm]{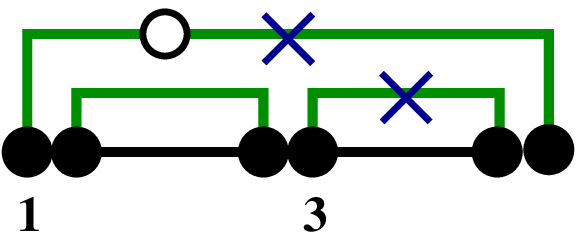}
}
+\delta\bar{\mu}_{12}
\raisebox{-0.75em}{
\includegraphics[height=1cm]{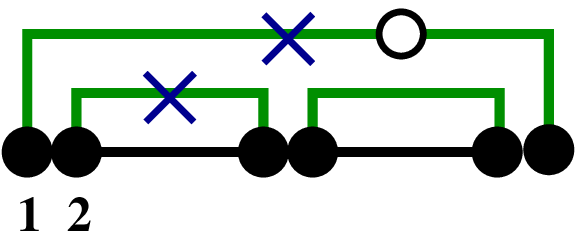}
}
+\mathcal{O}\left(G^4\right).
\end{align}
At zero temperature, the diagrams containing a contraction with a circle and a
cross vanish, cf. the remark after Eqs.~\eqref{eq:L_rg_3} and~\eqref{eq:G_rg_3}.
Writing the diagrammatic Eq.~\eqref{eq:L_rg_V3} explicitly yields
\begin{align}
\nonumber
\frac{\partial}{\partial E}\delta L(E)&=
\frac{1}{2}\delta{\bar\mu}_{12} G_{12}(E)
\frac{f'(\bar{\omega}_1)f'(\bar{\omega}_2)}{\bar{\omega}_{12}+\chi(E_{12})}Z(E_{12})
G_{\bar{2}\bar{1}}(E_{12})\\
\nonumber
&\hspace{1cm}-\frac{1}{2}G_{12}(E)
\frac{f'(\bar{\omega}_1)f(\bar{\omega}_2)}{\bar{\omega}_{12}+\chi(E_{12})}Z(E_{12})
\delta L(E_{12})
\frac{1}{\bar{\omega}_{12}+\chi(E_{12})}Z(E_{12})
G_{\bar{2}\bar{1}}(E_{12})\\
\nonumber
&\hspace{1cm}+\delta{\bar\mu}_{13}G_{12}(E){\mathcal{F}}(E_{12},\bar{\omega}_1)G_{\bar{2}3}(E_{12})
\frac{f'(\bar{\omega}_1)f'(\bar{\omega}_3)}{\bar{\omega}_{13}+\chi(E_{13})}
Z(E_{13})G_{\bar{3}\bar{1}}(E_{13})\\
\label{eq:L_rg_V_final}
&\hspace{1cm}
+\delta{\bar\mu}_{12}G_{12}(E)
\frac{f'(\bar{\omega}_1)f'(\bar{\omega}_2)}{\bar{\omega}_{12}+\chi(E_{12})} Z(E_{12})
G_{\bar{2}3}(E_{12}){\mathcal{F}}(E_{13},\bar{\omega}_1)
G_{\bar{3}\bar{1}}(E_{13}).
\end{align}
\end{widetext}
The initial condition for $\delta L(E)$ follows from Eq.~\eqref{eq:L_initial_final} as
\begin{align}
\label{eq:delta_L_initial_final}
\delta L(E)&=-ip\frac{\pi}{4}G_{12}^{(0)}G_{\bar{2}\bar{1}}^{(0)pp}\delta\bar{\mu}_{12},
\end{align}
and the corresponding initial condition for $\delta\Sigma_\alpha(E)$ by replacing the first vertex from the
left by the current vertex.

\section{RG equations for the Kondo model}
\label{sec:rg_kondo}
We now apply the RG-equations~\eqref{eq:L_rg_final},~\eqref{eq:G_rg_final}
and~\eqref{eq:L_rg_V_final} to the isotropic spin-$\frac{1}{2}$ Kondo model at zero
magnetic field. In that case, the Hamiltonian is $H_{\text{res}}+V$, where
[cf. Eq.~\eqref{eq:g_Kondo}]
\begin{align}
V&=\frac12 g_{11'}:a_1 a_{1'}:,\\
\label{eq:g_Kondo_definition}
g_{11'}&=
\frac12
\begin{cases}
J^{(0)}_{\alpha\alpha'}\underline{S}\cdot\underline{\sigma}_{\sigma\sigma'}
& \text{for $\eta=-\eta'=+$,} \\
-J^{(0)}_{\alpha'\alpha}\underline{S}\cdot\underline{\sigma}_{\sigma'\sigma}
& \text{for $\eta=-\eta'=-$,}\end{cases}
\end{align}
where $\underline{S}$ is the spin-$\frac12$ operator on the quantum dot,
$\underline{\sigma}$ is the vector of Pauli matrices, and the coupling
$J^{(0)}_{\alpha\alpha'}$ fulfills
$J^{(0)*}_{\alpha\alpha'}=J^{(0)}_{\alpha'\alpha}$. In the important case that
the Kondo model is derived using a Schrieffer-Wolff tranformation, the
couplings fulfill the additional constraint (see, e.g., Ref.~\onlinecite{korb_reininghaus_hs_koenig_PRB07}
for a detailed derivation)
\begin{align}
\label{eq:J_sw}
J^{(0)}_{\alpha\alpha'}&=2\sqrt{x_\alpha x_{\alpha'}}J_0, & &\text{where} &
\sum_\alpha x_\alpha=1.
\end{align}

\subsection{Initial condition for the vertex superoperators}

According to Eq.~\eqref{eq:G_definition}, the bare vertex superoperator
$G^{(0)pp'}_{11'}$ is defined in terms of $g_{11'}$ by its action on any
operator $b$:
\begin{align}
G^{(0)pp'}_{11'}b&=\delta_{pp'}
\begin{cases}
g_{11'}b & \text{for $p=+$,} \\
-bg_{11'} & \text{for $p=-$.}
\end{cases}
\end{align}
Since the operator $g_{11'}$ fulfills the symmetry property
\begin{align}
g_{11'}=-g_{1'1}
\end{align}
according to the definition~\eqref{eq:g_Kondo_definition}, $G^{(0)pp'}_{11'}$
also fulfills
\begin{align}
G^{(0)pp'}_{11'}=-G^{(0)p'p}_{1'1}.
\end{align}
Therefore, it is sufficient to consider the bare vertex for the
case $\eta=-\eta'=+$ in the following. We use the shorthand notation
\begin{align}
\widehat G^{(0)pp'}_{11'}&=G^{(0)pp'}_{+\alpha\sigma,-\alpha'\sigma'},
\end{align}
where the multiindices $1$, $1'$ on the left hand side only contain the
reservoir and spin indices, and not $\eta$, $\eta'$:
\begin{align}
1&\equiv\alpha\sigma, & 1'&\equiv\alpha'\sigma'.
\end{align}
Since the operator $g_{11'}$, which induces spin fluctuations on the quantum
dot, is proportional to the spin-$\frac12$ operator $\underline{S}$ [cf.
Eq.~\eqref{eq:g_Kondo}], we need superoperators which multiply an arbitrary
operator $b$ with $\underline{S}$ from the left and right to find a suitable
representation of the initial bare vertex $\widehat G^{(0)pp'}_{11'}$. We thus
define a vector superoperator $\underline{L}^p$ for $p=\pm$ by
\begin{align}
\underline{L}^pb=
\begin{cases}
\underline{S}b, & \text{for $p=+$}, \\
-b\underline{S}, & \text{for $p=-$}
\end{cases}
\end{align}
for any operator $b$. We can then write
\begin{align}
\widehat
G^{(0)pp'}_{11'}=\frac12\delta_{pp'}J^{(0)}_{\alpha\alpha'}\underline{L} ^p\cdot
\underline{\sigma}_{\sigma\sigma'}.
\end{align}
For the bare vertex averaged over the Keldysh indices [see
Eq.~\eqref{eq:vertex_averaged}],
we get
\begin{align}
\label{eq:vertex_G_initial_condition_L_plus_minus}
\widehat G^{(0)}_{11'}&=\sum_{p}\widehat G^{(0)pp}_{11'}=\frac12
J^{(0)}_{\alpha\alpha'} \left(\underline{L}^++\underline{L}^-\right)\cdot
\underline{\sigma}_{\sigma\sigma'}.
\end{align}
Similarly, according to Eq.~\eqref{eq:current_vertex}, the bare current vertex
averaged over
the Keldysh indices is
\begin{align}
\label{eq:vertex_I_initial_condition_L_plus_minus}
\widehat I^{\gamma(0)}_{11'}&=\sum_p \widehat I^{\gamma (0)pp}_{11'} =
\widehat{c}^\gamma_{11'}\sum_p p\,\widehat G^{(0)pp}_{11'} \nonumber \\
&=\frac12 \widehat{c}^\gamma_{11'}
J^{(0)}_{\alpha\alpha'} \left(\underline{L}^+-\underline{L}^-\right)
\cdot\underline{\sigma}_{\sigma\sigma'},
\end{align}
where
\begin{align}
\widehat{c}^\gamma_{11'}\equiv \widehat{c}^\gamma_{\alpha\alpha'}=
-\frac{1}{2}(\delta_{\alpha\gamma}-\delta_{\alpha'\gamma}).
\end{align}
We introduce the shorthand notation
\begin{align}
\widetilde G_{(0)11'}&=\sum_{pp'} p\,\widehat G^{(0)pp'}_{11'}
\end{align}
for the vertex which is first multiplied with the first Keldysh index and then
averaged over the Keldysh indices. The current vertex is just this vertex
multiplied with $\widehat{c}^\gamma_{11'}$:
\begin{align}
\widehat I^{\gamma(0)}_{11'}&=\widehat{c}^\gamma_{11'}\widetilde G^{(0)}_{11'}.
\end{align}
For the Kondo model, the bare vertex $\widetilde G^{(0)}_{11'}$ is
\begin{align}
\label{eq:vertex_Gtilde_initial_condition_L_plus_minus}
\widetilde
G^{(0)}_{11'}&=\frac12J^{(0)}_{\alpha\alpha'}\left(\underline{L}
^+-\underline{L}^-\right)\cdot\underline{\sigma}_{\sigma\sigma'}.
\end{align}

\subsection{Superoperator algebra}

\label{sec:superoperator_algebra}

Before we proceed with the parametrization of all superoperators, we define a
set of convenient basis superoperators which form a closed algebra.

\subsubsection{Vector superoperators}

We define the following vector basis superoperators:
\begin{align}
\underline{L}^1&=\phantom{-}\frac12\left(\underline{L}^+-\underline{L}^-\right)
-i\underline{L}^+\times\underline{L}^-, \\
\underline{L}^2&=-\frac12\left(\underline{L}^++\underline{L}^-\right), \\
\underline{L}^3&=\phantom{-}\frac12\left(\underline{L}^+-\underline{L}^-\right)
+i\underline{L}^+\times\underline{L}^-.
\end{align}
This means that the bare
vertex~\eqref{eq:vertex_G_initial_condition_L_plus_minus} can be expressed
using $\underline{L}^2$, and the current
vertex~\eqref{eq:vertex_I_initial_condition_L_plus_minus} using
$\left(\underline{L}^1+\underline{L}^3\right)$.

Note that this was the case even if we had not included the terms
$\sim\underline{L}^+\times\underline{L}^-$ in the definition of
$\underline{L}^{1,3}$. However, such terms are generated by the RG, and
including them in the basis superoperators makes the calculations simpler.

It should be noted that no other independent vector superoperators can be found
by combining $\underline{L}^+$ and $\underline{L}^-$ in an arbitrary way.

\subsubsection{Scalar superoperators}

We define the two scalar superoperators $L^a$ and $L^b$ by
\begin{align}
L^a&=\frac34+\underline{L}^+\cdot\underline{L}^-
&
L^b&=\frac14-\underline{L}^+\cdot\underline{L}^-.
\end{align}
These are the only independent scalar superoperators that can be formed from
$\underline{L}^+$ and $\underline{L}^-$.

\subsubsection{Trace of the basis superoperators}

The trace of some of the basis superoperators is zero. This means that applying
them to any operator $b$ will yield an operator with zero trace:
\begin{align}
\label{eq:algebra_trace}
\text{Tr}\,L^a&=0, & \text{Tr}\,\underline{L}^2&=0, &
\text{Tr}\,\underline{L}^3&=0.
\end{align}
For the other two basis superoperators $\underline{L}^1$ and $L^b$, there exist
operators $b$ for which $\text{Tr}\,\underline{L}^1b$ or $\text{Tr}L^bb$ are
non-zero. We note the properties
\begin{align}
\label{eq:trace_Lb_L1}
\text{Tr}\,L^b b&=\text{Tr}\,b , &
\text{Tr}\,\underline{L}^1 b&=\text{Tr}\,\underline{\sigma} b,
\end{align}
where $\underline{\sigma}$ acts on the quantum dot, and not on the reservoir
spins as in the rest of this paper.

Therefore, only $\underline{L}^1$ and $L^b$ are relevant for the
current vertex and the current kernel.

\subsubsection{Behavior of the basis superoperators under the $c$-transform}

\begin{align}
\left(L^{a,b}\right)^c&=L^{a,b}, &
\left(\underline{L}^{1,3}\right)^c&=\underline{L}^{1,3}, &
\left(\underline{L}^2\right)^c&=-\underline{L}^2.
\end{align}

\subsubsection{Products of scalar superoperators}

\begin{align}
\label{eq:algebra_products_LaLb}
\begin{aligned}
\left(L^a\right)^2&=L^a, & L^aL^b&=0, \\
L^bL^a&=0, & \left(L^b\right)^2&=L^b.
\end{aligned}
\end{align}

\subsubsection{Scalar multiplication of scalar and vector superoperators}

\label{sec:algebra_scalar_multiplication_with_LaLb}

The only non-zero products of scalar and vector basis superoperators are
\begin{align}
L^a\underline{L}^2&=\underline{L}^2, & \underline{L}^1L^a&=\underline{L}^1, \\
L^a\underline{L}^3&=\underline{L}^3, & \underline{L}^2L^a&=\underline{L}^2, \\
L^b\underline{L}^1&=\underline{L}^1, & \underline{L}^3L^b&=\underline{L}^3.
\end{align}

\subsubsection{Scalar products of vector superoperators}

\label{sec:algebra_scalar_products}

The only non-zero scalar products of the vector superoperators
$\underline{L}^{1,2,3}$ are
\begin{align}
\underline{L}^1\cdot\underline{L}^3&=3L^b, \\
\underline{L}^2\cdot\underline{L}^2&=\frac12L^a, \\
\underline{L}^3\cdot\underline{L}^1&=L^a.
\end{align}

\subsubsection{Vector products of vector superoperators}

\label{sec:algebra_vector_products}

The only non-zero vector products of the vector superoperators
$\underline{L}^{1,2,3}$ are
\begin{align}
i\underline{L}^1\times\underline{L}^2&=\underline{L}^1, &
i\underline{L}^2\times\underline{L}^2&=\frac12\underline{L}^2, \\
i\underline{L}^2\times\underline{L}^3&=\underline{L}^3, &
i\underline{L}^3\times\underline{L}^1&=2\underline{L}^2.
\end{align}
Closely related to these vector products are the commutator relations
\begin{align}
\label{eq:commutator_L2L123}
\left[L^2_i,L^{1,2,3}_j\right]&=-\frac{i}{2}\epsilon_{ijk}L^{1,2,3}_k.
\end{align}

\subsubsection{Extending the basis superoperators to the reservoir spin space}

The vector superoperators $\underline{L}^{1,2,3}$ and the scalar superoperators
$L^{a,b}$ act on operators of the local dot. In the superoperators
that are of interest in the context of the isotropic Kondo model, i.e., the
effective (current) vertex, the effective Liouvillian, and the current kernel,
they always appear together with the vector of the reservoir Pauli matrices or
the identity matrix in the reservoir spin space. The reason is that any other
combination of dot and reservoir superoperators would violate spin-rotational
invariance.

Therefore, it is convenient to define new superoperators, which act both on
operators of the local dot and the reservoir spin state, by
\begin{align}
\label{eq:definition_L123_sigma}
\widehat L^{1,2,3}_{\sigma\sigma'}&=\underline{L}^{1,2,3}
\cdot\underline{\sigma}_{\sigma\sigma'}, \\
\label{eq:definition_Lab_sigma}
\widehat L^{a,b}_{\sigma\sigma'}&=L^{a,b}\delta_{\sigma\sigma'}.
\end{align}
It will turn out that $\widehat L^{1,2,3,a,b}_{\sigma\sigma'}$ are sufficient
to describe not only the initial conditions of all superoperators, but also all
terms which are generated by the RG in leading and sub-leading order.

Multiplication of these superoperators is defined by
\begin{align}
\left(\widehat L^i \widehat L^j\right)_{\sigma\sigma'}&=
\sum_{\sigma_1}\widehat L^i_{\sigma\sigma_1} \widehat L^j_{\sigma_1\sigma'}.
\end{align}
The results of all such multiplications can be derived from the properties of
the superoperators $\underline{L}^{1,2,3}$ and $L^{a,b}$ and the property
\begin{align}
\sigma^i\sigma^j=\delta_{ij}+i\epsilon_{ijk}\sigma^k
\end{align}
of the Pauli matrices. They are summarized in Table~\ref{table:algebra}.
\begin{table}
\begin{tabular}{c|ccccc}
      & $\widehat L^a$ & $\widehat L^b$ & $\widehat L^1$        & $\widehat L^2$
                        & $\widehat L^3$  \\
\hline
$\widehat L^a$ & $\widehat L^a$ & $0$   & $0$          & $\widehat L^2$
               & $\widehat L^3$  \\
$\widehat L^b$ & $0$   & $\widehat L^b$ & $\widehat L^1$        & $0$
               & $0$    \\
$\widehat L^1$ & $\widehat L^1$ & $0$   & $0$          & $\widehat L^1$
               & $3\widehat L^b$ \\
$\widehat L^2$ & $\widehat L^2$ & $0$   & $0$          &
$\frac12\left(\widehat L^a+\widehat L^2\right)$ & $\widehat L^3$  \\
$\widehat L^3$ & $0$   & $\widehat L^3$ & $\widehat L^a+2\widehat L^2$   & $0$
               & $0$    \\
\end{tabular}

\caption{This table shows the result $\widehat L^i\widehat L^j$ of the
multiplication of any two basis superoperators $\widehat L^i$ and $\widehat
L^j$ from the superoperator algebra for the isotropic Kondo model, which
correspond to the row and column of the table, respectively.
\label{table:algebra}}
\end{table}

Sometimes, it is also necessary to multiply the Pauli matrices in reverse order
in the RG equations. To make this more convenient, we define
\begin{align}
\left(\widehat L^i\right)^T_{\sigma\sigma'}=\widehat L^i_{\sigma'\sigma}.
\end{align}
The result $\left[\left(\widehat L^i\right)^T\left(\widehat
L^j\right)^T\right]^T$ of the multiplication of these transposed superoperators
only differs in some minus signs from $\widehat L^i\widehat L^j$. The
results are summarized in Table~\ref{table:algebra_transposed}.
\begin{table}
\begin{tabular}{c|ccccc}
      & $\widehat L^a$ & $\widehat L^b$ & $\widehat L^1$        & $\widehat L^2$
                        & $\widehat L^3$  \\
\hline
$\widehat L^a$ & $\widehat L^a$ & $0$   & $0$          & $\widehat L^2$
               & $\widehat L^3$  \\
$\widehat L^b$ & $0$   & $\widehat L^b$ & $\widehat L^1$        & $0$
               & $0$    \\
$\widehat L^1$ & $\widehat L^1$ & $0$   & $0$          & $-\widehat L^1$
               & $3\widehat L^b$ \\
$\widehat L^2$ & $\widehat L^2$ & $0$   & $0$          &
$\frac12\left(\widehat L^a-\widehat L^2\right)$ & $-\widehat L^3$  \\
$\widehat L^3$ & $0$   & $\widehat L^3$ & $\widehat L^a-2\widehat L^2$   & $0$
               & $0$    \\
\end{tabular}

\caption{This table shows the result $\left[\left(\widehat
L^i\right)^T\left(\widehat L^j\right)^T\right]^T$ of the
multiplication of any two transposed basis superoperators $\widehat L^i$ and
$\widehat L^j$ from the superoperator algebra for the isotropic Kondo model,
which correspond to the row and column of the table, respectively.
\label{table:algebra_transposed}}
\end{table}

The trace over the reservoir spin indices only is denoted by $\text{Tr}_\sigma$.
We obtain
\begin{align}
\label{eq:trace_spin_indices}
\text{Tr}_\sigma\,\widehat{L}^{a,b} &= 2 L^{a,b}, &
\text{Tr}_\sigma\,\widehat{L}^{1,2,3} &= 0.
\end{align}

\subsection{Parametrization of the effective vertices, the effective
Liouvillian, and the current kernel}

Using the superoperator algebra defined in the previous subsection, the bare
vertex~\eqref{eq:vertex_G_initial_condition_L_plus_minus} can be written as
\begin{align}
\label{eq:bare_vertex_G_L2}
\widehat G^{(0)}_{11'}&=\sum_{p}\widehat G^{pp(0)}_{11'}=\frac12
J^{(0)}_{\alpha\alpha'} \left(\underline{L}^++\underline{L}^-\right)\cdot
\underline{\sigma}_{\sigma\sigma'} \nonumber \\
&=-J^{(0)}_{\alpha\alpha'}\widehat L^2_{\sigma\sigma'}.
\end{align}
Analogously, the bare current
vertex~\eqref{eq:vertex_I_initial_condition_L_plus_minus} is
\begin{align}
\widehat{I}^{\gamma(0)}_{11'}&=\frac12 \widehat{c}^\gamma_{11'} J^{(0)}_{\alpha\alpha'} \left(\widehat
L^1_{\sigma\sigma'}+\widehat L^3_{\sigma\sigma'}\right),
\end{align}
and similarly, the vertex $\widetilde G^{(0)}_{11'}$ [cf.
Eq.~\eqref{eq:vertex_Gtilde_initial_condition_L_plus_minus}] is
\begin{align}
\label{eq:bare_vertex_Gtilde_L_13}
\widetilde
G^{(0)}_{11'}&=\frac12J^{(0)}_{\alpha\alpha'}\left(\widehat
L^1_{\sigma\sigma'}+\widehat L^3_{\sigma\sigma'}\right).
\end{align}
However, the trace over $\underline{L}^3$ vanishes [cf.
Eq.~\eqref{eq:algebra_trace}], such that this part does not contribute to the
current. Therefore, when the trace is taken from the left, it is sufficient to include the term $\sim
\widehat L^1_{\sigma\sigma'}$ in the current vertex:
\begin{align}
\label{eq:bare_vertex_I_L1_trace}
\text{Tr}\,\widehat{I}^{\gamma(0)}_{11'}&=\frac12 \widehat{c}^\gamma_{11'}
J^{(0)}_{\alpha\alpha'} \text{Tr}\,\widehat L^1_{\sigma\sigma'}.
\end{align}
In the following, we will omit the trace when considering the current vertex or
the current kernel and always imply implicitly that it is taken from the left, i.e., we use
\begin{align}
\label{eq:bare_vertex_I_L1}
\widehat{I}^{\gamma(0)}_{11'}&\rightarrow\frac12 \widehat{c}^\gamma_{11'}
J^{(0)}_{\alpha\alpha'}\,\widehat L^1_{\sigma\sigma'}.
\end{align}

To find a convenient parametrization of all superoperators during the entire RG
flow, we use the symmetry properties [cf. Eqs.~\eqref{eq:G_permutation}
and~\eqref{eq:G_c_trafo}, which also apply for the current vertex and the
current kernel]
\begin{align}
\label{eq:symmetry_vertex_1}
G_{11'}(E)&=-G_{1'1}(E), & I^\gamma_{11'}(E)&=-I^\gamma_{1'1}(E), \\
G_{11'}(E)^c&=-G_{\bar1'\bar1}(-E^*), &
I^\gamma_{11'}(E)^c&=-I^\gamma_{\bar1'\bar1}(-E^*), \\
\label{eq:symmetry_L_Sigma}
L(E)^c&=-L(-E^*), & \Sigma_\gamma(E)^c&=-\Sigma_\gamma(-E^*),
\end{align}
the property that the effective Liouvillian and the effective vertex have zero
trace [Eq.~\eqref{eq:G_trace}],
\begin{align}
\text{Tr}\,L(E)&=0, & \text{Tr}\,G_{11'}(E)&=0,
\end{align}
charge conservation, and spin-rotational invariance.

Spin-rotational invariance limits the terms which contribute to the
superoperators to the basis superoperators $\widehat L^{a,b,1,2,3}$, which have
been introduced in the previous subsection.

From Eq.~\eqref{eq:symmetry_vertex_1} and charge conservation, we can deduce for
the effective vertex and the effective current vertex that
\begin{align}
G_{11'}(E)&\sim\delta_{\eta,-\eta'}, &
I^\gamma_{11'}(E)&\sim\delta_{\eta,-\eta'},
\end{align}
and that they can be described using $\widehat G_{11'}(E)$ and $\widehat
I^\gamma_{11'}(E)$, which depend only on the reservoir and spin indices (such
that $1\equiv\alpha\sigma$ on the right-hand side of the following equations):
\begin{align}
\label{eq:G_widehatG}
G_{11'}(E)&=
\begin{cases}
\widehat G_{11'}(E), & \text{for $\eta=-\eta'=+$}, \\
-\widehat G_{1'1}(E), & \text{for $\eta=-\eta'=-$}, \\
\end{cases}
\\
\label{eq:I_widehatI}
I^\gamma_{11'}(E)&=
\begin{cases}
\widehat I^\gamma_{11'}(E), & \text{for $\eta=-\eta'=+$}, \\
-\widehat I^\gamma_{1'1}(E), & \text{for $\eta=-\eta'=-$}. \\
\end{cases}
\end{align}
Because the trace of $G_{11'}(E)$ is zero, $\widehat G_{11'}(E)$ cannot contain
any terms $\sim \widehat L^1,L^b$:
\begin{align}
\label{eq:decomposition_G}
\widehat G_{11'}(E)&=\sum_{\chi=a,2,3}G^\chi_{\alpha\alpha'}(E) \widehat
L^\chi_{\sigma\sigma'}.
\end{align}
Comparing with the bare vertex~\eqref{eq:bare_vertex_G_L2} shows that
\begin{align}
\label{eq:initial_condition_G2}
G^{2(0)}_{\alpha\alpha'}(E)&=-J^{(0)}_{\alpha\alpha'}.
\end{align}
$G^2$ thus describes the exchange coupling and is related to $\underline{L}^+$
or $\underline{L}^-$, which multiply any operator with $\underline{S}$ either
from the left or from the right.

On the other hand, $G^3$ is generated from higher-order terms during the RG
flow, and the corresponding superoperator $\widehat L^3$ has a more complicated
matrix structure in Liouville space, which mixes all states. $G^3$ is
important for the generation of the current rate.

Finally, $G^a$ is a term that does not induce any spin flips, but can be
interpreted as potential scattering. It will turn out that no
contributions to $G^a$ will be generated by the RG even in next-to-leading
order.

For the current vertex, only the superoperators which have a non-zero trace are
of interest because the others do not contribute to the current. Therefore, we
can make the ansatz
\begin{align}
\label{eq:decomposition_I}
\widehat I^\gamma_{11'}(E)&=\sum_{\chi=b,1}I^{\gamma\chi}_{\alpha\alpha'}(E)
\widehat
L^\chi_{\sigma\sigma'}.
\end{align}
Comparing with the bare current vertex~\eqref{eq:bare_vertex_I_L1} yields
\begin{align}
\label{eq:initial_condition_I1}
\widehat{I}^{\gamma1(0)}_{\alpha\alpha'}&=\frac12 \widehat{c}^\gamma_{\alpha\alpha'}J^{(0)}_{\alpha\alpha'}
=-\frac14\left(\delta_{\alpha\gamma}-\delta_{\alpha'\gamma}\right)
J^{(0)}_{\alpha\alpha'}.
\end{align}
$I^{\gamma1}$ thus corresponds to an exchange coupling which is responsible for
the current flow. It will turn out that the coupling $I^{\gamma b}$ is not
important.

Similar considerations apply for the effective Liouvillian and the current
kernel: the former has zero trace, and for the latter, only superoperators with
non-zero trace are interesting. Moreover, only the scalar basis superoperators
$\widehat L^{a,b}$ are suitable for them. This motivates the ansatz
\begin{align}
\label{eq:Liouvillian_CurrentKernel_Kondo}
L(E)&=-i\Gamma(E)L^a, & \Sigma_\gamma(E)&=i\Gamma_\gamma(E)L^b.
\end{align}
Because of the symmetry
properties~\eqref{eq:symmetry_vertex_1}--\eqref{eq:symmetry_L_Sigma}, the
quantities $G^\chi_{\alpha\alpha'}(E)$,
$I^{\gamma\chi}_{\alpha\alpha'}(E)$, $\Gamma(E)$, and $\Gamma_\gamma(E)$ fulfill
\begin{align}
\Gamma(E)^*&=\Gamma(-E^*),
\\
\Gamma_\gamma(E)^*&=\Gamma_\gamma(-E^*),
\\
G^a_{\alpha\alpha'}(E)^*&=-G^a_{\alpha'\alpha}(-E^*),
\\
\label{eq:symmetry_G2}
G^2_{\alpha\alpha'}(E)^*&=G^2_{\alpha'\alpha}(-E^*),
\\
G^3_{\alpha\alpha'}(E)^*&=-G^3_{\alpha'\alpha}(-E^*),
\\
I^{\gamma b}_{\alpha\alpha'}(E)^*&=-I^{\gamma b}_{\alpha'\alpha}(-E^*),
\\
I^{\gamma1}_{\alpha\alpha'}(E)^*&=-I^{\gamma1}_{\alpha'\alpha}(-E^*).
\label{eq:symmetry_I1}
\end{align}

\subsection{Spin dynamics, current, and differential conductance}

The information about the physical observables is contained in $\Gamma(E)$ and
$\Gamma_\gamma(E)$. $\Gamma(E)$ is the spin relaxation/decoherence rate. Due to
Eq.~\eqref{eq:trace_Lb_L1}, the expectation value of the spin is given for any
local density
matrix $\rho$ by
\begin{align}
\left\langle \underline{S} \right\rangle&=\frac12\text{Tr}\left\{\underline{L}^1\rho\right\}.
\end{align}
Using Eq.~\eqref{eq:density_matrix_Fourier} and the representation
$L(E)=-i\Gamma(E)L^a$ for the Kondo model, we get
\begin{align}
\left\langle \underline{S} \right\rangle(E)&=
\frac12\text{Tr}\left\{\underline{L}^1\rho(E)\right
\}
\nonumber \\
&=
\frac12\text{Tr}\left\{\underline{L}^1\frac{i}{E+i\Gamma(E)L^a}\rho(t_0)\right
\}
\nonumber \\
&=
\frac12\frac{i}{E+i\Gamma(E)}\text{Tr}\left\{\underline{L}^1\rho(t_0)\right
\}
\nonumber \\
&=\frac{i}{E+i\Gamma(E)}
\left\langle \underline{S} \right\rangle(t_0),
\end{align}
where we have used that $\underline{L}^1 L^a=\underline{L}^1$.

To obtain an expression for the current, we substitute
Eq.~\eqref{eq:Liouvillian_CurrentKernel_Kondo} into Eq.~\eqref{eq:current_Fourier}:
\begin{align}
\left\langle I_\gamma \right\rangle (E) &=i\Gamma_\gamma(E)\text{Tr}\, L^b
\frac{1}{E+i\Gamma(E)L^a}\rho(t_0).
\end{align}
Using Eqs.~\eqref{eq:trace_Lb_L1} and~\eqref{eq:algebra_products_LaLb} yields
\begin{align}
\left\langle I_\gamma \right\rangle (E) &=\frac{i}{E}\Gamma_\gamma(E) =
2\frac{e}{h} \frac{i}{E}\pi \Gamma_\gamma(E),
\end{align}
where we have used that $e=\hbar=1$ in our units. The stationary current is
then [according to Eq.~\eqref{eq:stationary_current_1}]
\begin{align}
\label{eq:stationary_current_current_rate}
\left\langle I_\gamma \right\rangle^{\text{st}}&=\Gamma_{\gamma}\left(i0^+\right)
=2\frac{e}{h}\pi\Gamma_{\gamma}\left(i0^+\right).
\end{align}
To find a convenient description of the differential conductance, we express
variations of the current rate with a tensor $H^\gamma_{\alpha\alpha'}(E)$,
which is defined by
\begin{align}
\label{eq:tensorH_definition}
\pi\delta\Gamma_\gamma(E)&=\sum_{\alpha\alpha'} H^\gamma_{\alpha\alpha'}(E)
\left(\delta\mu_\alpha-\delta\mu_{\alpha'}\right), \\
H^\gamma_{\alpha\alpha'}(E)^*&=-H^\gamma_{\alpha'\alpha}(-E^*).
\end{align}
Current conservation implies that
\begin{align}
\sum_\gamma \left\langle I_\gamma \right\rangle (E) &=0
& &\Rightarrow & \sum_\gamma H^\gamma_{\alpha\alpha'}(E)&=0.
\end{align}
The conductance tensor $G^\gamma_{\alpha\alpha'}(E)$ is defined by
\begin{align}
\label{eq:tensorG_definition}
G^\gamma_{\alpha\alpha'}(E)&=H^\gamma_{\alpha\alpha'}(E)
-H^\gamma_{\alpha'\alpha}(E)
\end{align}
and fulfills
\begin{align}
\label{eq:G_prop_1}
G^\gamma_{\alpha\alpha'}(E)&=-G^\gamma_{\alpha'\alpha}(E). \\
\label{eq:G_prop_2}
G^\gamma_{\alpha\alpha'}(E)^*&=-G^\gamma_{\alpha'\alpha}(-E^*), \\
\label{eq:G_prop_3}
\sum_\gamma G^\gamma_{\alpha\alpha'}(E)&=0.
\end{align}
The conductance tensor permits us to write the variation of the current as
\begin{align}
\label{eq:cur_var_1}
\delta \left\langle I_\gamma \right\rangle (E) &=
\frac{i}{E}G_0\frac12\sum_{\alpha\alpha'}
G^\gamma_{\alpha\alpha'}(E)\left(\delta V_\alpha - \delta V_{\alpha'}\right) \\
\label{eq:cur_var_2}
&=
\frac{i}{E}G_0\sum_{\alpha<\alpha'}
G^\gamma_{\alpha\alpha'}(E)\left(\delta V_\alpha - \delta V_{\alpha'}\right) \\
\label{eq:cur_var_3}
&=
\frac{i}{E}G_0\sum_{\alpha}
\left\{\sum_{\alpha'}
G^\gamma_{\alpha\alpha'}(E)\right\}\delta V_\alpha,
\end{align}
where
\begin{align}
G_0&=2\frac{e^2}{h},
\end{align}
and $\mu_\alpha=eV_\alpha$.

\subsection{Shorthand notations}

Before we discuss the initial conditions for the RG flow, and the RG equations
for all couplings and rates, we summarize some shorthand notations which will
be useful in the folowing.

For the vertex $G$ (and similarly for the current vertex $I^\gamma$), we use
different notations, where the multiindex 1 always contains the reservoir
index, and optionally the spin index, and the index $\eta$:
\begin{align}
G_{11'}&\equiv G_{\eta\alpha\sigma,\eta'\alpha'\sigma'}, &
1&\equiv\eta\alpha\sigma, \\
\widehat G_{11'}&\equiv G_{\alpha\sigma,\alpha'\sigma'}, &
1&\equiv\alpha\sigma, \\
G^\chi_{11'}&\equiv G^\chi_{\alpha\alpha'}, &
1&\equiv\alpha.
\end{align}
A hat always indicates that the index $\eta$ is not included. Analogously, we
define energies which are shifted by the chemical potentials of the leads:
\begin{align}
\label{eq:shorthand_widehat_E}
\widehat E_{1\ldots n}&=E+\widehat\mu_{1\ldots n}, \\
\label{eq:shorthand_widehat_mu12}
\widehat \mu_{12}&=\mu_{\alpha_1}-\mu_{\alpha_2}, \\
\label{eq:shorthand_widehat_mu1234}
\widehat\mu_{1234}&=\widehat\mu_{12}+\widehat\mu_{34},\quad\text{etc.,}
\end{align}
We will replace the relevant vertex functions $G^2$, $G^3$, and $I^{\gamma 1}$
by the more convenient
\begin{align}
\label{eq:definition_J}
J_{11'}(E)&=-G^2_{11'}(E), \\
\label{eq:definition_K}
K_{11'}(E)&=-i\frac{2}{\pi}G^3_{11'}(\widehat E_{1'1}), \\
\label{eq:definition_I}
I^\gamma_{11'}(E)&=-4I^{\gamma1}_{11'}(E).
\end{align}
We note that the last symbol $I^\gamma_{11'}$ is not unambigious since it was
also defined for the full current vertex with $1\equiv\eta\alpha\sigma$. However,
in the context it will always be clear whether we consider the case $1\equiv\alpha$
or $1\equiv\eta\alpha\sigma$.

From the symmetry properties~\eqref{eq:symmetry_G2}--\eqref{eq:symmetry_I1} of
the original vertex functions, we can conclude
\begin{align}
J_{11'}(E)^*&=J_{1'1}(-E^*),\\
K_{11'}(E)^*&=K_{1'1}(-E^*),\\
I^\gamma_{11'}(E)^*&=-I^\gamma_{1'1}(-E^*).
\end{align}

\subsection{Initial conditions}

The initial conditions for the RG flow at high energies $E$ are determined
using a perturbative calculation as outlined in
Sec.~\ref{subsec:initial_conditions}. The idea is to find those terms in
lowest order in $J^{(0)}_{\alpha\alpha'}$ which are universal and not
logarithmically divergent.

\subsubsection{Vertex functions}

The initial values for $J_{\alpha\alpha'}(E)$ and $I^\gamma_{\alpha\alpha'}(E)$
are already known, cf. Eqs.~\eqref{eq:initial_condition_G2}
and~\eqref{eq:initial_condition_I1}:
\begin{align}
J_{\alpha\alpha'}=-G^2_{\alpha\alpha'}&\rightarrow J^{(0)}_{\alpha\alpha'}, \\
I^\gamma_{\alpha\alpha'}=-4I^{\gamma1}_{\alpha\alpha'} &\rightarrow
\left(\delta_{\alpha\gamma}-\delta_{\alpha'\gamma}\right)J^{(0)}_{\alpha\alpha'}
.
\end{align}

The initial condition for
$K_{\alpha\alpha'}(E)=-i\frac{2}{\pi}G^3_{\alpha\alpha'}\left(\widehat
E_{\alpha'\alpha}\right)$ is determined by considering the lowest order
diagrams for the effective vertex, and disregarding non-universal terms and
logarithmic terms. The result is [cf. Eq.~\eqref{eq:G_initial_final}]:
\begin{align}
\widehat G_{12}(E)&=
\widehat G_{12}^{(0)}-i\frac{\pi}{2}
\left\{\widehat G_{13}^{(0)}\widetilde G_{32}^{(0)}
- \widehat G_{32}^{(0)}\widetilde G_{13}^{(0)}
\right\}.
\end{align}
For the Kondo model, the bare vertices $\widehat G^{(0)}$ and
$\widetilde G^{(0)}$ are given by Eqs.~\eqref{eq:bare_vertex_G_L2}
and~\eqref{eq:bare_vertex_Gtilde_L_13}, respectively. Using the superoperator
algebra yields
\begin{align}
\widehat G_{13}^{(0)}\widetilde G_{32}^{(0)}
&=-\frac12 J^{(0)}_{\alpha_1\alpha_3} J^{(0)}_{\alpha_3\alpha_2}
\widehat L^2_{\sigma_1\sigma_3}\left(\widehat L^1_{\sigma_3\sigma_2}+\widehat
L^3_{\sigma_3\sigma_2}\right) \nonumber \\
&=
-\frac12 J^{(0)}_{\alpha_1\alpha_3} J^{(0)}_{\alpha_3\alpha_2}
\widehat L^3_{\sigma_1\sigma_2}
\end{align}
and
\begin{align}
&\widehat G_{32}^{(0)}\widetilde G_{13}^{(0)}\nonumber \\
&=-\frac12 J^{(0)}_{\alpha_3\alpha_2}\,J^{(0)}_{\alpha_1\alpha_3}\,
\left\{\left(\widehat L^2\right)^T
\left[\left(\widehat L^1\right)^T+\left(\widehat
L^3\right)^T\right]\right\}^T_{\sigma_1\sigma_2} \nonumber\\
&=
\frac12 J^{(0)}_{\alpha_1\alpha_3} J^{(0)}_{\alpha_3\alpha_2}
\widehat L^3_{\sigma_1\sigma_2}.
\end{align}
Finally, the initial condition for the vertex is
\begin{align}
\widehat G_{12}(E=iD)&=-J^{(0)}_{\alpha_1\alpha_2}\widehat L^2_{\sigma_1\sigma_2}
+i\frac{\pi}{2}J^{(0)}_{\alpha_1\alpha_3} J^{(0)}_{\alpha_3\alpha_2}
\widehat L^3_{\sigma_1\sigma_2}.
\end{align}
This means that the initial value for the vertex function $G^3$ is
\begin{align}
G^{3(0)}_{12}=i\frac{\pi}{2}J^{(0)}_{\alpha_1\alpha_3}
J^{(0)}_{\alpha_3\alpha_2}
\end{align}
and for the corresponding simplified function
\begin{align}
\label{eq:initial_condition_K}
K^{(0)}_{12}=J^{(0)}_{\alpha_1\alpha_3}
J^{(0)}_{\alpha_3\alpha_2}=\left[\left(J^{(0)}\right)^2\right]_{\alpha_1\alpha_2
} ,
\end{align}
where, in the last equation, we imply matrix multiplication w.r.t. the reservoir
indices.

\subsubsection{Effective Liouvillian and current kernel}

The perturbative solution for the effective Liouvillian for $T,V\ll|E|\ll D$
is given by Eq.~\eqref{eq:L_initial_2}. The condition \eqref{eq:condition_D} is
fulfilled,
\begin{align}
\nonumber
p G_{12}^{(0)}G_{\bar{2}\bar{1}}^{(0)pp}&=2\widehat{G}_{12}^{(0)}\widetilde{G}_{21}^{(0)}
=J^{(0)}_{\alpha_1\alpha_2}J^{(0)}_{\alpha_2\alpha_1}\text{Tr}_\sigma\,
\widehat{L}^2(\widehat{L}^1+\widehat{L}^3)\\
&=J^{(0)}_{\alpha_1\alpha_2}J^{(0)}_{\alpha_2\alpha_1}\text{Tr}_\sigma\,\widehat{L}^3=0,
\end{align}
such that all terms $\sim D$ vanish. Nevertheless, as already outlined in
Section~\ref{subsec:initial_conditions}, the problem with
the initial condition for the effective Liouvillian is that it contains non-universal terms
which cannot be neglected because they are proportional to the Fourier variable and hence
very large for $E=iD$. Therefore, we will use an alternative scheme to find the initial condition for
the Liouvillian, which will be discussed in
Sec.~\ref{sec:IC_finite_flow_parameter}.

However, for the derivative of the Liouvillian and for its variation, the linear terms in $E$
can be omitted, and we can use Eqs.~\eqref{eq:initial_Z}
and~\eqref{eq:delta_L_initial_final},
\begin{align}
\label{eq:Z_initial_kondo}
Z(E=iD)&=1,\\
\nonumber
\delta L(E=iD)&=-i\frac{\pi}{2}\widehat{G}_{12}^{(0)}\widetilde{G}_{21}^{(0)}\delta\hat{\mu}_{12}\\
&\hspace{-1cm}
=i\frac{\pi}{4}J^{(0)}_{\alpha_1\alpha_2}J^{(0)}_{\alpha_2\alpha_1}\delta\hat{\mu}_{\alpha_1\alpha_2}
\text{Tr}_\sigma\,\widehat{L}^2\left(\widehat{L}^1+\widehat{L}^3\right)=0.
\label{eq:variation_L_initial_kondo}
\end{align}
This gives
\begin{align}
\label{eq:variation_gamma_initial}
\delta\Gamma(E=iD)=0.
\end{align}

For the perturbative calculation of the current kernel $\Sigma_\gamma(E)$ for $T,V\ll|E|\ll D$, we have
to replace the first vertex $G_{12}^{(0)}$ of Eq.~\eqref{eq:L_initial_2} by the current
vertex $I^{\gamma(0)}_{12}$ and omit the first term $L^{(0)}$. Using $L^{(0)}=0$ for the Kondo model
without magnetic field, we obtain
\begin{align}
\Sigma_\gamma(E)&=\frac12 I^{(0)}_{12}\left[-ip\frac{\pi}{2}\bar{\mu}_{12}+
E_{12}\ln\frac{-iE_{12}}{D}\right.
\nonumber\\
&\quad\quad\quad\left. +
\frac14\left(\frac{\pi^2}{4}-3\right)E_{12}-p\frac{\pi}{2}(D+iE)\right]G^{(0)pp}_{\bar2\bar1}
\nonumber\\
&=-i\frac{\pi}{2}\widehat\mu_{12}\widehat I^{(0)}_{12} \widetilde G^{(0)}_{21}
-\frac{\pi}{2}(D+iE)\widehat I^{(0)}_{12} \widetilde G^{(0)}_{21}
\nonumber\\
\label{eq:kernel_initial_zw}
&\quad+\left[\widehat E_{12}\ln\frac{-i\widehat E_{12}}{D}
+\frac14\left(\frac{\pi^2}{4} -3\right)\widehat E_{12}\right]\widehat I^{(0)}_{12}
\widehat G^{(0)}_{21}.
\end{align}
Inserting Eqs.~\eqref{eq:bare_vertex_G_L2}, \eqref{eq:bare_vertex_Gtilde_L_13},
and~\eqref{eq:bare_vertex_I_L1} for the vertices, we find that, except for the first one,
all terms are zero due to
\begin{align}
\nonumber
&\widehat I^{\gamma(0)}_{12}\widetilde G_{21}^{(0)}=
\widehat c^{\gamma}_{12}\widetilde G^{(0)}_{12}\widetilde G^{(0)}_{21}\\
\nonumber
&=-\frac{1}{8}(\delta_{1\gamma}-\delta_{2\gamma})
J^{(0)}_{12}J^{(0)}_{21}
\text{Tr}_\sigma \widehat L^1 \left(\widehat L^1 + \widehat L^3\right)=0,\\
\nonumber
&W(\widehat E_{12})\widehat I^{\gamma(0)}_{12}\widehat G_{21}^{(0)}=
W(\widehat E_{12})\widehat c^\gamma_{12}\widetilde G^{(0)}_{12}\widehat G_{21}^{(0)}\\
\nonumber
&=-\frac{1}{2}W(\widehat E_{12})\widehat c^\gamma_{12}
J^{(0)}_{12}J^{(0)}_{21}
\text{Tr}_\sigma \widehat L^1 \widehat L^2=0,
\end{align}
where $1\leftrightarrow 2$ was used in the first relation, and
$\text{Tr}_\sigma \widehat L^1 \widehat L^2 = \text{Tr}_\sigma \widehat L^1 =0$ in
the second one. Here, $W(E)$ denotes any function of $E$.

The first term of Eq.~\eqref{eq:kernel_initial_zw} can be evaluated as
\begin{align}
\nonumber
\Sigma_\gamma(E=iD)\,&\,\\
&\hspace{-1cm}
=i\frac{\pi}{16}\widehat{\mu}_{12}\left(\delta_{1\gamma}-\delta_{2\gamma}\right)
J^{(0)}_{12}J^{(0)}_{21}
\text{Tr}_\sigma \widehat L^1\left(\widehat L^1+\widehat L^3\right)
\nonumber \\
&\hspace{-1cm}
=i\frac{3\pi}{8}\widehat{\mu}_{12}\left(\delta_{1\gamma}-\delta_{2\gamma}\right)
J^{(0)}_{12}J^{(0)}_{21} L^b,
\end{align}
where we have used
\begin{align}
\text{Tr}_\sigma \widehat L^1\left(\widehat L^1+\widehat L^3\right)
=\text{Tr}_\sigma 3\widehat L^b=6 L^b.
\end{align}
Using the representation $\Sigma_\gamma(E)=i\Gamma_\gamma(E)$ for the current
kernel, this results in the initial value
\begin{align}
\label{eq:initial_condition_current_rate}
\Gamma_\gamma(E=iD)&=\frac{3\pi}{8}\widehat\mu_{12}\left(\delta_{1\gamma}
-\delta_{2\gamma}\right) J^{(0)}_{12}J^{(0)}_{21}
\end{align}
for the current rate. The variation of the current rate is
\begin{align}
\delta\Gamma_\gamma(E=iD)&=\frac{3\pi}{8}\delta\widehat\mu_{12}\left(\delta_{1\gamma}
-\delta_{2\gamma}\right) J^{(0)}_{12}J^{(0)}_{21},
\end{align}
and the initial values of the tensors $H^\gamma_{12}$ and $G^\gamma_{12}$,
defined in Eqs.~\eqref{eq:tensorH_definition}
and~\eqref{eq:tensorG_definition}, respectively, are therefore
\begin{align}
H^\gamma_{12}(E=iD)&=\frac{3\pi^2}{8}\left(\delta_{1\gamma}
-\delta_{2\gamma}\right) J^{(0)}_{12}J^{(0)}_{21}, \\
G^\gamma_{12}(E=iD)&=\frac{3\pi^2}{4}\left(\delta_{1\gamma}
-\delta_{2\gamma}\right) J^{(0)}_{12}J^{(0)}_{21}.
\end{align}

\subsubsection{Summary of the initial conditions}

\label{sec:Summary-Initial-Conditions-Kondo}

At high energies, $E=iD$, we found the following initial conditions for the
isotropic Kondo model without magnetic field:
\begin{align}
\label{eq:initial_conditions_summary_begin}
G^2_{12}\left(E=iD\right)&=-J^{(0)}_{12}, \\
G^3_{12}\left(E=iD\right)&=i\frac{\pi}{2}J^{(0)}_{13}
J^{(0)}_{32}, \\
K_{12}\left(E=iD\right)&=J^{(0)}_{13}
J^{(0)}_{32}, \\
I^{\gamma 1}_{12}\left(E=iD\right)&=-\frac{1}{4}
\left(\delta_{1\gamma}-\delta_{2\gamma}\right)
J^{(0)}_{12}, \\
I^{\gamma}_{12}\left(E=iD\right)&=
\left(\delta_{1\gamma}-\delta_{2\gamma}\right)
J^{(0)}_{12}, \\
Z\left(E=iD\right)&=1, \\
\delta\Gamma\left(E=iD\right)&=0, \\
\Gamma_\gamma\left(E=iD\right)&=\frac{3\pi}{8}\widehat\mu_{12}
\left(\delta_{1\gamma}-\delta_{2\gamma}\right) J^{(0)}_{12}J^{(0)}_{21}, \\
\delta\Gamma_\gamma\left(E=iD\right)&=\frac{3\pi}{8} \delta\widehat\mu_{12}
\left(\delta_{1\gamma}-\delta_{2\gamma}\right) J^{(0)}_{12}J^{(0)}_{21}, \\
H^\gamma_{12}\left(E=iD\right)&=\frac{3\pi^2}{8}
\left(\delta_{1\gamma}-\delta_{2\gamma}\right) J^{(0)}_{12}J^{(0)}_{21}, \\
G^\gamma_{12}\left(E=iD\right)&=\frac{3\pi^2}{4}
\left(\delta_{1\gamma}-\delta_{2\gamma}\right) J^{(0)}_{12}J^{(0)}_{21}.
\label{eq:initial_conditions_summary_end}
\end{align}

\subsection{RG equations}

We will now discuss how the generic RG equations for the effective
Liouvillian~\eqref{eq:L_rg_final}, for the effective
vertex~\eqref{eq:G_rg_final}, the variation of the effective
Liouvillian~\eqref{eq:L_rg_V_final}, and the corresponding equations for the
current vertex and the current kernel are evaluated for the isotropic Kondo
model.

\subsubsection{Strategy for the selection and evaluation of diagrams}

\label{sec:strategy-diagram-selection}

It has been shown in Sec.~\ref{sec:Summary-Initial-Conditions-Kondo} that
the initial conditions for the vertex functions at $E=iD$ fulfill
\begin{align}
J\left(E=iD\right)&\sim G^2\left(E=iD\right)\sim J^{(0)}, \\
K\left(E=iD\right)&\sim G^3\left(E=iD\right)\sim\left[J^{(0)}\right]^2, \\
I^\gamma\left(E=iD\right)&\sim I^{\gamma1}\left(E=iD\right)\sim J^{(0)},
\end{align}
where we have omitted the reservoir indices. For arbitrary $E$, we use
\begin{align}
J_{12}(E)&=-G^2_{12}(E)
\end{align}
as a reference scale for the RG flow, in the sense that we calculate all
quantities, such as vertex functions and rates, in leading and subleading order in
$J_{12}(E)$. We will see that the relations
\begin{align}
K(E)&\sim \left[J(E)\right]^2, & I^\gamma(E)&\sim J(E)
\end{align}
still hold during the RG flow.

It will be shown later [cf. Eq.~\eqref{eq:G2_rg_leading_order}] that the
behavior of the scale $J$ at large $E$ is (note that we omit the reservoir
indices and the Fourier variable for simplicity)
\begin{align}
\frac{\partial J}{\partial E}\sim \frac{1}{E}J^2,
\end{align}
which implies
\begin{align}
\frac{\partial }{\partial E}J^n\sim \frac{1}{E}J^{n+1}.
\end{align}
This means that we have to include the following terms on the right-hand side for
$J$ and other quantities which are $\sim J$ at large $E$:
\begin{align}
&\sim \frac{1}{E}J^2 & &\text{(leading order), or} \\
&\sim \frac{1}{E}J^3 & &\text{(subleading order).}
\end{align}
This includes the current vertex $I^\gamma$, which is $\sim J$ according to the
initial condition~\eqref{eq:initial_condition_I1}, and the rate $\Gamma$, which
is $\sim J$ because we will show that the right-hand side of its RG equation is
$\sim\frac{1}{E}J^2$ in leading order [cf. Eq.~\eqref{eq:L_rg_leading_order}].
In the RG equations for these quantities, we discard terms such as
\begin{align}
&\frac{1}{E}J^4, & &\frac{\Delta}{E^2}J^3,
\end{align}
which are of higher order in $J$ or have a prefactor $\frac{\Delta}{E}$,
where $\Delta$ is an energy scale, like, e.g., the voltage $V$, which
fulfills $|\Delta|\ll |E|$.

On the other hand, $K$ and $\Gamma_\gamma$ are $\sim J^2$ according to the
initial conditions~\eqref{eq:initial_condition_K}
and~\eqref{eq:initial_condition_current_rate}. Therefore, we have to consider
all terms
\begin{align}
&\sim \frac{1}{E}J^3 & &\text{(leading order), or} \\
&\sim \frac{1}{E}J^4 & &\text{(subleading order)}
\end{align}
on the right-hand side of their RG equations in order to cover all important
contributions.

There are still two special cases which have to be considered separately,
namely, the coupling $G^a$ and the variation $\delta\Gamma$ of the rate
$\Gamma$:
\begin{itemize}
\item
We will show later [cf. Eq.~\eqref{eq:Ga-leading-order}] that the part
$G^a$ of the vertex is $G^a\sim\frac{V}{E}J^2$ in leading order.
Therefore, its contribution to the right-hand side of the effective vertex, the current
vertex, and the effective Liouvillian (which are all $\sim J$ in leading
order) is at least of the order
\begin{align}
\frac{1}{E}JG^a\sim\frac{V}{E^2}J^3
\end{align}
and can thus be neglected. Therefore, it is not necessary to include $G^a$ in
the calculations.
\item
The leading order term on the right-hand side of $\delta\Gamma$
[Eq.~\eqref{eq:L_rg_V_final}] is $\sim\frac{1}{E}J^2$, which makes
$\delta\Gamma$ itself $\sim J$ in leading order. Considering subleading terms
($\sim\frac{1}{E}J^3$ on the right-hand side, which cause contributions $\sim J^2$ to
$\delta\Gamma$) is not necessary because it would only result in additional
terms $\sim \frac{1}{E}J^4$ beyond the subleading order on the right-hand side of
Eq.~\eqref{eq:L_rg_V_final}, and, as we will see later [cf.
Eq.~\eqref{eq:SigmaI_rg_V_terms_beyond_subleading_order}], terms
$\sim\frac{1}{E}J^5$ beyond the subleading order in the analogous RG equation
for $\delta\Gamma_\gamma$.
\end{itemize}

An overview of the orders up to which the terms on the right-hand side have to be
considered for the different vertex couplings and rates is shown in
Table~\ref{table:diagram_selection_strategy}.
\begin{table}
\begin{tabular}{c|c|c}
 & r.h.s. (leading order) & r.h.s. (subleading order) \\
\hline
$J$ & $\sim\frac{1}{E}J^2$ & $\sim\frac{1}{E}J^3$ \\
$K$ & $\sim\frac{1}{E}J^3$ & $\sim\frac{1}{E}J^4$ \\
$G^a$ & --- & --- \\
$I^\gamma$ & $\sim\frac{1}{E}J^2$ &
$\sim\frac{1}{E}J^3$ \\
\hline
$\Gamma$ & $\sim\frac{1}{E}J^2$ &
$\sim\frac{1}{E}J^3$ \\
$\delta\Gamma$ & $\sim\frac{1}{E}J^2$ &
--- \\
$\delta\Gamma_\gamma$ & $\sim\frac{1}{E}J^3$ &
$\sim\frac{1}{E}J^4$
\end{tabular}

\caption{This table shows the order in $J$ up to which the terms on the
right-hand side of the RG equations have to be considered for all relevant vertex
couplings and rates.
\label{table:diagram_selection_strategy}}
\end{table}

\subsubsection{Propagators and frequency integrals}

For the isotropic Kondo model, the effective Liouvillian $L(E)=-i\Gamma(E)L^a$
has a threefold degenerate eigenvalue $-i\Gamma(E)$. The fourth eigenvalue zero
cannot occur in resolvents between vertices (according to
Ref.~\onlinecite{RTRG_FS}, this follows from the fact that only the vertices
which are averaged over the Keldysh indices appear in the RG equations).

Therefore, we can always replace the quantities $\chi(E)$ and $Z(E)$, defined
in Eq.~\eqref{eq:chi_Z}, which are superoperators in Liouville space, according
to
\begin{align}
\chi(E)=Z(E)\left[E-L(E)\right] & \rightarrow
Z(E)\left[E+i\Gamma(E)\right], \\
Z(E)=\frac{1}{1-\frac{\partial}{\partial E}L(E)} & \rightarrow
\frac{1}{1+i\frac{\partial}{\partial E}\Gamma(E)}
\end{align}
by complex numbers. We use the shorthand notations
\begin{align}
\chi_{1\ldots n}&=\chi\left(E_{1\dots n}\right), \\
Z_{1\ldots n}&=Z\left(E_{1\dots n}\right)
\end{align}
for these in the following.

Consequently, also the propagator
\begin{align}
\Pi_{1\ldots n}=
\Pi\left(E_{1\ldots n}\right)=\frac{Z_{1\ldots n}}{\bar\omega_{1\ldots
n}+\chi_{1\ldots n}}
\end{align}
is a complex number.

This permits us to simplify the RG equations~\eqref{eq:L_rg_final},
\eqref{eq:G_rg_final}, and~\eqref{eq:L_rg_V_final} by factoring out all
frequency-dependent parts, and separating the frequency-integrations from the
evaluation of the frequency-independent vertex superoperators in Liouville space.

The frequency integrals which are required for the evaluation of the RG
equations are
\begin{align}
\label{eq:F1_TV}
F^{(1)}_{12} &=
Z_{12}\int d\omega \int d\omega' \frac{f'(\omega)f'(\omega')}{
\omega+\omega'+\chi_{12}}, \\
\label{eq:F1_2_TV}
F^{(1)}_{12,34} &= Z_{12} Z_{34} \int d\omega \int d\omega'
\frac{f'(\omega)f^a(\omega')}{
(\omega+\omega'+\chi_{12})(\omega+\omega'+\chi_{34})}, \\
\label{eq:F2_TV}
F^{(2)}_{12,34} &= Z_{12}\int d\omega \int d\omega'
\frac{{\mathcal{F}}_{34}(\omega)f'(\omega)f'(\omega')}{
\omega+\omega'+\chi_{12}}, \\
\label{eq:F3_TV}
F^{(3)}_{12} &= -Z_{12}\int d\omega \frac{f'(\omega)}{\omega+\chi_{12}}, \\
\label{eq:F4_TV}
F^{(4)}_{12,34} &= Z_{12}\int d\omega
\frac{{\mathcal{F}}_{34}(\omega)f'(\omega)}{
\omega+\chi_{12}},
\end{align}
where $\mathcal{F}_{1\ldots n}(\omega)$ is a shorthand notation for
\begin{align}
\mathcal{F}_{1\ldots n}(\omega)&=\mathcal{F}\left(E_{1\ldots n},\omega\right),
\end{align}
which has been defined in Eq.~\eqref{eq:integrated_prop_difference_frequency}.
The evaluation of these integrals will be discussed in
Appendix~\ref{sec:integrals}.

Note that the first two of these integrals fulfill the relation
\begin{align}
Z_{12}F^{(1)}_{12}&=F^{(1)}_{12,12},
\end{align}
which can be shown using integration by parts.

Using these integrals,the RG equations~\eqref{eq:L_rg_final},
\eqref{eq:G_rg_final}, and~\eqref{eq:L_rg_V_final} can be written as
\begin{widetext}
\begin{align}
\frac{\partial^2}{\partial E^2}L(E)&=
\frac{1}{2}G_{12}(E)
G_{\bar{2}\bar{1}}(E_{12})
F^{(1)}_{12}
+G_{12}(E)G_{\bar{2}3}(E_{12}) G_{\bar{3}\bar{1}}(E_{13})
\left[F^{(2)}_{13,12}+F^{(2)}_{12,13}\right],
\label{eq:L_rg_Kondo}
\\
\nonumber
\frac{\partial}{\partial E}G_{12}(E)&=
\left[G_{13}(E) G_{\bar{3}2}(E_{13})F^{(3)}_{13}
+G_{34}(E)G_{\bar{4}1}(E_{34})G_{\bar{3}2}(E_{13}) F^{(4)}_{13,34}
\right.\\
\nonumber
&\left.\hspace{1cm}
+G_{13}(E) G_{24}(E_{13}) G_{\bar{4}\bar{3}}(E_{1234})F^{(4)}_{13,1234}
-(1\leftrightarrow 2)\right]\\
\label{eq:G_rg_Kondo}
&\hspace{1cm}
-\frac{1}{2}G_{34}(E)
G_{12}(E_{34}) G_{\bar{4}\bar{3}}(E_{1234}) F^{(1)}_{1234,34},
\\
\nonumber
\frac{\partial}{\partial E}\delta L(E)&=
\frac{1}{2}\delta{\bar\mu}_{12} G_{12}(E) G_{\bar{2}\bar{1}}(E_{12})
F^{(1)}_{12}
-\frac{1}{2}G_{12}(E)
Z_{12}\delta L(E_{12})
G_{\bar{2}\bar{1}}(E_{12})
F^{(1)}_{12}
\\
&\hspace{1cm}+G_{12}(E) G_{\bar{2}3}(E_{12})
G_{\bar{3}\bar{1}}(E_{13})
\left[
\delta{\bar\mu}_{13}F^{(2)}_{13,12}
+\delta{\bar\mu}_{12}F^{(2)}_{12,13}\right].
\label{eq:L_rg_V_Kondo}
\end{align}
\end{widetext}
Analogous equations can be found for the current kernel and the current vertex
by replacing the leftmost effective vertex $G$ with a current vertex $I^\gamma$.

\subsubsection{Summation over $\eta$ indices}

To perform the sum over the $\eta$ indices in the RG equations, we use
Eqs.~\eqref{eq:G_widehatG} and~\eqref{eq:I_widehatI} and the shorthand
notations~(\ref{eq:shorthand_widehat_E}--\ref{eq:shorthand_widehat_mu1234}) for
the quantities with a hat, which do not depend on the $\eta$ indices any more.
We adopt the same notation for $Z_{12}$, $\chi_{12}$, and the integrals
$F^{i}_{12}$ and $F^{i}_{12,34}$, and define that $\widehat Z_{12}$,
$\widehat\chi_{12}$, $\widehat F^{i}_{12}$, and $\widehat F^{i}_{12,34}$ do not
depend on the $\eta$ indices any more, and that always
$\eta_1=-\eta_2=\eta_3=-\eta_4=+$.

We now perform the summation in the different terms in the RG equations for
$L(E)$, $\delta L(E)$, and $G_{12}(E)$.

\paragraph{Terms in the RG equation for $L(E)$:}
The leading order terms are
\begin{align}
&\phantom{=}\frac{1}{2}G_{12}(E) G_{\bar{2}\bar{1}}(E_{12}) F^{(1)}_{12}
\nonumber \\
&=\frac{1}{2}\widehat G_{12}(E) \widehat
G_{21}(\widehat E_{12}) \widehat F^{(1)}_{12}
\nonumber \\
&\phantom{=}+\frac{1}{2}\left[-\widehat G_{21}(E)\right]\left[-\widehat
G_{12}(\widehat E_{21})\right] \widehat F^{(1)}_{21}
\nonumber \\
&=\widehat G_{12}(E) \widehat G_{\bar{2}\bar{1}}(\widehat E_{12})
\widehat
F^{(1)}_{12},
\end{align}
where we have exchanged the indices $1$ and $2$ in the second term to merge it
with the first one.

The subleading order terms are
\begin{align}
&\phantom{=} G_{12}(E)G_{\bar{2}3}(E_{12}) G_{\bar{3}\bar{1}}(E_{13})
\left[F^{(2)}_{13,12}+F^{(2)}_{12,13}\right]
\nonumber \\
&=\widehat G_{12}(E)\widehat G_{23}(\widehat E_{12})
\widehat G_{31}(\widehat E_{13})
\left[\widehat F^{(2)}_{13,12}+\widehat F^{(2)}_{12,13}\right]
\nonumber \\
&\phantom{=}-\widehat G_{21}(E)\widehat G_{32}(\widehat E_{21})
\widehat G_{13}(\widehat E_{31})
\left[\widehat F^{(2)}_{31,21}+\widehat F^{(2)}_{21,31}\right]
\nonumber \\
&=\widehat G_{12}(E)\widehat G_{23}(\widehat E_{12})
\widehat G_{31}(\widehat E_{13})
\left[\widehat F^{(2)}_{13,12}+\widehat F^{(2)}_{12,13}\right]
\nonumber \\
&\phantom{=}-\widehat G_{12}(E)\widehat G_{31}(\widehat E_{12})
\widehat G_{23}(\widehat E_{32})
\left[\widehat F^{(2)}_{32,12}+\widehat F^{(2)}_{12,32}\right].
\end{align}

\paragraph{Terms in the RG equation for $\delta L(E)$:}
Very similar considerations apply to the terms that contribute to the
renormalization of $\delta L(E)$, even though some of them contain an
additional factor $\delta\bar\mu_{12}$. We find
\begin{multline}
\frac{1}{2}\delta{\bar\mu}_{12} G_{12}(E) G_{\bar{2}\bar{1}}(E_{12})
F^{(1)}_{12}\\
=\delta{\widehat\mu}_{12} \widehat G_{12}(E)\widehat
G_{21}(\widehat E_{12}) \widehat F^{(1)}_{12}
\end{multline}
for the leading order terms,
\begin{multline}
-\frac{1}{2} G_{12}(E)
Z_{12}\delta L(E_{12})G_{\bar{2}\bar{1}}(E_{12})
F^{(1)}_{12}\\
=-\widehat G_{12}(E) \widehat Z_{12}\delta
L(\widehat E_{12}) \widehat
G_{21}(\widehat E_{12}) \widehat F^{(1)}_{12}
\end{multline}
for the terms which contain $\delta L$, and
\begin{align}
&\phantom{=}G_{12}(E) G_{\bar{2}3}(E_{12}) G_{\bar{3}\bar{1}}(E_{13})
\left[\delta{\bar\mu}_{13}F^{(2)}_{13,12}
+\delta{\bar\mu}_{12}F^{(2)}_{12,13}\right]
\nonumber \\
&=\widehat G_{12}(E) \widehat G_{23}(\widehat E_{12})
\widehat  G_{31}(\widehat E_{13})
\left[\delta{\widehat \mu}_{13}\widehat F^{(2)}_{13,12}
+\delta{\widehat \mu}_{12}\widehat F^{(2)}_{12,13}\right]
\nonumber \\
&\phantom{=}
-\widehat G_{21}(E) \widehat G_{32}(\widehat E_{21})
\widehat  G_{13}(\widehat E_{31})
\left[\delta{\widehat \mu}_{31}\widehat F^{(2)}_{31,21}
+\delta{\widehat \mu}_{21}\widehat F^{(2)}_{21,31}\right]
\nonumber \\
&=\widehat G_{12}(E) \widehat G_{23}(\widehat E_{12})
\widehat  G_{31}(\widehat E_{13})
\left[\delta{\widehat \mu}_{13}\widehat F^{(2)}_{13,12}
+\delta{\widehat \mu}_{12}\widehat F^{(2)}_{12,13}\right]
\nonumber \\
&\phantom{=}
-\widehat G_{12}(E) \widehat G_{31}(\widehat E_{12})
\widehat  G_{23}(\widehat E_{32})
\left[\delta{\widehat \mu}_{32}\widehat F^{(2)}_{32,12}
+\delta{\widehat \mu}_{12}\widehat F^{(2)}_{12,32}\right].
\end{align}
for the subleading terms.

\paragraph{Terms in the RG equation for $G_{12}(E)$:}
We only consider the case $\eta_1=-\eta_2=+$ here, i.e., we consider the
renormalization of $\widehat G_{12}(E)$.

For the first leading order term, we get
\begin{align}
G_{13}(E)G_{\bar32}(E_{13}) F^{(3)}_{13}
=\widehat G_{13}(E)\widehat G_{32} (\widehat E_{13})
\widehat F^{(3)}_{13},
\end{align}
and for the term where the indices $1$ and $2$ are exchanged,
\begin{multline}
-G_{23}(E)G_{\bar31}(E_{23}) F^{(3)}_{23} \\
=-\widehat G_{32}(E)\widehat G_{13} (\widehat E_{32})
\widehat F^{(3)}_{32}.
\end{multline}
For the first subleading term that contributes to the renormalization of the
effective vertex, we get
\begin{align}
&\phantom{=}G_{34}(E)G_{\bar{4}1}(E_{34})G_{\bar{3}2}(E_{13}) F^{(4)}_{13,34}
-(1\leftrightarrow2)
\nonumber \\
&=\widehat G_{43}(E)\widehat G_{14}(\widehat E_{43}) \widehat
G_{32}(\widehat E_{13}) \widehat F^{(4)}_{13,43}
\nonumber \\
&\phantom{=}+\widehat G_{34}(E) \widehat G_{42}(\widehat E_{34})
\widehat G_{13}(\widehat E_{32}) \widehat F^{(4)}_{32,34}
\end{align}
(note that $\eta_1=-\eta_2=+$ implies $\eta_3=-\eta_4=-$ in the first term, and
$\eta_3=-\eta_4=+$ in the second one, where $1$ and $2$ are interchanged).

Interchanging $3\leftrightarrow4$ in the first term permits us to merge both terms
to
\begin{multline}
\widehat G_{34}(E)\left[\widehat G_{13}(\widehat E_{34}) \widehat
G_{42}(\widehat E_{14}) \widehat F^{(4)}_{14,34} \right.
\\
\left.+\widehat G_{42}(\widehat E_{34})
\widehat G_{13}(\widehat E_{32}) \widehat F^{(4)}_{32,34}\right].
\end{multline}

The second subleading term is
\begin{align}
&\phantom{=}G_{13}(E)G_{24}(E_{13})G_{\bar{4}\bar 3}(E_{1234})
F^{(4)}_{13,1234}
-(1\leftrightarrow2)
\nonumber \\
&=\widehat G_{13}(E)\widehat G_{42}(\widehat E_{13}) \widehat
G_{34}(\widehat E_{1243}) \widehat F^{(4)}_{13,1243}
\nonumber \\
&\phantom{=}+\widehat G_{32}(E) \widehat G_{14}(\widehat E_{32})
\widehat G_{43}(\widehat E_{1234}) \widehat F^{(4)}_{32,1234}
\end{align}
(note that $\eta_1=-\eta_2=+$ implies $\eta_3=-\eta_4=-$ in the first term, and
$\eta_3=-\eta_4=+$ in the second one, where $1$ and $2$ are interchanged).

The third subleading term becomes
\begin{multline}
-\frac12 G_{34}(E)G_{12}(E_{34})G_{\bar{4}\bar 3}(E_{1234})
F^{(1)}_{1234,34}\\
=-\widehat G_{34}(E)\widehat G_{12}(\widehat E_{34})\widehat
G_{43} (\widehat E_{1234}) \widehat F^{(1)}_{1234,34},
\end{multline}
where we interchanged the indices $3$ and $4$ in the term with
$\eta_3=-\eta_4=-$ to merge both terms.

This term can be merged with the first subleading term to
\begin{align}
-\widehat G_{34}(E)&\left[\widehat G_{12}(\widehat E_{34})
\widehat G_{43} (\widehat E_{1234}) \widehat F^{(1)}_{1234,34}
\right.
\nonumber \\
&-\widehat G_{13}(\widehat E_{34}) \widehat
G_{42}(\widehat E_{14}) \widehat F^{(4)}_{14,34}
\nonumber \\
&\left.-\widehat G_{42}(\widehat E_{34})
\widehat G_{13}(\widehat E_{32}) \widehat F^{(4)}_{32,34}\right].
\end{align}

\paragraph{Summary:}
After the summation over the $\eta$ indices, the RG equations take the form
\begin{widetext}
\begin{align}
\frac{\partial^2}{\partial E^2}L(E)&=
\widehat G_{12}(E) \widehat G_{21}(\widehat E_{12})
\widehat F^{(1)}_{12}
\nonumber \\
&\phantom{=}
+
\widehat G_{12}(E)\widehat G_{23}(\widehat E_{12})
\widehat G_{31}(\widehat E_{13})
\left[\widehat F^{(2)}_{13,12}+\widehat F^{(2)}_{12,13}\right]
\nonumber \\
&\phantom{=}-\widehat G_{12}(E)\widehat G_{31}(\widehat E_{12})
\widehat G_{23}(\widehat E_{32})
\left[\widehat F^{(2)}_{32,12}+\widehat F^{(2)}_{12,32}\right],
\label{eq:L_rg_eta_summation}
\\
\frac{\partial}{\partial E}\delta L(E)
&=\delta{\widehat\mu}_{12} \widehat G_{12}(E)\widehat
G_{21}(\widehat E_{12}) \widehat F^{(1)}_{12}
\nonumber \\
&\phantom{=}-\widehat G_{12}(E) \widehat Z_{12}\delta
L(\widehat E_{12}) \widehat
G_{21}(\widehat E_{12}) \widehat F^{(1)}_{12},
\nonumber \\
&\phantom{=}+\widehat G_{12}(E) \widehat G_{23}(\widehat E_{12})
\widehat  G_{31}(\widehat E_{13})
\left[\delta{\widehat \mu}_{13}\widehat F^{(2)}_{13,12}
+\delta{\widehat \mu}_{12}\widehat F^{(2)}_{12,13}\right]
\nonumber \\
&\phantom{=}
-\widehat G_{12}(E) \widehat G_{31}(\widehat E_{12})
\widehat  G_{23}(\widehat E_{32})
\left[\delta{\widehat \mu}_{32}\widehat F^{(2)}_{32,12}
+\delta{\widehat \mu}_{12}\widehat F^{(2)}_{12,32}\right]
\label{eq:L_rg_V_eta_summation}
\\
\frac{\partial}{\partial E}\widehat G_{12}(E)&=
\widehat G_{13}(E)\widehat G_{32} (\widehat E_{13})
\widehat F^{(3)}_{13}
-\widehat G_{32}(E)\widehat G_{13} (\widehat E_{32})
\widehat F^{(3)}_{32}
\nonumber \\
&\phantom{=}
-\widehat G_{34}(E)
\left[\widehat G_{12}(\widehat E_{34})\widehat
G_{43} (\widehat E_{1234}) \widehat F^{(1)}_{1234,34}
-\widehat G_{13}(\widehat E_{34}) \widehat
G_{42}(\widehat E_{14}) \widehat F^{(4)}_{14,34} \right.
\nonumber \\
&\phantom{=}\hspace{6em}\left.
-\widehat G_{42}(\widehat E_{34})
\widehat G_{13}(\widehat E_{32}) \widehat F^{(4)}_{32,34}\right]
\nonumber \\
&\phantom{=}
+\widehat G_{13}(E)\widehat G_{42}(\widehat E_{13}) \widehat
G_{34}(\widehat E_{1243}) \widehat F^{(4)}_{13,1243}
+\widehat G_{32}(E) \widehat G_{14}(\widehat E_{32})
\widehat G_{43}(\widehat E_{1234}) \widehat F^{(4)}_{32,1234}.
\label{eq:G_rg_eta_summation}
\end{align}
\end{widetext}
The corresponding equations for the current kernel, its variation, and the
current vertex can be obtained by replacing the first effective vertex by a
current vertex in these RG equations.

\subsubsection{Summation over the reservoir spin indices}

To perform the summation over the spin index $\sigma$ in $1\equiv\alpha\sigma$
in Eqs.~(\ref{eq:L_rg_eta_summation}--\ref{eq:G_rg_eta_summation}),
we use the decompositions~\eqref{eq:decomposition_G},
\eqref{eq:decomposition_I},
and~\eqref{eq:Liouvillian_CurrentKernel_Kondo}, and the properties of the
basis superoperators which are summarized in
Sec.~\ref{sec:superoperator_algebra}.

\paragraph{Terms in the RG equation for $L(E)$:}
The leading order term in Eq.~\eqref{eq:L_rg_eta_summation} contains the
following product of effective vertices (note that we frequently omit the
Fourier argument in this section to improve the readability of the equations):
\begin{align}
\widehat G_{12}\widehat G_{21}
&=
\sum_{\chi,\chi'=a,2,3}G^\chi_{12}G^{\chi'}_{21}
\text{Tr}_\sigma\left(\widehat L^{\chi}\widehat L^{\chi'}\right).
\end{align}
According to the multiplication Table~\ref{table:algebra}, the only
combinations which yield a non-zero trace over the spin degree of freedom in
this equation are $\chi=\chi'=a$ and $\chi=\chi'=2$:
\begin{align}
\widehat G_{12}\widehat G_{21}&=
G^a_{12}G^a_{21} \underbrace{\text{Tr}_\sigma\left(\widehat L^{a}\widehat
L^{a}\right)}_{=\text{Tr}_\sigma\widehat L^a}
+
G^2_{12}G^2_{21}
\underbrace{\text{Tr}_\sigma\left(\widehat L^{2}\widehat L^{2}\right)}_{=\frac12
\text{Tr}_\sigma\left(\widehat L^a+\widehat L^2\right)}
\nonumber \\
&=\left(2G^a_{12}G^a_{21} + G^2_{12}G^2_{21}\right) L^a.
\end{align}
It will be shown later that $G^a\sim \frac{V}{E}J^2$, such that it only
contributes to the renormalization of $L(E)$ beyond the subleading order.
Therefore, these terms can be omitted.

Including the Fourier arguments and the $F$-integral, the leading order
contribution in Eq.~\eqref{eq:L_rg_eta_summation} is thus
\begin{align}
\label{eq:L_rg_spin_summation_lo}
G^2_{12}(E)G^2_{21}(\widehat E_{12})\widehat F^{(1)}_{12} L^a.
\end{align}

In the products $\sim \widehat G\widehat G\widehat G$ that contribute to
$L(E)$, only the contribution $\sim \widehat L^2$ of the effective vertex is
needed. Including the term $\sim\widehat L^a$ would lead to terms in
$\Gamma(E)$ which are beyond the subleading order in $J$, and the term
$\sim\widehat L^3$ cannot be included in any product of three vertices which is
non-zero when summed over all spin degrees of freedom according to the
multiplication Tables~\ref{table:algebra} and~\ref{table:algebra_transposed}
and the property $\text{Tr}_\sigma\widehat L^{2,3}=0$.

Therefore, the products of three vertices which appear
in~\eqref{eq:L_rg_eta_summation} are
\begin{align}
\widehat G_{12}\widehat G_{23}\widehat G_{31}
&=G^2_{12}G^2_{23}G^2_{31} \text{Tr}_\sigma\left(\widehat L^2 \widehat
L^2 \widehat L^2\right), \\
-\widehat G_{12}\widehat G_{31}\widehat G_{23}
&=-G^2_{12}G^2_{31}G^2_{23} \text{Tr}_\sigma\left(\widehat L^{2,T} \widehat
L^{2,T} \widehat L^{2,T}\right).
\end{align}
According to the Tables~\ref{table:algebra} and~\ref{table:algebra_transposed},
we get
\begin{align}
\widehat L^2 \widehat L^2 \widehat L^2&=\frac12\widehat L^2 \left(\widehat
L^a+\widehat L^2\right)
=\frac34 \widehat L^2+\frac14\widehat L^a, \\
\widehat L^{2,T} \widehat L^{2,T} \widehat L^{2,T} &=
\frac12\widehat L^{2,T}\left(\widehat L^{a,T}  -\widehat L^{2,T}\right)
\nonumber \\
&=\frac34 \widehat L^{2,T}-\frac14\widehat L^{a,T}.
\end{align}
When performing the trace over the spin degree of freedom, this yields $\frac12
L^a$ and $-\frac12 L^a$, respectively.

The sum of the subleading terms in Eq.~\eqref{eq:L_rg_eta_summation}, including
the Fourier variables and the $F$-integrals, is thus
\begin{multline}
\label{eq:L_rg_spin_summation_nlo}
\frac12 G^2_{12}(E)G^2_{23}(\widehat E_{12})G^2_{31}(\widehat E_{13})
\left[\widehat F^{(2)}_{13,12}+\widehat F^{(2)}_{12,13}\right]L^a
\\
+
\frac12 G^2_{12}(E)G^2_{31}(\widehat E_{12})G^2_{23}(\widehat E_{32})
\left[\widehat F^{(2)}_{32,12}+\widehat F^{(2)}_{12,32}\right]
L^a.
\end{multline}

\paragraph{Terms in the RG equation for $\delta\Gamma(E)$:}

As discussed earlier, we only need the leading order term for $\delta\Gamma(E)$
(cf. Table~\ref{table:diagram_selection_strategy}). Therefore, only the
first term from the RG equation~\eqref{eq:L_rg_V_eta_summation} is required. It
differs from the one in the corresponding equation~\eqref{eq:L_rg_eta_summation}
only in the additional factor $\delta\widehat\mu_{12}$. The spin summation can
thus be done similarly, and the result, analogous to
Eq.~\eqref{eq:L_rg_spin_summation_lo}, is
\begin{align}
\label{eq:deltaL_rg_spin_summation_lo}
\delta\widehat\mu_{12}G^2_{12}(E)G^2_{21}(\widehat E_{12})\widehat
F^{(1)}_{12} L^a.
\end{align}

\paragraph{Current kernel:}

The RG equation for the current kernel and for its variation can be obtained
from the respective Eqs.~\eqref{eq:L_rg_eta_summation}
and~\eqref{eq:L_rg_V_eta_summation} for the effective Liouvillian and its
variation by replacing the first vertex by a current vertex.

The leading order term is
\begin{align}
\widehat I^\gamma_{12}\widehat G_{21} &=\sum_{\chi=b,1}\sum_{\chi'=a,2,3}
I^{\gamma\chi}_{12}G^{\chi'}_{21} \underbrace{\text{Tr}_\sigma \left(\widehat
L^\chi\widehat
L^{\chi'}\right)}_{\text{$\neq0$ only for $\chi=1$, $\chi'=3$}}
\nonumber \\
\label{eq:product_IG_trace}
&=6 I^{\gamma1}_{12} G^3_{21} L^b.
\end{align}
For the variation of the current kernel, we also need products of the form
$\widehat I^\gamma_{12}\delta L \widehat G_{21}$. Using $\delta
L(E)=-i\delta\Gamma(E)L^a$, we get
\begin{align}
\widehat I^\gamma_{12}\delta L\widehat G_{21}&=
-i\delta\Gamma \sum_{\chi=b,1}\sum_{\chi'=a,2,3}
I^{\gamma\chi}_{12}G^{\chi'}_{21} \underbrace{\text{Tr}_\sigma \widehat
L^\chi \widehat L^a\widehat
L^{\chi'}}_{\text{$\neq0$ only for $\chi\chi'=13$}}
\nonumber \\
\label{eq:product_IdLG_trace}
&=-6i\,\delta\Gamma I^{\gamma1}_{12} G^3_{21} L^b.
\end{align}
Combining these terms and including the Fourier arguments and $F$-integrals
yields the leading order contribution to the variation of the current kernel:
\begin{align}
\label{eq:SigmaI_rg_V_spin_summation_lo}
6 I^{\gamma1}_{12}(E) G^3_{21}(\widehat E_{12})\left[\delta\widehat \mu_{12} + i\widehat
Z_{12}\delta\Gamma(\widehat E_{12})\right] \widehat F^{(1)}_{12} L^b.
\end{align}

The subleading terms for the current kernel and its variation contain the
products
\begin{align}
&\widehat I^{\gamma}_{12} \widehat G_{23} \widehat G_{31},
&
&-\widehat I^{\gamma}_{12} \widehat G_{31} \widehat G_{23}.
\end{align}
We are only interested in contributions which have a non-zero trace. Therefore,
only the term $\sim\widehat L^1$ in the current vertex, the term $\sim\widehat
L^2$ in the first vertex $\widehat G$, and the term $\sim\widehat L^3$ in the
last vertex $\widehat G$ are relevant (cf. Tables~\ref{table:algebra}
and~\ref{table:algebra_transposed}). The required superoperator products are
therefore
\begin{align}
\widehat L^1\widehat L^2 \widehat L^3&=\widehat L^1\widehat L^3=3\widehat L^b,
\\
\widehat L^{1,T} \widehat L^{2,T} \widehat L^{2,T}
&=-\widehat L^{1,T} \widehat L^{3,T}=-3\widehat L^b.
\end{align}
When performing the trace over the spin degree of freedom, this yields $6L^b$
and $-6L^b$, respectively. Consequently, the subleading terms which contribute
to the variation of the current kernel are
\begin{align}
&6 I^{\gamma1}_{12}(E) G^2_{23}(\widehat E_{12})
G^3_{31}(\widehat E_{13})
\nonumber \\
&\quad\quad\quad\quad\times
\left[\delta\widehat\mu_{13}\widehat F^{(2)}_{13,12}
+\delta\widehat\mu_{12}\widehat F^{(2)}_{12,13}\right]L^b
\nonumber \\
&+6 I^{\gamma1}_{12}(E)G^2_{31}(\widehat E_{12})
G^3_{23}(\widehat E_{32})
\nonumber \\
&\quad\quad\quad\quad\times
\left[\delta\widehat\mu_{32}\widehat F^{(2)}_{32,12}
+\delta\widehat\mu_{12}\widehat F^{(2)}_{12,32}\right]L^b.
\label{eq:SigmaI_rg_V_spin_summation_nlo}
\end{align}

\paragraph{Terms in the RG equation for $G_{12}(E)$:}

First, we consider the leading order terms in
Eq.~\eqref{eq:G_rg_eta_summation}. The product of two effective vertices
$\widehat G$ is
\begin{align}
\widehat G_{13}\widehat G_{32}&=
\sum_{\chi,\chi'=a,2,3}G^\chi_{13}G^{\chi'}_{32} \underbrace{\widehat
L^{\chi}\widehat L^{\chi'}}_{\text{$\neq0$ only for $\chi\neq3$}}
\nonumber \\
&=\left(\tfrac12 G^2_{13}G^2_{32} + G^a_{13}G^2_{32} + G^2_{13}G^a_{32}\right)
\widehat L^2
\nonumber \\
&\phantom{=}+\left(G^2_{13}G^3_{32}+G^a_{13}G^3_{32}\right) \widehat
L^3
\nonumber \\
\label{eq:product_GG}
&\phantom{=}+\left(\tfrac12 G^2_{13}G^2_{32} + G^a_{13}G^a_{32} \right)
\widehat L^a.
\end{align}
Note that the spin indices are contained in the matrices
$\widehat L^{a,2,3}$, which are defined in
Eqs.~\eqref{eq:definition_L123_sigma} and~\eqref{eq:definition_Lab_sigma}, on
the right-hand side. All matrices $\widehat L^{a,b,1,2,3}$ which appear in the final
results on the right-hand side of equations in this section have the spin indices
$\sigma_1$ and $\sigma_2$, which are left out here to improve the readability,
i.e.,
\begin{align}
\widehat L^{a,b,1,2,3}&\equiv \widehat L^{a,b,1,2,3}_{\sigma_1\sigma_2}.
\end{align}
If the spin indices are reversed, we find (note that an overall minus sign has
been added for convenience because it also appears in the RG equations where
the spin indices are interchanged)
\begin{align}
-\widehat G_{32}\widehat G_{13}&=
-\sum_{\chi,\chi'=a,2,3}G^\chi_{32}G^{\chi'}_{13}
\underbrace{\left[\left(\widehat
L^{\chi}\right)^T\left(\widehat L^{\chi'}\right)^T \right]^T}_{\text{$\neq0$
only for $\chi\neq3$}}
\nonumber \\
&=\left(\tfrac12 G^2_{32}G^2_{13} + G^a_{32}G^2_{13} + G^2_{32}G^a_{13}\right)
\widehat L^2
\nonumber \\
&\phantom{=}+\left(G^2_{32}G^3_{13}+G^a_{32}G^3_{13}\right) \widehat L^3
\nonumber \\
\label{eq:product_GG_reversed}
&\phantom{=}-\left(\tfrac12 G^2_{32}G^2_{13} + G^a_{32}G^a_{13} \right)
\widehat L^a.
\end{align}
We will first discuss why the terms containing $G^a$ can be omitted here and in
all other RG equations. We get the leading order part of the RG equation for
$G^a_{12}(E)$ by including the Fourier arguments in the contributions
$\sim\widehat L^a$ from Eqs.~\eqref{eq:product_GG}
and~\eqref{eq:product_GG_reversed} and adding the $F$-integrals:
\begin{align}
&\phantom{=}
\left.\frac{\partial}{\partial E}G^a_{12}(E)\right|_{\text{l.o.}}
\nonumber \\
&=
\left[\tfrac12 G^2_{13}(E)G^2_{32}(\widehat E_{13})
+G^a_{13}G^a_{32}(E)(\widehat E_{13})\right]
\widehat F^{(3)}_{13}
\nonumber \\
&\phantom{=}
-\left[\tfrac12 G^2_{32}(E)G^2_{13}(\widehat E_{32}) +
G^a_{32}(E)G^a_{13}(\widehat E_{32}) \right]
\widehat F^{(3)}_{32}.
\end{align}
In the case $V=0$, all Fourier arguments and $F$-integrals are equal, and the
right-hand side is thus zero. For $V\ll|E|$, an expansion of the effective vertices and
the $F$-integrals yields
\begin{align}
G^2_{12}(\widehat E_{12})&=  G^2_{12}(E)+
\mathcal{O}\left\{\frac{V}{E}\left[J(E)\right]^2\right\}, \\
\widehat F^{(3)}_{12}-\widehat
F^{(3)}_{32}&=\mathcal{O}\left(\frac{V}{E^2}\right).
\end{align}
This means that the leading contribution to the renormalization of
$G^a_{12}(E)$ is
\begin{align}
\left.\frac{\partial}{\partial E}G^a_{12}(E)\right|_{\text{l.o.}}
&\sim \frac{V}{E^2}J^2
\sim \frac{\partial}{\partial E}\left(\frac{V}{E}J^2\right)
+\mathcal{O}\left(\frac{V}{E^2}J^3\right)
\end{align}
(where we have used $\frac{\partial J}{\partial E}\sim \frac1{E}J^2$ in leading
order), and that the leading contribution to $G^a_{12}(E)$ itself is thus
\begin{align}
\label{eq:Ga-leading-order}
G^a_{12}(E) &\sim \frac{V}{E}J^2.
\end{align}
As discussed in Sec.~\ref{sec:strategy-diagram-selection}, this observation
allows us to omit all terms which contain $G^a$, because they would only cause
contributions beyond the subleading order to the RG equations of the physical
observables.

In the subleading terms in Eq.~\eqref{eq:G_rg_eta_summation}, different
products of effective vertices occur, which are evaluated in
Appendix~\ref{sec:subleading_terms_G}. The final results are [cf.
Eqs.~\eqref{eq:product_GGG_final_1_appendix}--
\eqref{eq:product_GGG_final_5_appendix}]
\begin{align}
\widehat G_{34}\widehat G_{12}\widehat G_{43}
&=
\frac12 G^2_{34} G^2_{12} G^2_{43} \widehat L^2_{12}
-G^2_{34} G^2_{12} G^3_{43} \widehat L^3_{12},
\label{eq:product_GGG_final_1}
\\
\widehat G_{34}\widehat G_{13}\widehat G_{42}
&=
\frac14 G^2_{34} G^2_{13} G^2_{42} \widehat L^2_{12},
\\
\widehat G_{34}\widehat G_{42}\widehat G_{13}
&=
\frac14 G^2_{34} G^2_{42} G^2_{13} \widehat L^2_{12},
\\
\widehat G_{13}\widehat G_{42}\widehat G_{34}
&=
\frac14 G^2_{13} G^2_{42} G^2_{34} \widehat L^2_{12}
-G^2_{13} G^2_{42} G^3_{34} \widehat L^3_{12},
\\
\widehat G_{32}\widehat G_{14}\widehat G_{43}
&=
\frac14 G^2_{32} G^2_{14} G^2_{43} \widehat L^2_{12}
-G^2_{32} G^2_{14} G^3_{43} \widehat L^3_{12}.
\label{eq:product_GGG_final_5}
\end{align}

\paragraph{Terms in the RG equation for the current vertex $I^\gamma_{12}(E)$:}

We have to replace the first vertex in each of the terms on the right-hand side of
Eq.~\eqref{eq:G_rg_eta_summation} by a current vertex in order to obtain the RG
equation for the current vertex. The leading order terms are
\begin{align}
\widehat I^\gamma_{13}\widehat G_{32} &=\sum_{\chi=b,1}\sum_{\chi'=a,2,3}
I^{\gamma\chi}_{13}G^{\chi'}_{32} \underbrace{\widehat L^\chi\widehat
L^{\chi'}}_{\text{$\neq0$ for $\chi\neq b$}}
\nonumber \\
\label{eq:product_IG}
&=I^{\gamma1}_{13}\left(G^a_{32}+G^2_{32}\right) \widehat L^1 +3
I^{\gamma1}_{13} G^3_{32} \widehat L^b,
\end{align}
and
\begin{align}
-\widehat I^\gamma_{32}\widehat G_{13} &=-\sum_{\chi=b,1}\sum_{\chi'=a,2,3}
I^{\gamma\chi}_{32}G^{\chi'}_{13} \underbrace{\left[ \left(\widehat
L^\chi\right)^T\left(\widehat
L^{\chi'}\right)\right]^T}_{\text{$\neq0$ for $\chi\neq b$}}
\nonumber \\
\label{eq:product_IG_reversed}
&=I^{\gamma1}_{32}\left(-G^a_{13}+G^2_{13}\right) \widehat L^1 -3
I^{\gamma1}_{32} G^3_{13} \widehat L^b.
\end{align}
According to Eqs.~\eqref{eq:SigmaI_rg_V_spin_summation_lo}
and~\eqref{eq:SigmaI_rg_V_spin_summation_nlo}, the part $\sim \widehat L^b$ of
the current vertex does not contribute to the current kernel and can therefore
be neglected. The only relevant leading order contribution to the
renormalization of the current vertex is thus
\begin{align}
\label{eq:I_rg_spin_summation_lo}
I^{\gamma1}_{13}(E) G^2_{32}(\widehat E_{13}) \widehat F^{(3)}_{13}
+I^{\gamma1}_{32}(E) G^2_{13}(\widehat E_{32}) \widehat F^{(3)}_{32}.
\end{align}
The subleading terms are evaluated in Appendix~\ref{sec:subleading_terms_I},
cf. Eqs.~\eqref{eq:product_IGG_final_1_appendix}--\eqref{eq:product_IGG_final_5_appendix}:
\begin{align}
\widehat I^\gamma_{34}\widehat G_{12}\widehat G_{43}
&=
-I^{\gamma1}_{34} G^2_{12} G^2_{43} \widehat L^1_{12},
\label{eq:product_IGG_final_1}
\\
\widehat I^\gamma_{34}\widehat G_{13}\widehat G_{42}
&=
-I^{\gamma1}_{34} G^2_{13} G^2_{42} \widehat L^1_{12},
\\
\widehat I^\gamma_{34}\widehat G_{42}\widehat G_{13}
&=
-I^{\gamma1}_{34} G^2_{42} G^2_{13} \widehat L^1_{12},
\\
\widehat I^\gamma_{13}\widehat G_{42}\widehat G_{34}
&=0,
\\
\widehat I^\gamma_{32}\widehat G_{14}\widehat G_{43}
&=0.
\label{eq:product_IGG_final_5}
\end{align}

\paragraph{Summary:} After the summation over the spin indices, we get the
following RG equations for the rate $\Gamma(E)$, which is contained in
the effective Liouvillian $L(E)=-i\Gamma(E)L^a$ [by adding the contributions
from Eqs.~\eqref{eq:L_rg_spin_summation_lo}
and~\eqref{eq:L_rg_spin_summation_nlo}], the variation of the rate
$\delta\Gamma(E)$ [which we only need the leading order
contribution~\eqref{eq:deltaL_rg_spin_summation_lo} for], the variation of the
current kernel $\delta\Sigma_\gamma(E)=i\delta\Gamma_\gamma(E)L^b$ [by adding
the
contributions from Eqs.~\eqref{eq:SigmaI_rg_V_spin_summation_lo}
and~\eqref{eq:SigmaI_rg_V_spin_summation_nlo}], the effective vertex
[by collecting the terms $\sim\widehat L^{2,3}$ which do not contain $G^a$ from
Eqs.~\eqref{eq:product_GG}, \eqref{eq:product_GG_reversed}
and~(\ref{eq:product_GGG_final_1}--\ref{eq:product_GGG_final_5})], and the
current vertex [by considering the terms from
Eqs.~\eqref{eq:I_rg_spin_summation_lo}
and~\eqref{eq:product_IGG_final_1}--\eqref{eq:product_IGG_final_5})]
\begin{widetext}
\begin{align}
\frac{\partial^2}{\partial E^2}\Gamma(E)&=
i G^2_{12}(E)G^2_{21}(\widehat E_{12}) \widehat F^{(1)}_{12}
+i
\frac12 G^2_{12}(E)G^2_{23}(\widehat E_{12})G^2_{31}(\widehat E_{13})
\left[\widehat F^{(2)}_{13,12}+\widehat F^{(2)}_{12,13}\right]
\nonumber \\
&\phantom{=
i G^2_{12}(E)G^2_{21}(\widehat E_{12}) \widehat F^{(1)}_{12}}
+
i\frac12 G^2_{12}(E)G^2_{31}(\widehat E_{12})G^2_{23}(\widehat E_{32})
\left[\widehat F^{(2)}_{32,12}+\widehat F^{(2)}_{12,32}\right],
\\
\frac{\partial}{\partial E}\delta\Gamma(E)&=
i\delta\widehat\mu_{12}G^2_{12}(E)G^2_{21}(\widehat E_{12})\widehat
F^{(1)}_{12},
\\
\frac{\partial}{\partial E}\delta\Gamma_\gamma(E)
&=-6i I^{\gamma1}_{12}(E) \left\{
G^3_{21}(\widehat E_{12})\left[\delta\widehat \mu_{12} + i\widehat
Z_{12}\delta\Gamma(\widehat E_{12})\right] \widehat F^{(1)}_{12}\right.
\nonumber \\
&\phantom{=}\hspace{6em}
+G^2_{23}(\widehat E_{12})G^3_{31}(\widehat E_{13})
\left[\delta\widehat\mu_{13}\widehat F^{(2)}_{13,12}
+\delta\widehat\mu_{12}\widehat F^{(2)}_{12,13}\right]
\nonumber \\
&\left.\phantom{=}\hspace{6em}
+G^2_{31}(\widehat E_{12})G^3_{23}(\widehat E_{32})
\left[\delta\widehat\mu_{32}\widehat F^{(2)}_{32,12}
+\delta\widehat\mu_{12}\widehat F^{(2)}_{12,32}\right]\right\},
\\
\frac{\partial}{\partial E}G^2_{12}(E)&=
\frac12 G^2_{13}(E)G^2_{32}(\widehat E_{13})\widehat F^{(3)}_{13}
+\frac12 G^2_{32}(E)G^2_{13}(\widehat E_{32})\widehat F^{(3)}_{32}
\nonumber \\
&\phantom{=}
+\frac12 G^2_{34}(E)\left\{
-G^2_{12}(\widehat E_{34}) G^2_{43}(\widehat E_{1234})\widehat F^{(1)}_{1234,34}
+\frac12 G^2_{13}(\widehat E_{34}) G^2_{42}(\widehat E_{14})\widehat F^{(4)}_{14,34}\right. \nonumber \\
&\phantom{=}\left.\hspace{12em}
+\frac12 G^2_{42}(\widehat E_{34}) G^2_{13}(\widehat E_{32})\widehat F^{(4)}_{32,34}\right\}
\nonumber \\
&\phantom{=}
+\frac14 G^2_{13}(E)G^2_{42}(\widehat E_{13}) G^2_{34}(\widehat E_{1243})
\widehat F^{(4)}_{13,1243}
+\frac14 G^2_{32}(E)G^2_{14}(\widehat E_{32}) G^2_{43}(\widehat E_{1234})
\widehat F^{(4)}_{32,1234},
\\
\frac{\partial}{\partial E}G^3_{12}(E)&=
G^2_{13}(E)G^3_{32}(\widehat E_{13})\widehat F^{(3)}_{13}
+G^2_{32}(E)G^3_{13}(\widehat E_{32})\widehat F^{(3)}_{32}
+G^2_{34}(E)G^2_{12}(\widehat E_{34}) G^3_{43}(\widehat E_{1234})
\widehat F^{(1)}_{1234,34}
\nonumber \\
&\phantom{=}
-G^2_{13}(E)G^2_{42}(\widehat E_{13}) G^3_{34}(\widehat E_{1243})
\widehat F^{(4)}_{13,1243}
-G^2_{32}(E)G^2_{14}(\widehat E_{32}) G^3_{43}(\widehat E_{1234})
\widehat F^{(4)}_{32,1234},
\\
\frac{\partial}{\partial E}I^{\gamma1}_{12}(E)&=
I^{\gamma1}_{13}(E) G^2_{32}(\widehat E_{13}) \widehat F^{(3)}_{13}
+I^{\gamma1}_{32}(E) G^2_{13}(\widehat E_{32}) \widehat F^{(3)}_{32}
+I^{\gamma1}_{34}(E)G^2_{12}(\widehat E_{34})
G^2_{43}(\widehat E_{1234}) \widehat F^{(1)}_{1234,34}
\nonumber \\
&\phantom{=}
-I^{\gamma1}_{34}(E)G^2_{13}(\widehat E_{34})
G^2_{42}(\widehat E_{14}) \widehat F^{(4)}_{14,34}
-I^{\gamma1}_{34}(E)G^2_{42}(\widehat E_{34})
G^2_{13}(\widehat E_{32}) \widehat F^{(4)}_{32,34}.
\end{align}
With $Z(E)=1/\left[1+i\frac{\partial}{\partial E}\Gamma(E)\right]$,
$\frac{\partial}{\partial E}Z(E)=-iZ(E)^2\frac{\partial^2}{\partial E^2}\Gamma(E)$,
and using the shorthand notations~(\ref{eq:definition_J}--\ref{eq:definition_I}),
we get the final RG equations
\begin{align}
\frac{\partial}{\partial E}\Gamma(E)&=-i\left[Z(E)^{-1}-1\right]
\label{eq:rg_gamma_TV}
\\
\frac{\partial}{\partial E}Z(E)&=
 Z(E)^2 J_{12}(E)J_{21}(\widehat E_{12}) \widehat F^{(1)}_{12}
-\frac12Z(E)^2 J_{12}(E)J_{23}(\widehat E_{12})J_{31}(\widehat E_{13})
\left[\widehat F^{(2)}_{13,12}+\widehat F^{(2)}_{12,13}\right]
\nonumber \\
&\phantom{=
Z(E)^2 J_{12}(E)J_{21}(\widehat E_{12}) \widehat F^{(1)}_{12}}
-\frac12 Z(E)^2 J_{12}(E)J_{31}(\widehat E_{12})J_{23}(\widehat E_{32})
\left[\widehat F^{(2)}_{32,12}+\widehat F^{(2)}_{12,32}\right],
\label{eq:rg_Z_TV}
\\
\frac{\partial}{\partial E}\delta\Gamma(E)&=
i\delta\widehat\mu_{12}J_{12}(E)J_{21}(\widehat E_{12})\widehat F^{(1)}_{12},
\label{eq:rg_deltaL_TV}
\\
\frac{\partial}{\partial E}[\pi\delta\Gamma_\gamma(E)]
&=
-\frac{3\pi^2}{4} I^{\gamma}_{12}(E) \left\{K_{21}\left(E\right)\left[\delta\widehat \mu_{12} +
i\widehat Z_{12}\delta\Gamma(\widehat E_{12})\right] \widehat F^{(1)}_{12}\right.
\nonumber \\
&\phantom{=}
- J_{23}(\widehat E_{12}) K_{31}(E)
\left[\delta\widehat\mu_{13}\widehat F^{(2)}_{13,12}
+\delta\widehat\mu_{12}\widehat F^{(2)}_{12,13}\right] \nonumber \\
&\left.\phantom{=}
- J_{31}(\widehat E_{12}) K_{23}(E)
\left[\delta\widehat\mu_{32}\widehat F^{(2)}_{32,12}
+\delta\widehat\mu_{12}\widehat F^{(2)}_{12,32}\right]\right\},
\label{eq:rg_SigmaI_TV}
\\
\frac{\partial}{\partial E}J_{12}(E)&=
-\frac12 J_{13}(E)J_{32}(\widehat E_{13})\widehat F^{(3)}_{13}
-\frac12 J_{32}(E)J_{13}(\widehat E_{32})\widehat F^{(3)}_{32}
\nonumber \\
&\phantom{=}
-\frac12 J_{34}(E)\left\{J_{12}(\widehat E_{34}) J_{43}(\widehat E_{1234})
\widehat F^{(1)}_{1234,34}
-\frac12 J_{13}(\widehat E_{34}) J_{42}(\widehat E_{14})\widehat F^{(4)}_{14,34}\right.
\nonumber \\
&\phantom{=}\hspace{12em}\left.
-\frac12 J_{42}(\widehat E_{34}) J_{13}(\widehat E_{32})\widehat F^{(4)}_{32,34}\right\}
\nonumber \\
&\phantom{=}
+\frac14 J_{43}(\widehat E_{1234})
\left\{J_{14}(E)J_{32}(\widehat E_{14}) \widehat F^{(4)}_{14,1234}
+ J_{32}(E)J_{14}(\widehat E_{32}) \widehat F^{(4)}_{32,1234}\right\},
\label{eq:rg_G2_TV}
\\
\frac{\partial}{\partial E}K_{12}(E)&=
-J_{13}(\widehat E_{21})K_{32}\left(E\right)\widehat F^{(3)}_{23}
-J_{32}(\widehat E_{21})K_{13}\left(E\right)\widehat F^{(3)}_{31}
+J_{34}(\widehat E_{21})J_{12}(\widehat E_{2134})
K_{43}\left(E \right)\widehat F^{(1)}_{34,2134}
\nonumber \\
&\phantom{=}
-J_{13}(\widehat E_{21})J_{42}(\widehat E_{23})
K_{34}\left(E \right)\widehat F^{(4)}_{23,43}
-J_{32}(\widehat E_{21})J_{14}(\widehat E_{31})
K_{43}\left(E \right)\widehat F^{(4)}_{31,34},
\label{eq:rg_G3_TV}
\\
\frac{\partial}{\partial E}I^{\gamma}_{12}(E)&=
-I^{\gamma}_{13}(E) J_{32}(\widehat E_{13}) \widehat F^{(3)}_{13}
-I^{\gamma}_{32}(E) J_{13}(\widehat E_{32}) \widehat F^{(3)}_{32}
+I^{\gamma}_{34}(E)J_{12}(\widehat E_{34})
J_{43}(\widehat E_{1234}) \widehat F^{(1)}_{1234,34}
\nonumber \\
&\phantom{=}
-I^{\gamma}_{34}(E)J_{13}(\widehat E_{34})
J_{42}(\widehat E_{14}) \widehat F^{(4)}_{14,34}
-I^{\gamma}_{34}(E)J_{42}(\widehat E_{34})
J_{13}(\widehat E_{32}) \widehat F^{(4)}_{32,34}.
\label{eq:rg_I_TV}
\end{align}

\end{widetext}
Using these equations and their behavior at large values of $|E|$, we can now
justify the strategy for the selection of diagrams which has been presented in
section~\ref{sec:strategy-diagram-selection}.

For large $|E|$, the integrals which appear in the leading order terms of
the RG equations can be approximated by
\begin{align}
\widehat F^{(1)}_{12}&\approx\widehat F^{(3)}_{12}\approx\frac{1}{E}.
\end{align}
Therefore, we find that the right-hand side of the RG equations of couplings and rates
take the form (we leave out reservoir indices because we are only interested in
the overall scale)
\begin{align}
\frac{\partial^2}{\partial E^2}\Gamma&\sim \frac{1}{E} J^2,
\label{eq:L_rg_leading_order}
\\
\frac{\partial}{\partial E}\delta\Gamma&\sim \frac{1}{E} J^2,
\label{eq:L_rg_V_leading_order}
\\
\frac{\partial}{\partial E}\delta\Gamma_\gamma&\sim \frac{1}{E} I^\gamma K,
\label{eq:SigmaI_rg_V_leading_order}
\\
\frac{\partial}{\partial E}J&\sim \frac{1}{E} J^2,
\label{eq:G2_rg_leading_order}
\\
\frac{\partial}{\partial E}K&\sim \frac{1}{E} J K,
\label{eq:G3_rg_leading_order}
\\
\frac{\partial}{\partial E}I^{\gamma}&\sim \frac{1}{E} I^\gamma J
\label{eq:I_rg_leading_order}
\end{align}
for large values of $|E|$.

Comparing Eqs.~\eqref{eq:L_rg_leading_order}
and~\eqref{eq:L_rg_V_leading_order} with Eq.~\eqref{eq:G2_rg_leading_order}
shows that
\begin{align}
\label{eq:L_leading_order_behavior}
\Gamma(E)&\sim J(E), & \delta\Gamma(E)&\sim J(E)
\end{align}
at large $|E|$.

Equation~\eqref{eq:G2_rg_leading_order} confirms that the RG
equations~\eqref{eq:G3_rg_leading_order} and~\eqref{eq:I_rg_leading_order} are
consistent with the behavior
\begin{align}
\label{eq:K_I_leading_order_behavior}
K(E)&\sim \left[J(E)\right]^2, & I^\gamma(E)&\sim J(E)
\end{align}
of $K_{12}(E)$ and $I^\gamma_{12}(E)$ at large $|E|$.

Finally, substituting Eq.~\eqref{eq:K_I_leading_order_behavior}
into~\eqref{eq:SigmaI_rg_V_leading_order} results in
\begin{align}
\label{eq:SigmaI_rg_V_leading_order_behavior}
\frac{\partial}{\partial E}\delta\Gamma_\gamma(E)&\sim \frac{1}{E}J^3.
\end{align}
On the other hand, considering subleading terms $\sim J^2$ for
$\delta\Gamma(E)$ in the right-hand side of Eq.~\eqref{eq:rg_SigmaI_TV}
would add terms
\begin{align}
\label{eq:SigmaI_rg_V_terms_beyond_subleading_order}
\frac{1}{E}I^\gamma K J^2\sim \frac{1}{E} J^5
\end{align}
to $\frac{\partial}{\partial E}\delta\Gamma_\gamma(E)$, which is two orders in $J$
higher than the leading
contribution~\eqref{eq:SigmaI_rg_V_leading_order_behavior}. These terms beyond
the subleading order can be neglected.

To solve the RG equations \eqref{eq:rg_gamma_TV}--\eqref{eq:rg_I_TV}
numerically we use the initial conditions
\eqref{eq:initial_conditions_summary_begin}--\eqref{eq:initial_conditions_summary_end},
\begin{align}
\label{eq:ini_cond_Z}
Z\left(E=iD\right)&=1, \\
\label{eq:ini_cond_delta_Gamma}
\delta\Gamma\left(E=iD\right)&=0, \\
\label{eq:ini_cond_delta_Gamma_gamma}
\pi\delta\Gamma_\gamma\left(E=iD\right)&=\frac{3\pi^2}{8} \delta\widehat\mu_{12}
\left(\delta_{1\gamma}-\delta_{2\gamma}\right) J^{(0)}_{12}J^{(0)}_{21}, \\
\label{eq:ini_cond_J}
J_{12}\left(E=iD\right)&=J^{(0)}_{12}, \\
\label{eq:ini_cond_K}
K_{12}\left(E=iD\right)&=J^{(0)}_{13}J^{(0)}_{32}, \\
\label{eq:ini_cond_I_gamma}
I^{\gamma}_{12}\left(E=iD\right)&=\left(\delta_{1\gamma}-\delta_{2\gamma}\right)J^{(0)}_{12}.
\end{align}
Furthermore, we use the special form \eqref{eq:J_sw} for $J_{12}^{(0)}$ and consider
(for simplicity) the case of two reservoirs with $\alpha\equiv L/R\equiv \pm$ and
symmetric coupling $x_L=x_R=\frac{1}{2}$,
\begin{align}
\label{eq:J_symmetric_2_reservoirs}
J^{(0)}_{\alpha\alpha'}&=J_0.
\end{align}
The chemical potentials are written as $\mu_{\alpha}=\alpha \frac{V}{2}$, where $V$ denotes the bias
voltage. In this case, we get from Eqs.~\eqref{eq:tensorH_definition}, \eqref{eq:tensorG_definition},
\eqref{eq:G_prop_3} and \eqref{eq:cur_var_3}
\begin{align}
\label{eq:delta_SigmaI_2_reservoirs}
\pi \delta\Gamma_\gamma(E) &= G^\gamma_{LR}(E)(\delta\mu_L-\delta\mu_R)
= \gamma \frac{G(E)}{G_0} \delta V,\\
\label{eq:delta_I_2_reservoirs}
\delta \left\langle I_\gamma \right\rangle (E) &=\gamma \frac{i}{E}G(E)\delta V,
\end{align}
where $G^L_{LR}(E)=G(E)/G_0$ is the conductance $G(E)$ in units of the universal
conductance $G_0=\frac{2e^2}{h}$. The variation of the stationary current follows from
\begin{align}
\label{eq:stationary_I_2_reservoirs}
\delta \left\langle I_\gamma \right\rangle^{\text{st}}
= \gamma G^{\text{st}} \delta V,
\end{align}
where $G^{\text{st}}=G(0)$ is the stationary conductance.

For the special case of two reservoirs with symmetric couplings, we obtain the
initial conditions
\begin{align}
\label{eq:ini_cond_Z_2_res}
Z\left(E=iD\right)&=1, \\
\label{eq:ini_cond_delta_Gamma_2_res}
\delta\Gamma\left(E=iD\right)&=0, \\
\label{eq:ini_cond_delta_Gamma_gamma_2_res}
\pi\delta\Gamma_\gamma\left(E=iD\right)&=\gamma\frac{3\pi^2}{4} (J_0)^2 \delta V, \\
\label{eq:ini_cond_G_2_res}
G(E=iD) &= G_0\frac{3\pi^2}{4}(J_0)^2,\\
\label{eq:ini_cond_J_2_res}
J_{12}\left(E=iD\right)&=J_0, \\
\label{eq:ini_cond_K_2_res}
K_{12}\left(E=iD\right)&=2 (J_0)^2, \\
\label{eq:ini_cond_I_gamma_2_res}
I^{\gamma}_{12}\left(E=iD\right)&=\left(\delta_{1\gamma}-\delta_{2\gamma}\right)J_0,
\end{align}

We note that these initial conditions do not contain nonuniversal terms of
$O(\frac{1}{D})$ and higher orders in the bare coupling $J_0$. Therefore, to
extract only the universal part of the solution up to subleading order, one has
to use the scaling limit \eqref{eq:scaling_limit}. Furthermore, as explained in
Section~\ref{subsec:initial_conditions}, the missing initial
condition for $\Gamma(E)$ at $E=iD$ has to be determined from another reference
point since this energy scale is related to the Kondo temperature and not universal.
As shown in the next section, one can set up the scaling limit and the initial
condition for $\Gamma$ by studying the analytic solution for $T=V=0$.

\subsection{Analytic solution for $T=V=0$, the scaling limit
and the initial condition for $\Gamma$}
\label{sec:RG_T0_V0}

In the special case of zero temperature and zero voltage, the RG equations
can be solved analytically (except for $\Gamma$). We set $E=i\Lambda$ and start the RG flow at
$\Lambda=\Lambda_0\equiv D$. In accordance with the initial conditions
\eqref{eq:ini_cond_J_2_res}, \eqref{eq:ini_cond_K_2_res} and \eqref{eq:ini_cond_I_gamma_2_res},
we find that the vertices $J_{12}$ and $K_{12}$ do not depend on the lead indices,
\begin{align}
J_{12}&=J, & K_{12}&=K,
\end{align}
[we omit the variable $E=i\Lambda$ in all quantities for simplicity here],
and that the current vertex can be parametrized by
\begin{align}
I^\gamma_{12}&=\left(\delta_{1\gamma}-\delta_{2\gamma}\right) J_I.
\end{align}
By substituting the
integrals~\eqref{eq:integral_F1_TV0}--\eqref{eq:integral_F4_TV0} into the RG
equations~\eqref{eq:rg_gamma_TV}--\eqref{eq:rg_I_TV} we obtain
\begin{align}
\label{eq:rg_Gamma_T=V=0}
\frac{d}{d\Lambda}\Gamma &= \frac{1}{Z}-1,\\
\label{eq:rg_Z_T=V=0}
\frac{d}{d\Lambda}Z &= \frac{1}{\Lambda+\Gamma}4Z^2J^2, \\
\label{eq:rg_cond_T=V=0}
\frac{d}{d\Lambda}G&=
-G_0\frac{3\pi^2}{2}\frac{1}{\Lambda+\Gamma} J_I K ,\\
\label{eq:rg_J_T=V=0}
\frac{d}{d\Lambda}J &= -\frac{1}{\Lambda+\Gamma}2J^2\left(1+ZJ\right), \\
\label{eq:rg_K_T=V=0}
\frac{d}{d\Lambda}K &= -\frac{1}{\Lambda+\Gamma} 4KJ\left(1-ZJ\right), \\
\label{eq:rg_JI_T=V=0}
\frac{d}{d\Lambda}J_I &= -\frac{1}{\Lambda+\Gamma}2J_IJ.
\end{align}
We define
\begin{align}
\widetilde J&=ZJ, & \widetilde J_I&=ZJ_I
\end{align}
and a new flow parameter
\begin{align}
\lambda=\Lambda+\Gamma,
\end{align}
which implies
\begin{align}
\frac{d\lambda}{d\Lambda}=\frac{1}{Z}.
\end{align}
Transforming the RG equations to the flow parameter $\lambda$ yields
\begin{align}
\label{eq:rg_tilde_Gamma_T=V=0}
\frac{d}{d\lambda}\Gamma &= 1-Z,\\
\frac{d}{d\lambda}Z &= \frac{1}{\lambda}4Z\widetilde J^2, \\
\frac{d}{d\lambda}G &= -G_0\frac{3\pi^2}{2}\frac{1}{\lambda}\widetilde J_IK,\\
\label{eq:rg_tilde_J_T=V=0}
\frac{d}{d\lambda}\widetilde J &=
-\frac{1}{\lambda}2\widetilde J^2\left(1-\widetilde J\right), \\
\frac{d}{d\lambda}K &= -\frac{1}{\lambda} 4K\widetilde
J\left(1-\widetilde J\right), \\
\frac{d}{d\lambda}\widetilde J_I &= -\frac{1}{\lambda}2\widetilde
J_I\widetilde J\left(1-2\widetilde J\right).
\end{align}
Integrating Eq.~\eqref{eq:rg_tilde_J_T=V=0}, we obtain the invariant
\begin{align}
\label{eq:T0_V0_TK}
T_K\equiv(\Lambda+\Gamma)\left(\frac{\widetilde{J}}{1-\widetilde{J}}\right)^{1/2}
e^{-\frac{1}{2\widetilde{J}}}.
\end{align}
The nonuniversal invariant $T_K$ sets the low energy scale. If not written
explicitly, we will use the energy unit $T_K=1$ in the following. Note that this
invariant is identical to the Kondo temperature defined in Eq.~\eqref{eq:TK_pms},
since in the limit $\Lambda\equiv D\rightarrow\infty$ we get $\tilde{J}=J_0\rightarrow 0$.

Taking ratios one can eliminate $\lambda$ from some of the RG equations and obtains
\begin{align}
\label{eq:rg_tilde_Z_T=V=0}
\frac{dZ}{d\widetilde{J}} &= -2\frac{Z}{1-\widetilde{J}}, \\
\label{eq:rg_tilde_G_T=V=0}
\frac{dG}{d\widetilde{J}} &= G_0\frac{3\pi^2}{2}
\frac{\widetilde{J}_I K}{2\widetilde{J}^2(1-\widetilde{J})},\\
\label{eq:rg_tilde_K_T=V=0}
\frac{dK}{d\tilde{J}} &= 2\frac{K}{\widetilde{J}} ,\\
\label{eq:rg_tilde_JI_T=V=0}
\frac{d\widetilde J_I}{d\widetilde{J}} &= \widetilde J_I
\left(\frac{1}{\widetilde J}-\frac{1}{1-\widetilde J}\right).
\end{align}
Integrating the RG equations for $Z$, $K$ and $J_I$, we obtain another set of
invariants
\begin{align}
\label{eq:invariant_ZKI}
c_Z\equiv\frac{Z}{\left(1-\widetilde{J}\right)^2}
\quad,\quad
c_K\equiv\frac{K}{2\widetilde{J}^2}
\quad,\quad
c_I\equiv\frac{\widetilde{J}_I}{\widetilde{J}(1-\widetilde{J})}.
\end{align}
Inserting this solution for $K$ and $\widetilde{J}_I$ into the RG equation
\eqref{eq:rg_tilde_G_T=V=0} for $G$ and integrating it, we find another invariant
\begin{align}
\label{eq:invariant_G}
c_G \equiv G - c_I c_K G_0 \frac{3\pi^2}{4}\widetilde{J}^2.
\end{align}
The invariants $c_Z$, $c_K$, $c_I$ and $c_G$ are fixed by comparing with the initial
conditions \eqref{eq:ini_cond_Z_2_res}, \eqref{eq:ini_cond_J_2_res}, \eqref{eq:ini_cond_K_2_res},
\eqref{eq:ini_cond_I_gamma_2_res} and \eqref{eq:ini_cond_G_2_res} in the scaling limit
\eqref{eq:scaling_limit}. We obtain $c_Z=c_K=c_I=1$ and $c_G=0$, leading to the
universal results
\begin{align}
\label{eq:T0_V0_Z}
Z&=\left(1-\widetilde J\right)^2, \\
\label{eq:T0_V0_K}
K&=2\widetilde J^2, \\
\label{eq:T0_V0_JI}
\widetilde J_I&=\widetilde J\left(1-\widetilde J\right),\\
\label{eq:T0_V0_G}
G&=G_0\frac{3\pi^2}{4}\widetilde{J}^2.
\end{align}
So far we have solved all RG equations analytically except for the RG equation
\eqref{eq:rg_tilde_Gamma_T=V=0} for $\Gamma$. The solution of this equation is needed
to determine $\widetilde{J}$ as function of $\Lambda$ via \eqref{eq:T0_V0_TK}.
Taking the ratio of \eqref{eq:rg_tilde_Gamma_T=V=0} and \eqref{eq:rg_tilde_J_T=V=0},
and using \eqref{eq:T0_V0_TK} and \eqref{eq:T0_V0_Z}, we obtain
\begin{align}
\label{eq:rg_Gamma_tilde_J}
\frac{d}{d\widetilde{J}}\left(\frac{\Gamma}{T_K}\right)=
-\frac{1-\frac{1}{2}\widetilde{J}}{\widetilde{J}\sqrt{\widetilde{J}(1-\widetilde{J})}}
e^{\frac{1}{2\widetilde J}}.
\end{align}
This is a complicated differential equation and cannot be solved analytically.
Furthermore, since only the ratio $\Gamma/T_K$ is universal, it is
impossible to set up a universal initial condition at high energies from a
perturbative calculation of $\Gamma$. Therefore, we study the numerical solution
of the differential equation for $\Gamma$ by starting at $\Lambda=0$ and using
the exact and universal result $G(0)/G_0=1$ as boundary condition. Using
Eqs.~\eqref{eq:T0_V0_G} and \eqref{eq:T0_V0_TK}, the initial condition for
$\Gamma(0)/T_K$ can then be calculated from
\begin{align}
\label{eq:Gamma_initial}
\frac{\Gamma(0)}{T_K}=\left(\frac{1-\widetilde J(0)}{\widetilde J(0)}\right)^{1/2}
e^{\frac{1}{2\widetilde J(0)}}
\quad,\quad
\widetilde{J}(0)=\frac{2}{\pi\sqrt{3}}.
\end{align}
With this starting point, one can solve Eq.~\eqref{eq:rg_Gamma_tilde_J} numerically and
can determine the universal ratio $\Gamma(i\Lambda)/T_K$ for any
value $\Lambda$ of the flow parameter. With this solution, one can determine
$\widetilde{J}$ as function of $\Lambda/T_K$ from Eq.~\eqref{eq:T0_V0_TK},
and $Z$, $K$, $J_I$ and $G$ follow from Eqs.~\eqref{eq:T0_V0_Z}, \eqref{eq:T0_V0_K},
\eqref{eq:T0_V0_JI} and \eqref{eq:T0_V0_G}. In this way, one can find a universal
result for all quantities for any flow parameter $\Lambda$.

At finite temperature $T$ or bias voltage $V$, we start at a finite flow parameter
$\Lambda=\Lambda_0\gg T,V$ and use the result of the $T=V=0$ solution as initial condition.
The stability of the solution can be checked by increasing $\Lambda_0$ by several orders of magnitude.

Although this procedure by first solving the $T=V=0$ RG equations from
$\Lambda=0$ up to $\Lambda_0$ and, subsequently, at finite $T$ or $V$, solving backwards from
$\Lambda_0$ down to $\Lambda=0$, is in principle possible, it is numerically not
the most accurate one. The reason is that, at $T=V=0$, the universal conductance is
not precisely reproduced numerically by the two subsequent steps. Therefore, we
describe in the following section a numerically more precise procedure.

\subsection{Initial conditions for a finite flow parameter}
\label{sec:IC_finite_flow_parameter}

For given $T,V$ and given symmetric Kondo coupling $J_0=J^{(0)}_{\alpha\alpha'}$
we use the following strategy to set up the initial conditions at $E=i\Lambda_0$
(we use the energy unit $T_K=1$ in the following).
\begin{enumerate}
\item We calculate the initial values from the $T=V=0$ solutions~(\ref{eq:T0_V0_Z})-(\ref{eq:T0_V0_G}) according to
\begin{align}
J(i\Lambda_0)&=J_0,\\
Z(i\Lambda_0)&=\frac{1+2J_0-\sqrt{1+4J_0}}{2J_0^2}, \\
\widetilde{J}(i\Lambda_0)&=\widetilde{J}_0=Z(i\Lambda_0)J_0,\\
K(i\Lambda_0)&=2\widetilde{J}_0^2, \\
J_I(i\Lambda_0)&=J_0\left(1-\widetilde J_0\right), \\
G(i\Lambda_0)&=G_0\frac{3\pi^2}{4}\widetilde J_0^2.
\end{align}
\item For any arbitrary initial value $\Gamma_0=\Gamma(i\Lambda_0)$, we can
calculate the initial flow parameter $\Lambda_0\left(\Gamma_0\right)$
from Eq.~\eqref{eq:T0_V0_TK} (with $T_K\equiv 1$),
\begin{align}
\Lambda_0&=-\Gamma_0+e^{\frac{1}{2\widetilde J_0}}\sqrt{\frac{1-\widetilde
J_0}{\widetilde J_0}},
\end{align}
and then solve the $T=V=0$ flow equations from $\Lambda=\Lambda_0$
to $\Lambda=0$ to obtain the final differential conductance $G(0)$.
This procedure defines a function $G_{\Gamma_0}$, which maps the initial value
of $\Gamma_0$ to the differential conductance at $E=T=V=0$.
\label{itm:calculate_conductance_for_given_Gamma0}
\item The initial value $\Gamma_0$ is chosen such that the differential
conductance at $E=T=V=0$, calculated as described in
step~\ref{itm:calculate_conductance_for_given_Gamma0}, is equal to the unitary
value $G_0=2\frac{e^2}{h}$.
\item We use the initial values $Z(i\Lambda_0)$, $K(i\Lambda_0)$, $J_I(i\Lambda_0)$,
$G(i\Lambda_0)$, $\Gamma_0$ and
$\Lambda_0$, which were calculated for the case $T=V=0$, also for finite $T$ and
$V$. This is justified if the initial flow parameter $\Lambda_0$ is much larger
than $T$ and $V$ because the renormalization of all quantities does not depend
significantly on energy scales which are much smaller than the current flow
parameter.
\item
The RG equations~\eqref{eq:rg_gamma_TV}--\eqref{eq:rg_I_TV} couple vertex
couplings and rates to each other whose Fourier variables $E$ differ by
the voltage $V$. Therefore, we also need initial values for
\begin{align}
J\left(i\Lambda_0+nV\right), Z\left(i\Lambda_0+nV\right), \ldots,
\end{align}
where $n$ is an integer number. These are calculated from
$J\left(i\Lambda_0\right)$, $Z\left(i\Lambda_0\right)$, etc., by solving
the RG equations (\ref{eq:rg_Gamma_T=V=0}-\ref{eq:rg_JI_T=V=0}) along the path
parallel to the real axis.
\end{enumerate}
As already explained in Sec.~\ref{sec:E_flow_idea}, the last point is very essential
and is an improvement compared to Ref.~\onlinecite{pletyukhov_hs_PRL12}. Only by
including it one finds that it is sufficient to take a ratio of $\frac{\Lambda_0}{T_K}\sim 10^6$
[i.e., $\tilde{J}\equiv J_0\sim 0.04$, cf. Eq.~\eqref{eq:T0_V0_TK}] to find a stable solution
of the RG equations for $T,V\ll T_K$.

\subsection{RG equations and initial conditions in second order}

As reference, we will state here how the corresponding initial conditions are
determined when we consider all RG equations only in leading second order.
From the third order RG equations~(\ref{eq:rg_gamma_TV})-(\ref{eq:rg_I_TV}), one
can easily obtain the second order equations by leaving out all summands which
contain three vertices $J$, $K$, or $I^\gamma$.

The solution for $T=V=0$ and the initial conditions for a finite flow parameter
have to be modified acccordingly. We only mention the results here.
\begin{itemize}
\item Solution for $T=V=0$:
\begin{align}
\frac{\Lambda+\Gamma}{T_K}&=\frac{1}{J}e^{\frac{1}{2\widetilde J}}, \\
Z&=\frac{1}{1+2J}, \\
K&=2J^2, \\
J_I&=J, \\
G&=G_0\frac{3\pi^2}{4}J^2.
\end{align}
\item Initial condition for finite flow parameter (with $T_K\equiv 1$):
\begin{align}
\Lambda_0&=-\Gamma_0+\frac{1}{J_0}e^{\frac{1}{2J_0}}, \\
Z(i\Lambda_0)&=\frac{1}{1+2J_0}, \\
K(i\Lambda_0)&=2J_0^2, \\
J_I(i\Lambda_0)&=J_0, \\
G(i\Lambda_0)&=G_0\frac{3\pi^2}{4}J_0^2,
\end{align}
where $\Gamma_0$ is again determined such that the differential conductance $G$
at $E=0$ becomes unitary.
\end{itemize}

\section{Results}
\label{sec:results}
\subsection{Differential conductance at finite temperature and voltage}
\label{sec:cond_finite_TV}
\begin{figure}[htbp!]
  \includegraphics[width=\linewidth]{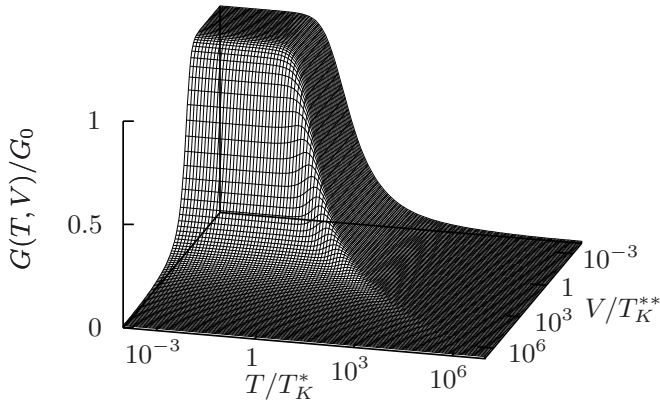}
  \caption{The differential conductance $G(T,V)$ as function of temperature and
voltage, scaled by the unitary conductance $G_0=G(T=V=0)=2e^2/h$. All RG
equations were considered in third order truncation and solved numerically.}
\label{fig:G-TV-3d}
\end{figure}

We have solved the RG equations which have been derived in the previous section
for the second and third order truncation schemes numerically. Thus we have
obtained results for the differential conductance
$G(T, V)$ for transport through a Kondo quantum dot at finite temperature and
voltage. Figure~\ref{fig:G-TV-3d} shows a three-dimensional plot of $G(T, V)$
calculated in third order.
The temperature and voltage are scaled by $T_K^*$ and $T_K^{**}$, respectively,
as defined by Eqs.~\eqref{eq:TK*_definition} and \eqref{eq:TK**_definition}.
The plateau in the upper left corner of Fig.~\ref{fig:G-TV-3d} corresponds to
the unitary conductance $G_0=\frac{2e^2}{h}$ which is reached if both temperature and voltage are
several orders of magnitude smaller than the Kondo temperature.

\begin{figure}[htbp!]
  \includegraphics[width=\linewidth]{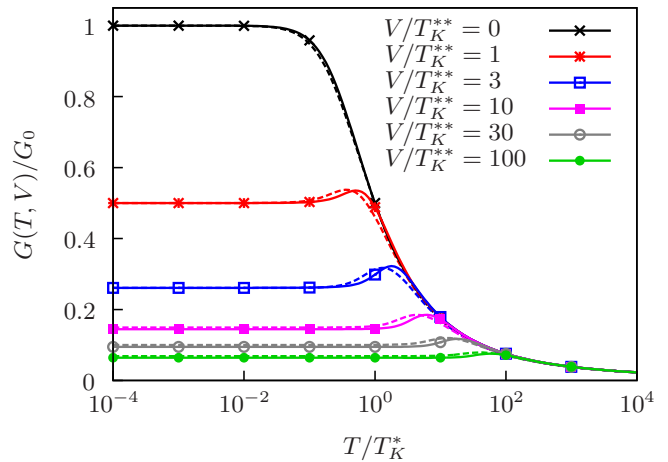}
  \caption{(Color online) The differential conductance $G_V(T)$ as function of temperature for
    constant voltage. The solid (dashed) lines show the results of the
    calculations in third (second) order truncation.}
\label{fig:G-T-peak}
\end{figure}
Figure~\ref{fig:G-T-peak} shows the $T$-dependence of the differential
conductance for six different fixed values of $V$.
For $V=0$, the temperature dependence of the conductance has already been
compared to numerically exact NRG calculations in Ref.~\onlinecite{pletyukhov_hs_PRL12},
where a deviation $\lesssim 3\%$ has been found in the whole temperature regime
independent of the truncation order. For finite voltage, the results for the second
and third order truncation schemes agree quite well if the temperature and the
voltage are scaled by the corresponding values of $T_K^*$ and $T_K^{**}$,
respectively. In the range
\begin{align}
10^{-4}\le \frac{T}{T_K^*},\frac{V}{T_K^{**}}\le 10^7,
\end{align}
which is plotted in Fig.~\ref{fig:G-TV-3d}, the maximal deviation in the
differential conductance $G(T, V)$ between both truncation schemes is less than
$15\%$. In contrast, the ratio $\frac{T_K^*}{T_K^{**}}$ depends crucially on the
truncation order, which seems to hold also for other ratios of low-energy scales
within our method when applied to the strong coupling regime (see the next section).
We obtain $\frac{T_K^*}{T_K^{**}}\approx 1.044$ in second order truncation, and
\begin{align}
\label{eq:ratio_TK_*_**}
\frac{T_K^*}{T_K^{**}}\approx 0.623
\end{align}
in third order truncation. Since the result in third order truncation is expected
to lie closer to the correct result (see also the next section), the last result may
serve as a guideline for more precise calculations in the future.
We note that our prediction is quite close to the result
$T_K^*/T_K^{**}\sim 0.66$ obtained in Ref.~\onlinecite{spataru_PRB10}, where
the $GW$ approximation within the $\sigma G \sigma W$ formalism has been used for
the symmetric Anderson model. Taking our result,
the Fermi liquid coefficients $c_T^{**}$ and $c_V^{**}$ can be calculated from
Eqs.~\eqref{eq:c_TV*} and \eqref{eq:formula_cT_cV_**}, leading to the prediction
\begin{align}
\label{eq:cT_cV_**_prediction}
c_T^{**} \approx 16.95,\quad
c_V^{**} \approx 2.58.
\end{align}

Figure~\ref{fig:G-T-peak} clearly shows that the differential conductance
$G_V(T)\equiv G(T,V)$ is in general not a monotonous function of $T$ if $V$ is fixed.
\begin{figure}[htbp!]
  \includegraphics[width=\linewidth]{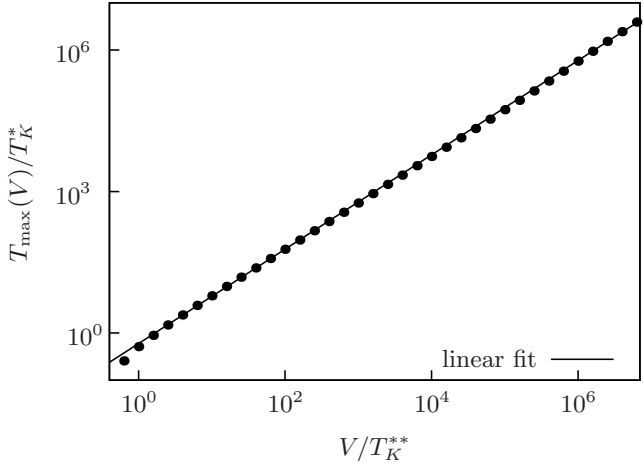}
  \caption{The position $T_{\text{max}}$ of the peak in the differential
    conductance $G_V(T)$, calculated within the third order truncation scheme, at
    constant voltage $V$, cf. the solid lines in Fig.~\ref{fig:G-T-peak}. The
    solid line corresponds to the linear fit
    $\frac{T}{T_K^*}=0.597653\frac{V}{T_K^{**}}$ at large voltages.}
\label{fig:G-T-peak-position}
\end{figure}
\begin{figure}[htbp!]
  \includegraphics[width=\linewidth]{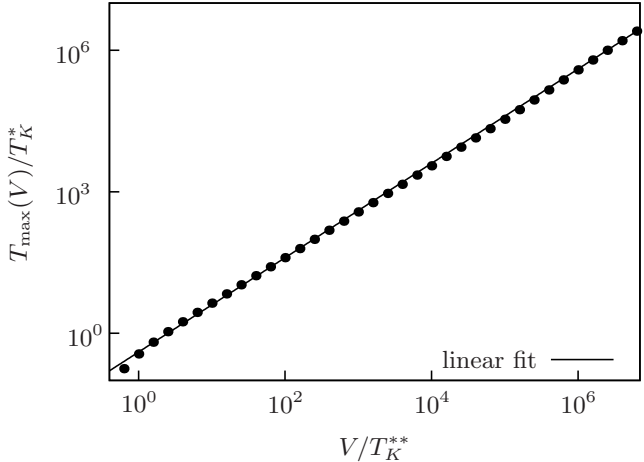}
  \caption{The position $T_{\text{max}}$ of the peak in the differential
    conductance $G_V(T)$, calculated within the second order truncation scheme, at
    constant voltage $V$, cf. the dashed lines in Fig.~\ref{fig:G-T-peak}. The
    solid line corresponds to the linear fit
    $\frac{T}{T_K^*}=0.399717\frac{V}{T_K^{**}}$ at large voltages.}
\label{fig:G-T-peak-position-1-loop}
\end{figure}
For $V\gtrsim T_K^{**}$, there is a pronounced maximum.
Figure~\ref{fig:G-T-peak-position} shows how the position $T_{\text{max}}(V)$ of
this maximum depends on the voltage $V$ (if the RG equations are solved using
the third order truncation). For $V\gtrsim T_K^*$, i.e., in the regime
where there is a pronounced maximum, the function $T_{\max}(V)$ can be
approximated quite well by a linear fit over six orders of magnitude in the
ratio $V/T_K^{**}$.
Figure~\ref{fig:G-T-peak-position-1-loop} shows that the behavior is very similar
if the conductance is calculated within the second order truncation scheme.

\begin{figure}[htbp!]
  \includegraphics[width=\linewidth]{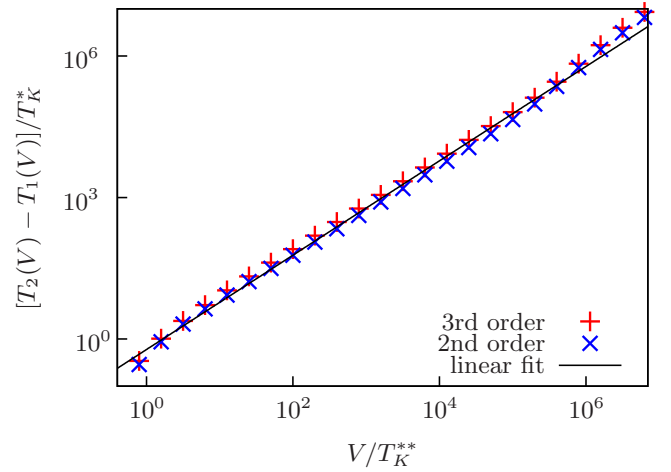}
  \caption{(Color online) The width $T_2(V)-T_1(V)$ of the peak in the differential conductance
    $G_V(T)$ at finite voltage $V$, cf. Fig.~\ref{fig:G-T-peak}. $T_1(V)$ and $T_2(V)$
    are the temperatures for which
    $G_V[T_i(V)]-G_V(T=0)=\frac12[G_V(T=0)+G_V(T_{\text{max}}(V))]$, $i=1,2$. The solid
    line corresponds to the linear fit
    $\frac{T}{T_K^*}=0.597653\frac{V}{T_K^{**}}$ that has been used in
    Fig.~\ref{fig:G-T-peak-position}.}
\label{fig:G-T-peak-width}
\end{figure}
This linear fit from Fig.~\ref{fig:G-T-peak-position} appears to be a
reasonable approximation also for the width of the peak of the function $G_V(T)$ at
fixed $V$, see Fig.~\ref{fig:G-T-peak-width}. We define the peak width as
the difference $T_2(V)-T_1(V)$, where $T_i(V)$, $i=1,2$, are chosen such that
$G_V[T_i(V)]$ is the average of the zero-temperature conductance $G_V(T=0)$
and the maximum value $G_V[T_{\text{max}}(V)]$. This means that the excess
conductance $G_V(T)-G_V(T=0)$ reaches exactly half its maximum value at
$T_i(V)$, $i=1, 2$.

At first sight, it might seem counter-intuitive that the conductance can increase for
increasing temperature since temperature acts as a cutoff of the RG-flow, as can be seen
from the form of the integrals in Eqs.~\eqref{eq:F1_TV}-\eqref{eq:F4_TV}. However, by
looking at the final RG equations \eqref{eq:rg_gamma_TV}-\eqref{eq:rg_I_TV}, one can
see that the cutoff provided by the voltage $V$ is very different and by no means an
overall cutoff for all quantities. In particular, the quantities for $E=i\Lambda$ and
for $E=\pm V + i\Lambda$ have a considerably different flow as function of $\Lambda$ and are coupled to each
other in a complicated way. For finite $V\gtrsim T_K^{**}$ and $T=0$, this leads to the effect that
the conductance shows a maximum for $E=i\Lambda \sim i T_K^{**}$ as function of the flow
parameter $\Lambda$. This in turn leads to a maximum for the temperature dependence
of the conductance since temperature is an overall cutoff for all quantities. Besides this
subtle technical explanation, a more physical interpretation can be given in terms of
the spectral function of the Kondo model, which is believed to have side-peaks at
$\omega \sim \pm V$.\cite{Meir_etal_PRL93,Koenig_etal_PRL96} Therefore, when temperature
is of the order of the voltage, these side peaks can be reached and give rise to an
enhanced conductance.

\subsection{Expansion of the differential conductance for small temperature
    and/or voltage}
We now consider temperatures and voltages much smaller than the Kondo
temperature and calculate numerical approximations for the coefficients $c_T^*$
and $c_V^*$, which appear in the Fermi liquid result~\eqref{eq:Expansion_cT_cV_*},
using the differential conductance obtained in the second and third order
truncation schemes.
We note again that due to our improved scheme for determining
the initial condition of the RG flow at finite voltage (see Section~\ref{sec:IC_finite_flow_parameter}),
we obtain here an improved result for the Fermi liquid coefficient $c_V^*$ in comparison
to Ref.~\onlinecite{pletyukhov_hs_PRL12}. Our result for the coefficient $c_T^*$ is the same as in
Ref.~\onlinecite{pletyukhov_hs_PRL12}, but since recent NRG calculations \cite{merker_PRB13}
have obtained an improved value for $c_T^{*}$, the quality of our results has to
be revisited also for this quantity.

\begin{figure}[htbp!]
  \includegraphics[width=\linewidth]{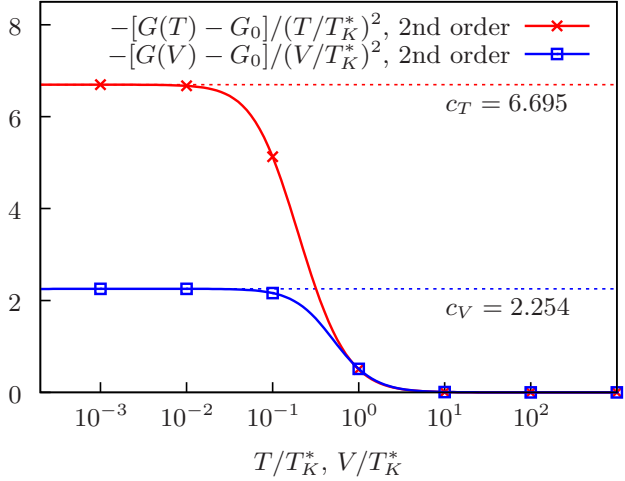}
  \caption{(Color online) Quadratic fit for the differential conductance $G(T)\equiv G_{V=0}(T)$ in second
    order truncation at low
    temperatures and $G(V)\equiv G_{T=0}(V)$ at low voltages, respectively. The differential
    conductance is plotted in such a way that the coefficients $c_T^*$ and $c_V^*$
    in the expansions $G(T)/G_0\approx 1-c_T^*\left(\frac T{T_K^*}\right)^2$ and
    $G(V)/G_0\approx 1-c_V^*\left(\frac V{T_K^*}\right)^2$ can be identified easily.}
\label{fig:G0-expansion-1-loop}
\end{figure}

\begin{figure}[htbp!]
  \includegraphics[width=\linewidth]{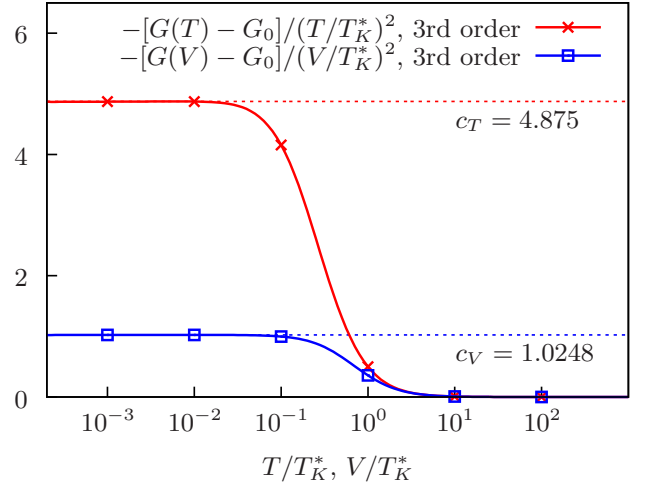}
  \caption{(Color online) Quadratic fit for the differential conductance $G(T)\equiv G_{V=0}(T)$ in third
    order truncation at low
    temperatures and $G(V)\equiv G_{T=0}(V)$ at low voltages, respctively, cf.
    Fig.~\ref{fig:G0-expansion-1-loop}.}
\label{fig:G0-expansion-2-loop}
\end{figure}

Figures~\ref{fig:G0-expansion-1-loop} and~\ref{fig:G0-expansion-2-loop} visualize
how the coefficients $c_T^*$ and $c_V^*$ can be determined in second and third
order, respectively, using a suitable plot of the differential conductance.
When comparing the results in second and third order to the known results
\eqref{eq:ratio_cV_cT} and \eqref{eq:c_TV*}, it is important to note that the
Fermi liquid coefficients are ratios of various low-energy scales in different
energy regimes
\begin{align}
\label{eq:c_TV_*_ratio}
c_T^* = \left(\frac{T_K^*}{T_K^\prime}\right)^2,\quad
c_V^* = \left(\frac{T_K^*}{T_K^{\prime\prime}}\right)^2,
\end{align}
where the scales $T_K^\prime,T_K^{\prime\prime}$ characterize the curvature of the
function $H(T,V)=1-G(T,V)/G_0$ w.r.t. temperature and voltage at $T=V=0$,
\begin{align}
\label{eq:curvature}
\left(\frac{1}{T_K^\prime}\right)^2 = \frac{1}{2}\frac{\partial^2 H}{\partial T^2}\bigg|_{T=V=0},\quad
\left(\frac{1}{T_K^{\prime\prime}}\right)^2 = \frac{1}{2}\frac{\partial^2 H}{\partial V^2}\bigg|_{T=V=0}.
\end{align}
Analogously, one can write the Fermi liquid coefficients $c_T^{**}$ and $c_V^{**}$,
defined by Eq.~\eqref{eq:Expansion_cT_cV_*}, as
\begin{align}
\label{eq:c_TV_**_ratio}
c_T^{**} = \left(\frac{T_K^{**}}{T_K^\prime}\right)^2,\quad
c_V^{**} = \left(\frac{T_K^{**}}{T_K^{\prime\prime}}\right)^2.
\end{align}
As a consequence, the absolute values of the Fermi liquid coefficients relate energy scales close
to $T_K$ (represented by $T_K^*$ or $T_K^{**}$) to energy scales very far below $T_K$ (represented
by $T_K^\prime$ or $T_K^{\prime\prime}$), i.e. they probe the correlation between different energy
regimes. In contrast, the ratio of the Fermi liquid coefficients
\begin{align}
\label{eq:ratio_cV_cT_TK*prime}
\frac{c_V}{c_T} =
\frac{c_V^*}{c_T^*} = \frac{c_V^{**}}{c_T^{**}} = \left(\frac{T_K^{\prime}}{T_K^{\prime\prime}}\right)^2
\end{align}
probes only the energy regime very far below the Kondo temperature, analog to the ratio
$\frac{T_K^*}{T_K^{**}}$, which probes only the energy regime close to $T_K$.

Since only the ratios of the various energy scales $T_K^*, T_K^{**}, T_K^\prime$ and $T_K^{\prime\prime}$
are universal, conclusions about the reliability of the method are not unambigious when comparing
the known results \eqref{eq:ratio_cV_cT} and \eqref{eq:c_TV*}
\begin{align}
\label{eq:known_results_cT*,cV*}
\sqrt{c_T^*} &= \frac{T_K^*}{T_K^\prime} \approx 2.57,\quad
\sqrt{c_V^*} = \frac{T_K^*}{T_K^{\prime\prime}} \approx 1.00,\\
\label{eq:known_results_ratio_cV_cT}
\sqrt\frac{c_V^*}{c_T^*} &= \frac{T_K^{\prime}}{T_K^{\prime\prime}} = \frac{\sqrt{3}}{\pi\sqrt{2}}
\approx 0.39
\end{align}
to the results in second order truncation
\begin{align}
\label{eq:2nd_cT*,cV*}
\frac{T_K^*}{T_K^\prime} &\approx 2.59,\quad
\frac{T_K^*}{T_K^{\prime\prime}} \approx 1.5,\quad
\frac{T_K^{\prime}}{T_K^{\prime\prime}} \approx 0.58,
\end{align}
or to the ones in third order truncation
\begin{align}
\label{eq:3nd_cT*,cV*}
\frac{T_K^*}{T_K^\prime} &\approx 2.21,\quad
\frac{T_K^*}{T_K^{\prime\prime}} \approx 1.01,\quad
\frac{T_K^{\prime}}{T_K^{\prime\prime}} \approx 0.46.
\end{align}
The fact that $\frac{T_K^*}{T_K^\prime}$ is quite close to the correct result in second
order truncation, whereas $\frac{T_K^*}{T_K^{\prime\prime}}$ is very precise in
third order truncation, might be an accident since the
ratio of two energy scales can have a significantly different error than the energy
scales themselves. Overall we observe that all ratios depend significantly on the
truncation order and, if the ratio of two energy scales from the same energy regime is
taken, the result improves when increasing the truncation order. For example, the ratio
$\frac{T_K^\prime}{T_K^{\prime\prime}}$ improves from $49\%$ deviation in second order to
$18\%$ error in third order truncation. Furthermore, we expect
that a perturbative truncation of the RG equations should lead to a better
improvement for larger energy scales when increasing the truncation order. Therefore,
we speculate that our result \eqref{eq:ratio_TK_*_**}
for $\frac{T_K^*}{T_K^{**}}$ in third order truncation might have an even
better quality than the corresponding result for the ratio $\frac{T_K^\prime}{T_K^{\prime\prime}}$.
This is also in accordance with the fact that the result \eqref{eq:GV_2_3} is in agreement with
experiment and in agreement with another recent effective action
method \cite{smirnov_grifoni_PRB03}. Moreover, as already mentioned in
Sec.~\ref{sec:cond_finite_TV}, our result \eqref{eq:ratio_TK_*_**} for the ratio
$T_K^*/T_K^{**}$ is very close to the result of Ref.~\onlinecite{spataru_PRB10}. This provides
evidence that our results in third order truncation are quite reliable for temperatures
and voltages close to the Kondo temperature.
Furthermore, speculating that $T_K^*$ and $T_K^{**}$ are approximately correct in third order truncation, the
precise result for $\frac{T_K^*}{T_K^{\prime\prime}}$ in the same order indicates that
$T_K^{\prime\prime}$
is quite reliable, i.e. it seems that the voltage dependence can also be trusted for small voltages
$V\ll T_K$. In contrast, the poor result for $\frac{T_K^*}{T_K^\prime}$
in third order truncation indicates that $T_K^{\prime}$ deviates significantly from the correct result,
i.e. the temperature dependence is not so well described for $T\ll T_K$.
Therefore, it seems that the rather precise result for $\frac{T_K^*}{T_K^\prime}$ in second order truncation
is an accident and originates from the fact that both $T_K^*$ and $T_K^\prime$ are incorrect
by approximately the same factor, whereas, in third order truncation, $T_K^*$ is more precise than $T_K^\prime$,
leading to the counterintuitive effect that the quality of $\frac{T_K^*}{T_K^\prime}$
decreases with increasing truncation order.

In summary, we have presented arguments that the voltage dependence of the conductance seems to
be quite reliable in third order truncation, whereas the temperature dependence needs to be
improved in the regime $T\ll T_K$. Nevertheless, our arguments are partially based on speculations
and need further substantiation by improved calculations for the nonequilibrium Kondo model
in the strong coupling regime. Furthermore, we note that the third order RG scheme
is in principle capable of reproducing the Fermi liquid result
$\frac{c_V}{c_T}=\frac3{2\pi^2}$ exactly when an analytical solution of the RG
equations is done for either $V=0$ and $T\ll T_K$ or $T=0$ and $V\ll T_K$, and
$\frac{c_V}{c_T}$ is expanded systematically in orders of $J$, see
Ref.~\onlinecite{pletyukhov_hs_PRL12} for details. The numerical
solution contains all terms up to a certain order in $J$, but higher order
terms which are not treated consistently cause deviations from the exact value
for $\frac{c_V}{c_T}$.

We note that our improved scheme to determine
$c_V^*$ can also be applied to the $S=1$ Kondo model and to the calculation of
the Fermi liquid coefficients of the static magnetic susceptibility
$\chi/\chi_0 = 1-a_T (T/T_K)^2-a_V(V/T_K)^2$, where the authors of
Ref.~\onlinecite{hoerig_mora_schuricht_PRB14} report the following changes to their results in
third order truncation \cite{hoerig_privat}
\begin{align}
\left.\frac{c_V}{c_T}\right|_{S=1} &= 0.179, \\
\left.\frac{a_V}{a_T}\right|_{S=1/2} &= 0.11, &
\left.\frac{a_V}{a_T}\right|_{S=1} &= 0.10.
\end{align}
For $S=1$, their results for $\frac{c_V}{c_T}$ deviate by $\sim 9\%$ from the exact
value $\frac{c_V}{c_T}=\frac{3}{2\pi^2}\frac{4+10S}{5+8S}\approx 0.164$, as derived in
Ref.~\onlinecite{hoerig_mora_schuricht_PRB14}.

\subsection{Comparison with experiments}
We have compared our calculations with experimental results obtained by
Kretinin \textit{et al.}\cite{kretinin_PRB85} They measured the differential
conductance at finite temperature and voltage in an InAs nanowire-based quantum
dot. This system can be described by the Kondo model, provided that both
temperature and voltage are sufficiently small to suppress charge fluctuations.
In previous publications, only the results for $G(T=0,V)$ and $G(T,V=0)$
have been compared between theory and experiment, here we present the comparison where
{\it both} temperature and bias voltage are finite.

\begin{figure}[htbp!]
  \includegraphics[width=\linewidth]{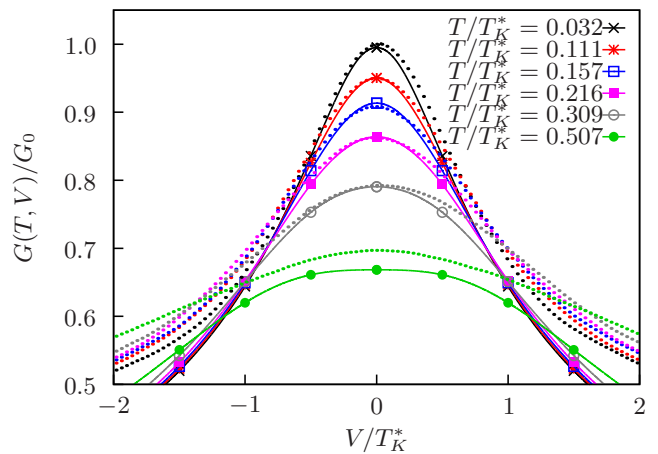}
  \caption{(Color online) Comparison of the differential conductance $G_T(V)$ for fixed $T$
    calculated with RTRG (solid lines) with experimental data from
    Ref.~\onlinecite{kretinin_PRB85} (dots).}
\label{fig:G-V-experiment}
\end{figure}
\begin{figure}[htbp!]
  \includegraphics[width=\linewidth]{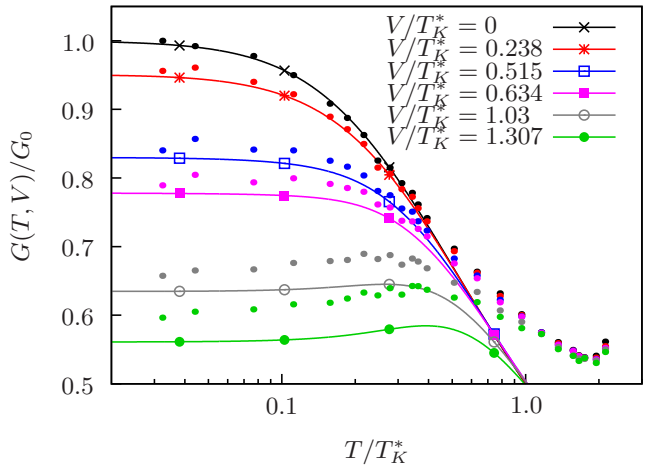}
  \caption{(Color online) Comparison of the differential conductance $G_V(T)$ for fixed $V$
    calculated with RTRG (solid lines) with experimental data from
    Ref.~\onlinecite{kretinin_PRB85} (dots).}
\label{fig:G-T-experiment}
\end{figure}
In Figs.~\ref{fig:G-V-experiment} and~\ref{fig:G-T-experiment}, either the
temperature or the voltage is fixed, and the differential conductance $G$ is
plotted as function of the other quantity. Both figures compare the results
for six different fixed values of the temperature or the voltage, respectively.

We find good agreement between our calculations and the experiment if
temperature and voltage are much smaller than the Kondo temperature. If either
of these quantities is too large, charge fluctuations become important,
which cannot be described properly by the Kondo model that our
calculations are based on.

In an earlier publication,\cite{klochan_scaling_kondo_ZBP} results obtained with
the method presented here had been used to determine the Kondo temperature of an
experimental device and the temperature at which the experiment had been
performed.

\section{Summary and outlook}
\label{sec:summary}

In this paper, we presented a real time renormalization group approach that
extends the flow scheme introduced in Ref.~\onlinecite{pletyukhov_hs_PRL12},
which uses the Fourier variable as the flow parameter. We showed how universal
RG equations can be set up in all orders and that only an expansion in the frequency-independent
effective two-point vertex is needed to guarantee convergence of all frequency integrals.
The RG equations can be solved in various truncation orders providing a consistency check
for the reliability of the results. Whereas in this paper the RG equations
have been solved explicitly up to third order truncation for the Kondo model, we have also outlined the
procedure how to determine all terms in fourth order truncation. This might be helpful
for future applications to test the reliability of the results even further.

We have shown that universality can be achieved for the Kondo model by using
appropriate boundary conditions including the universal stationary conductance at
zero temperature and zero bias voltage. With our procedure it is possible to arrive
at stable results already for initial cutoffs which are about six orders of magnitude
larger than the Kondo temperature. This is a significant improvement compared to other
methods trying to solve directly for the universal properties of the Kondo model
instead of the more involved Anderson impurity model.

We applied the method to the nonequilibrium spin-$\frac{1}{2}$ Kondo model at zero magnetic
field but arbitrary temperature and voltage. We found that the temperature-dependent
conductance $G_V(T)$ at fixed voltage $V$ exhibits non-monotonic behavior. The height and
width of the appearing local peak were shown to scale linearly with the applied voltage over
approximately six orders of magnitude in units of the Kondo temperature. We compared
our results to recent experiments and found good agreement in the regime where
the Kondo model is expected to describe the experimental system accurately.

To characterize the temperature and voltage dependence of the conductance in different energy
regimes close and far below the Kondo temperature, we have defined four different energy scales
$T_K^*$, $T_K^{**}$, $T_K^\prime$, and $T_K^{\prime\prime}$. The scales
$T_K^\prime$/$T_K^{\prime\prime}$ are defined from the curvature of $G_{V=0}(T)$/$G_{T=0}(V)$
at $T=0$/$V=0$, and the scales $T_K^*$/$T_K^{**}$ by the half width at half maximum of the peak
of $G_{V=0}(T)$/$G_{T=0}(V)$ at $T=0$/$V=0$. All these energy scales are proportional to the
Kondo temperature $T_K$. We found that the shape of the conductance
$G(T,V)$ is independent of the truncation order when $T$ and $V$ are scaled in units of
$T_K^*$ and $T_K^{**}$, respectively, providing evidence for the reliability of our result
for the temperature and voltage dependence of the conductance. However, an interesting issue is
the determination of the three independent universal ratios of the four characteristic energy scales,
which turn out to depend crucially on the truncation order.
The ratio $\frac{T_K^\prime}{T_K^{\prime\prime}}=\frac{\sqrt{3}}{\pi\sqrt{2}}\approx 0.39$
is known exactly from Fermi liquid relations,
relating the temperature and voltage dependence for $T,V\ll T_K$. Numerically exact results
exist for the ratio $\frac{T_K^*}{T_K^\prime}\approx 2.57$ from recent NRG calculations, relating the temperature
dependence for $T\sim T_K$ to the one for $T\ll T_K$. Our method predicts in third order truncation the
result $\frac{T_K^*}{T_K^{**}}\approx 0.62$ for the remaining unknown ratio, relating the temperature and voltage
dependence at energies close to $T_K$. We presented evidence for the reliability of this result in third order
approximation, based on the result $G_{T=0}(V=T_K^*)\approx \frac{2}{3}G_0$, which has been confirmed
experimentally and by another recent effective action method. From a comparison of our results for
the other two ratios in third order truncation with the exact ones we obtained evidence that
our results for the voltage dependence of the conductance are quite accurate for all voltages,
whereas the ones for the temperature dependence need to be improved for $T\ll T_K$.

Concerning future directions, the $E$-flow scheme offers a systematic method to avoid $\frac{1}{E^n}$ and
logarithmic divergencies by resumming self-energy insertions and vertex corrections. Since
approximation schemes in different truncation orders can be defined, its reliability can be tested by
itself, in particular in those regimes where the vertices start to grow. So far, applications were
successful for $2$-level models where the dynamics of the local system is driven by spin fluctuations
(Kondo model), energy fluctuations (spin-boson model) or charge fluctuations (interacting resonant level model).
In the future, it is of particular interest to understand the interplay between these fluctuations, as described,
e.g., by the Anderson impurity model (spin and charge fluctuations) or by quantum dots coupled to a bosonic
environment (charge/spin and energy fluctuations). In particular, the Anderson impurity model is
expected to be a suitable model to extract the universal behavior in the Kondo regime without
resorting to the boundary condition of universal conductance, as used in this paper. This
is motivated by recent NRG studies,\cite{hanl_PRB14} where it was shown that universality is reached
much faster for the Anderson impurity model compared to the Kondo model. Furthermore, the Anderson
impurity model allows for the study of potential scattering terms away from the particle-hole symmetric point
and logarithmic energy renormalizations in the mixed-valence regime. Other interesting applications for
the $E$-flow scheme are generic $n$-level quantum dots and models with quantum critical behavior, like,
e.g., the sub-Ohmic spin-boson model or multi-channel Kondo models.

\section*{Acknowledgments}
We thank C.~H\"orig, S.~Jakobs and A.~Kretinin for valuable discussions.
Moreover, we thank A.~Kretinin for providing us the raw data from Ref.~\onlinecite{kretinin_PRB85}
for Figs.~\ref{fig:G-V-experiment} and~\ref{fig:G-T-experiment}. Financial
support by the DFG-Forschergruppe 723 is greatly appreciated.

\begin{appendix}

\section{Evaluation of the integrals in the $E$-flow equations for the
isotropic Kondo model}
\label{sec:integrals}

We will now discuss how the integrals~\eqref{eq:F1_TV}--\eqref{eq:F4_TV}, which
occur in the $E$-flow RG equations for the Kondo model, can be evaluated:
\begin{align}
F^{(1)}_{12} &=Z_{12}\int d\omega \int d\omega' \frac{f'(\omega)f'(\omega')}{
\omega+\omega'+\chi_{12}}, \\
F^{(1)}_{12,34} &= Z_{12} Z_{34} \nonumber \\
&\ \int d\omega \int d\omega'
\frac{f'(\omega)f^a(\omega')}{
(\omega+\omega'+\chi_{12})(\omega+\omega'+\chi_{34})}, \\
F^{(2)}_{12,34} &= Z_{12}\int d\omega \int d\omega'
\frac{{\mathcal{F}}_{34}(\omega)f'(\omega)f'(\omega')}{
\omega+\omega'+\chi_{12}}, \\
F^{(3)}_{12} &= -Z_{12}\int d\omega \frac{f'(\omega)}{\omega+\chi_{12}}, \\
F^{(4)}_{12,34} &= Z_{12}\int d\omega
\frac{{\mathcal{F}}_{34}(\omega)f'(\omega)}{
\omega+\chi_{12}},
\end{align}
where the integral
$\mathcal{F}_{12}(\omega)=\mathcal{F}\left(E_{12},\omega\right)$ is given
by Eq.~\eqref{eq:integrated_prop_difference_frequency},
\begin{align}
\mathcal{F}_{12}(\omega)&=
Z_{12}\int d\omega'f^a\left(\omega'\right)\\
&\quad\quad\quad\quad
\left[\frac{1}{\omega+\omega'+\chi_{12}}
-\frac{1}{\omega'+\chi_{12}}\right].
\nonumber
\end{align}
We note that $\mathcal{F}_{12}(0)=0$.

\subsection{Integrals at zero temperature}

For $T=0$, we can use
\begin{align}
\label{eq:Fermi_function_T0}
f(\omega)&=1-\Theta(\omega)\quad\Rightarrow\quad f'(\omega)=-\delta(\omega)
\end{align}
to evaluate the integrals and get
\begin{align}
F^{(1)}_{12} &=
\frac{Z_{12}}{\chi_{12}}=\frac{1}{E_{12}+i\Gamma\left(E_{12}\right)}, \\
F^{(1)}_{12,34} &=
\begin{cases}
\frac{Z_{12}^2}{\chi_{12}}=Z_{12}F^{(1)}_{12} & \text{if $E_{12}=E_{34}$,} \\
\frac{Z_{12}Z_{34}}{\chi_{12}-\chi_{34}}(\ln{\chi_{12}}-\ln{\chi_{34}}) &
\text{otherwise,}
\end{cases}
\\
F^{(2)}_{12,34} &= Z_{12}\frac{\mathcal{F}_{34}(0)}{\chi_{12}}=0, \\
F^{(3)}_{12} &=
\frac{Z_{12}}{\chi_{12}}=\frac{1}{E_{12}+i\Gamma\left(E_{12}\right)}, \\
F^{(4)}_{12,34} &= -Z_{12}\frac{\mathcal{F}_{34}(0)}{\chi_{12}}=0.
\end{align}
In the special case that no bias voltage is applied, i.e., all chemical
potentials are the same, and
\begin{align}
E_{12}&=E, & Z_{12}&=Z, & \chi_{12}&=\chi,
\end{align}
the integrals are
\begin{align}
\label{eq:integral_F1_TV0}
F^{(1)}_{12} &= \frac{1}{E+i\Gamma\left(E\right)}, \\
F^{(1)}_{12,34} &= \frac{Z}{E+i\Gamma\left(E\right)}, \\
F^{(2)}_{12,34} &= 0, \\
F^{(3)}_{12} &= \frac{1}{E+i\Gamma\left(E\right)}, \\
\label{eq:integral_F4_TV0}
F^{(4)}_{12,34} &= 0.
\end{align}

\subsection{Integrals at finite temperature}

\subsubsection{Prerequisites}

For finite $T$, the Fermi function, its antisymmetric part, and its derivative
are
\begin{align}
f(\omega)&=\frac12-T\sum_{n\in\mathbb{Z}}\frac1{\omega-i\omega_n},
\label{eq:fermiFiniteT} \\
f^a(\omega)&=\phantom{\frac12}-T\sum_{n\in\mathbb{Z}}\frac1{\omega-i\omega_n},
\label{eq:fermiAFiniteT}\\
f'(\omega)&=\phantom{\frac12-}
T\sum_{n\in\mathbb{Z}}\frac1{(\omega-i\omega_n)^2},
\label{eq:fermiDiffFiniteT}
\end{align}
where
\begin{align}
\omega_n=2\pi T\left(n+\frac12\right)
\end{align}
is a Matsubara frequency for any integer number $n$. The integrals are
calculated by closing the integration path in the upper half of the complex
plane and applying the residue theorem. Note that $\chi_{ij}$ is an analytic
function in the upper half of the complex plane, such that the only poles in the upper half
plane are the poles of the Fermi function or its derivative, which are first and
second order poles, respectively.

In the following, we will often use the shorthand notations
\begin{align}
Z&=Z_{12}, \\
\bar Z&=Z_{34}, \\
\gamma&=\frac{\chi_{12}}{2\pi i T}, \\
\bar\gamma&=\frac{\chi_{34}}{2\pi i T}.
\end{align}

\subsubsection{Polygamma Functions}
The polygamma functions are derivatives of the logarithm of the Gamma function
$\Gamma(z)$. They can be used to evaluate series of resolvents that contain
Matsubara frequencies. The first polygamma function, also called digamma
function, is given by
\begin{align}
 \psi(z)&=\frac{d}{dz}\ln\Gamma(z)=-\widehat\gamma+\sum_{n=0}^\infty\left(\frac1
{n+1}-\frac1{n+z}\right),
\end{align}
where $\widehat\gamma$ is the Euler-Mascheroni constant. The digamma function
permits us to evaluate series of the form
\begin{align}
 \sum_{n=0}^\infty\left(\frac1{n+z_1}-\frac1{n+z_2}\right)=\psi(z_2)-\psi(z_1).
\end{align}
Further types of series can be evaluated using the derivatives of the digamma
function,
\begin{align}
 \psi'(z)&=\phantom{-}\sum_{n=0}^\infty\frac1{(n+z)^2},
 \\
 \psi''(z)&=-\sum_{n=0}^\infty\frac2{(n+z)^3}.
\end{align}

\subsubsection{Integrals which depend only on $E_{12}$}
For the integral $F^{(1)}_{12}$, we get
\begin{align}
F^{(1)}_{12}&=Z\int d\omega\int d\omega'\frac{f'(\omega)f'(\omega')}
{\omega+\omega'+\chi} \nonumber\\
&=Z\,T^2\sum_{n\in\mathbb{Z}} \sum_{m\in\mathbb{Z}}\int d\omega\int
d\omega'
\nonumber\\
&\quad\quad\quad\quad\quad\frac1{(\omega- i\omega_n)^2}
\frac1{(\omega'- i\omega_m)^2}
\frac1{\omega+\omega'+\chi} \nonumber\\
&=Z\,(2\pi i T)T\sum_{n=0}^\infty \sum_{m\in\mathbb{Z}} \int d\omega'
\frac1{(\omega'- i\omega_m)^2}\nonumber\\
&\quad\quad\quad\quad\quad
\times\frac{-1}{( i\omega_n+\omega'+\chi)^2} \nonumber\\
&=Z\,(2\pi i T)^2\sum_{n=0}^\infty \sum_{m=0}^\infty
\frac{2}{( i\omega_n+ i\omega_m+\chi)^3}.
\end{align}
Using $i\omega_n=2\pi iT\left(n+\frac12\right)$ and $\chi=(2\pi iT)\gamma$, this
yields
\begin{align}
F^{(1)}_{12}
&=\frac{Z}{2\pi i T}\sum_{n=0}^\infty \sum_{m=0}^\infty
\frac{2}{(n+m+\gamma+1)^3} \nonumber\\
&\overset{N=n+m}{=}2
\frac{Z}{2\pi i T}\sum_{N=0}^\infty \sum_{m=0}^N
\frac{1}{(N+\gamma+1)^3}.
\end{align}
Note that there are $N+1$ equal terms in the $m$-sum:
\begin{align}
F^{(1)}_{12}
&=2\frac{Z}{2\pi i T}\sum_{N=0}^\infty
\frac{N+1}{(N+\gamma+1)^3} \nonumber\\
&=2\frac{Z}{2\pi i T}\sum_{N=0}^\infty
\frac{N+\gamma+1-\gamma}{(N+\gamma+1)^3} \nonumber\\
&=2\frac{Z}{2\pi i T}\sum_{N=0}^\infty
\left[\frac1{(N+\gamma+1)^2}-\gamma\frac1{(N+\gamma+1)^3}
\right]
\nonumber\\
&=\frac{Z}{2\pi i
T}\left[2\psi'(\gamma+1)+\gamma\psi''(\gamma+1)\right].
\end{align}

Evaluating the integral $F^{(3)}_{12}$ yields
\begin{align}
F^{(3)}_{12} &=
-Z\int d\omega\frac{f'(\omega)}{\omega+\chi}\nonumber\\
&=-Z\,T\sum_{n\in\mathbb{Z}}
\int d\omega
\frac1{(\omega- i\omega_n)^2}\frac1{\omega+\chi}\nonumber\\
&=Z(2\pi i
T)\sum_{n=0}^\infty\frac1{( i\omega_n+\chi)^2}\nonumber\\
&=\frac Z{2\pi i T}\sum_{n=0}^\infty\frac1{\left(n+\gamma+\frac12\right)^2}
\nonumber\\
&=\frac Z{2\pi i T}\psi'\left(\gamma+\frac12\right).
\end{align}

\subsubsection{Integrals which depend on $E_{12}$ and $E_{34}$}
The evaluation of the integrals that depend on two $E$-arguments can be done
analogously. The calculations are straightforward, but lengthy. Therefore, we
only list the results here.

In the special case that both $E$-arguments are equal, we get
\begin{widetext}
\begin{align}
F^{(1)}_{12,12}&=Z\,F^{(1)}_{12}, \\
F^{(2)}_{12,12}&=\frac{Z^2}{2\pi i T}\left\{
\left[2\psi'(\gamma+1)+\gamma\psi''(\gamma+1)\right]
\left[\psi(\gamma+1)-\psi\left(\gamma+\frac12\right)-1\right]
+\gamma\left[\psi'(\gamma+1)\right]^2-\frac12\psi''(\gamma+1)\right\}, \\
F^{(4)}_{12,12}&=\frac{Z^2}{2\pi i T}
\left\{
\psi'\left(\gamma+\frac12\right)
\left[\psi\left(\gamma+\frac12\right)-\frac12\psi\left(\gamma+1\right)
+1\right]
-\psi'(\gamma+1)\left[\frac12\psi\left(\gamma+\frac12\right)+1\right]
\right.\nonumber\\
&\quad\quad\quad\quad\left.+
\frac12\sum_{k=0}^\infty\frac{d}{dk}\left[
\frac{\psi(k+\gamma+1)}{k+\gamma+\frac12}
-
\frac{\psi(k+\gamma+\frac12)}{k+\gamma+1}
\right]
\right\}.
\end{align}
If $E_{12}\neq E_{34}$, we get
\begin{align}
F^{(1)}_{12,34} &=
\frac{Z\bar Z}{2\pi i
T(\bar\gamma-\gamma)}\left[\psi(\bar\gamma+1)-\psi(\gamma+1)
+\bar\gamma\psi'(\bar\gamma+1)-\gamma\psi'(\gamma+1)\right]
\end{align}
and
\begin{align}
F^{(4)}_{12,34}
 &=\frac{Z\bar Z}{2\pi i T}
\left\{
\psi'\left(\gamma+\frac12\right)
\left[\psi\left(\bar\gamma+\frac12\right)-\frac12\psi\left(\bar\gamma+1\right)
+\frac1{1-2(\bar\gamma-\gamma)}\right]\right.\nonumber\\
&\quad\quad
-\psi'(\bar\gamma+1)\left[\frac12\psi\left(\gamma+\frac12\right)+\frac1{
1-2(\bar\gamma-\gamma)}\right]\nonumber\\
&\quad\quad\left.+
\frac12\sum_{k=0}^\infty\frac{d}{dk}\left[
\frac{\psi(k+\bar\gamma+1)}{k+\gamma+\frac12}
-
\frac{\psi(k+\gamma+\frac12)}{k+\bar\gamma+1}
\right]
\right\}.
\end{align}
\end{widetext}
No simple expression has been found for the integral $F^{(2)}_{12,34}$. We use
the approximation
\begin{align}
F^{(2)}_{12,34}&\approx\frac{(Z_{12})^2 F^{(2)}_{34,34} + (Z_{34})^2 F^{(2)}_{12,12}}{2Z_{12}Z_{34}},
\end{align}
which, for $|E|\gg T,V$, neglects only contributions of $O(\frac{T,V}{|E|^2}J^3)$
in the RG equations which are beyond subleading order, consistent with the
strategy described in Section~\ref{sec:strategy-diagram-selection}.
It has been verified that replacing the exact expressions for the integrals
$F^{(1)}_{12,34}$ and $F^{(4)}_{12,34}$ by similar approximations only has a
negligible effect on the results.

The approach used to evaluate the remaining series in $F^{(4)}_{12,34}$ is as follows:
\begin{itemize}
 \item Sum the terms from $k=0$ to $k=k_0-1$ explicitly.
\item Replace the remaining series, starting from $k=k_0$, by an integral.
\end{itemize}
We get
\begin{align}
 &\sum_{k=k_0}^\infty\frac{d}{dk}\left[
\frac{\psi(k+\bar\gamma+1)}{k+\gamma+\frac12}
-
\frac{\psi(k+\gamma+\frac12)}{k+\bar\gamma+1}
\right]\nonumber\\
\approx
&\int_{k_0-\frac12}^\infty\frac{d}{dk}\left[
\frac{\psi(k+\bar\gamma+1)}{k+\gamma+\frac12}
-
\frac{\psi(k+\gamma+\frac12)}{k+\bar\gamma+1}
\right]\\
=&
-\left[
\frac{\psi(k_0+\bar\gamma+\frac12)}{k_0+\gamma}
-
\frac{\psi(k_0+\gamma)}{k_0+\bar\gamma+\frac12}
\right].
\nonumber
\end{align}
The relative error caused by the approximation is
\begin{align}
 \epsilon&\sim \frac14 \frac1{a^2}, & &\text{where} &
a&=\min\left\{\left|k_0+\bar\gamma+\frac12\right|, |k_0+\gamma|\right\},
\end{align}
and can be made arbitrarily small by choosing a suitable $k_0$.

\section{Subleading terms in the RG equations for the vertices in the
Kondo model}

\subsection{Effective vertex $G_{12}(E)$}

\label{sec:subleading_terms_G}

In the subleading terms in Eq.~\eqref{eq:G_rg_eta_summation}, different
products of effective vertices occur, namely (without Fourier arguments and
$F$-integrals)
\begin{align}
&\widehat G_{34}\widehat G_{12}\widehat G_{43},
&
&\widehat G_{34}\widehat G_{13}\widehat G_{42},
&
&\widehat G_{34}\widehat G_{42}\widehat G_{13},
\nonumber \\
&\widehat G_{13}\widehat G_{42}\widehat G_{34},
&
&\widehat G_{32}\widehat G_{14}\widehat G_{43},
\label{eq:G_rg_subleading_terms}
\end{align}
which all have the form
\begin{multline}
\widehat G_{n_1n_2} \widehat G_{n_3n_4} \widehat G_{n_5n_6}
\\
=
\sum_{\chi\chi'\chi''}
G^{\chi}_{n_1n_2} G^{\chi'}_{n_3n_4} G^{\chi''}_{n_5n_6}
\widehat L^{\chi}_{n_1n_2} \widehat L^{\chi}_{n_3n_4} \widehat
L^{\chi}_{n_5n_6},
\end{multline}
where $n_i\in\{1,2,3,4\}$. Note that the $G^{\chi}_{n_1n_2}$ only depend on the
reservoir indices, and the $\widehat L^{\chi}_{n_1n_2}$ only on the spin
indices, and that we only consider $\chi,\chi',\chi''\in\{2,3\}$, such that
all occurring $\widehat L^\chi_{n_1n_2}$ contain the Pauli matrix
$\underline{\sigma}_{n_1n_2}$:
\begin{multline}
\widehat L^{\chi}_{n_1n_2} \widehat L^{\chi'}_{n_3n_4} \widehat
L^{\chi''}_{n_5n_6}
\\
=
\sum_{i,j,k\in\{x,y,z\}}
L^{\chi}_{i} L^{\chi'}_{j} L^{\chi''}_{k}
\sigma^i_{n_1n_2} \sigma^j_{n_3n_4} \sigma^k_{n_5n_6}.
\label{eq:product_LLL}
\end{multline}
To evaluate these subleading terms, we first evaluate the Pauli matrix products
using the multiplication rule
\begin{align}
\sigma^i_{13}\sigma^j_{32}=\delta_{ij}\delta_{12}+i\epsilon_{ijk}\sigma^k_{12}.
\end{align}
The results for the five different products in~\eqref{eq:G_rg_subleading_terms}
are:
\begin{align}
\sigma^i_{34}\sigma^j_{12}\sigma^k_{43}&=2\delta_{ik}\sigma^j_{12},
\label{eq:product_GGG_PauliMatrixProducts_1}
\\
\sigma^i_{34}\sigma^j_{13}\sigma^k_{42}
&=\left(\delta_{ij}\delta_{14}+i\epsilon_{jil}\sigma^l_{14}\right) \sigma^k_{42}
\nonumber \\
&=\delta_{ij}\sigma^k_{12}+i\epsilon_{jik}\delta_{12}
-\epsilon_{jil}\epsilon_{lkm}\sigma^m_{12}
\nonumber \\
&=\delta_{ij}\sigma^k_{12}-i\epsilon_{ijk}\delta_{12}
+\left(\delta_{ik}\delta_{jm}-\delta_{im}\delta_{jk}\right) \sigma^m_{12}
\nonumber \\
&=\delta_{ij}\sigma^k_{12}
+\delta_{ik}\sigma^j_{12}-\delta_{jk}\sigma^i_{12}-i\epsilon_{ijk}\delta_{12},
\\
\sigma^i_{34}\sigma^j_{42}\sigma^k_{13}
&=\delta_{ij}\sigma^k_{12}
+\delta_{ik}\sigma^j_{12}-\delta_{jk}\sigma^i_{12}
+i\epsilon_{ijk}\delta_{12},
\\
\sigma^i_{13}\sigma^j_{42}\sigma^k_{34}
&=\delta_{jk}\sigma^i_{12}
+\delta_{ik}\sigma^j_{12}-\delta_{ij}\sigma^k_{12}
-i\epsilon_{ijk}\delta_{12},
\\
\sigma^i_{32}\sigma^j_{14}\sigma^k_{43}
&=\delta_{jk}\sigma^i_{12}
+\delta_{ik}\sigma^j_{12}-\delta_{ij}\sigma^k_{12}
+i\epsilon_{ijk}\delta_{12}
\label{eq:product_GGG_PauliMatrixProducts_5}
\end{align}
(note that the last three results can be obtained by making suitable
permutations of the indices $i$, $j$, and $k$ in the second one). In the
following, we will discard all contributions $\sim\delta_{12}$ in the above
results because these would contribute to the renormalization of $G^a_{12}$,
which we have neglected.

To evaluate the products~\eqref{eq:product_LLL} further, we note that only the
third superoperator $L^{\chi''}_k$ can be $L^3_k$, because all products where
$L^3_k$ is multiplied with a component of $\underline{L}^2$ from the right are
zero according to the rules in Secs.~\ref{sec:algebra_scalar_products}
and~\ref{sec:algebra_vector_products}. Therefore, we only have to consider
products of the form
\begin{align}
L^2_i L^2_j L^{2,3}_k \sigma^i_{n_1n_2} \sigma^j_{n_3n_4} \sigma^k_{n_5n_6},
\end{align}
where the Pauli matrix products is one of
Eqs.~(\ref{eq:product_GGG_PauliMatrixProducts_1}--
\ref{eq:product_GGG_PauliMatrixProducts_5}), and we omit all terms
$\sim\delta_{12}$. This means that two out of the indices $i$, $j$, and $k$ are
always equal, and we can use the results from
Secs.~\ref{sec:algebra_scalar_multiplication_with_LaLb},
\ref{sec:algebra_scalar_products} and~\ref{sec:algebra_vector_products}, and the
commutator relations~\eqref{eq:commutator_L2L123}
to evaluate the frequently occurring products
\begin{align}
L^2_i L^2_i L^{2,3}_j \sigma^j_{12}&=\frac12 \widehat L^{2,3}_{12},
\\
L^2_i L^2_j L^2_j \sigma^i_{12}&=\frac12 \widehat L^2_{12},
\\
L^2_i L^2_j L^3_j&=0,
\\
L^2_i L^2_j L^2_i \sigma^j_{12}&=L^2_i L^2_i L^2_j \sigma^j_{12}
+L^2_i\left[L^2_j,L^2_i\right]\sigma^j_{12}
\nonumber \\
&=\frac12\widehat L^2_{12}-\frac{i}{2}\epsilon_{jik}L^2_i L^2_k \sigma^j_{12}
\nonumber \\
&=\frac12\widehat L^2_{12}-\frac12
\left(i\underline{L}^2\times\underline{L}^2\right)\cdot\underline{\sigma}_{12}
\nonumber \\
&=\frac14 \widehat L^2_{12},
\\
L^2_i L^2_j L^3_i \sigma^j_{12}&=L^2_i L^3_i L^2_j \sigma^j_{12}
+L^2_i\left[L^2_j,L^3_i\right]\sigma^j_{12}
\nonumber \\
&=-\frac{i}{2}\epsilon_{jik}L^2_i L^3_k \sigma^j_{12}
\nonumber \\
&=-\frac12
\left(i\underline{L}^2\times\underline{L}^3\right)\cdot\underline{\sigma}_{12}
\nonumber \\
&=-\frac12 \widehat L^3_{12}.
\end{align}
For the terms where the last superoperator in the product is a
component of $\underline{L}^2$, this yields:
\begin{align}
\widehat L^2_{34} \widehat L^2_{12} \widehat L^2_{43}
&=2 L^2_i L^2_j L^2_i \sigma^j_{12}
\nonumber \\
&=\frac12 \widehat L^2_{12},
\\
\widehat L^2_{34} \widehat L^2_{13} \widehat L^2_{42}
&=\widehat L^2_{34} \widehat L^2_{42} \widehat L^2_{13}
\nonumber \\
&= L^2_i L^2_i L^2_k \sigma^k_{12} + L^2_i L^2_j L^2_i \sigma^j_{12} - L^2_i
L^2_j L^2_j \sigma^i_{12}
\nonumber \\
&=\frac14 \widehat L^2_{12},
\\
\widehat L^2_{13} \widehat L^2_{42} \widehat L^2_{34}
&=\widehat L^2_{32} \widehat L^2_{14} \widehat L^2_{43}
\nonumber \\
&=L^2_i L^2_j L^2_j \sigma^i_{12} + L^2_i L^2_j L^2_i \sigma^j_{12} - L^2_i
L^2_i L^2_k \sigma^k_{12}
\nonumber \\
&=\frac14 \widehat L^2_{12},
\end{align}
and similarly for the products where the last factor is a component of
$\underline{L}^3$,
\begin{align}
\widehat L^2_{34} \widehat L^2_{12} \widehat L^3_{43}
&=2 L^2_i L^2_j L^3_i \sigma^j_{12}
\nonumber \\
&=-\widehat L^3_{12},
\\
\widehat L^2_{34} \widehat L^2_{13} \widehat L^3_{42}
&=\widehat L^2_{34} \widehat L^2_{42} \widehat L^3_{13}
\nonumber \\
&= L^2_i L^2_i L^3_k \sigma^k_{12} + L^2_i L^2_j L^3_i \sigma^j_{12} - L^2_i
L^2_j L^3_j \sigma^i_{12}
\nonumber \\
&=0,
\\
\widehat L^2_{13} \widehat L^2_{42} \widehat L^3_{34}
&=\widehat L^2_{32} \widehat L^2_{14} \widehat L^3_{43}
\nonumber \\
&=L^2_i L^2_j L^3_j \sigma^i_{12} + L^2_i L^2_j L^3_i \sigma^j_{12} - L^2_i
L^2_i L^3_k \sigma^k_{12}
\nonumber \\
&=-\widehat L^3_{12}.
\end{align}
The final result for the products in Eq.~\eqref{eq:G_rg_subleading_terms} is
\begin{align}
\widehat G_{34}\widehat G_{12}\widehat G_{43}
&=
\frac12 G^2_{34} G^2_{12} G^2_{43} \widehat L^2_{12}
-G^2_{34} G^2_{12} G^3_{43} \widehat L^3_{12},
\label{eq:product_GGG_final_1_appendix}
\\
\widehat G_{34}\widehat G_{13}\widehat G_{42}
&=
\frac14 G^2_{34} G^2_{13} G^2_{42} \widehat L^2_{12},
\\
\widehat G_{34}\widehat G_{42}\widehat G_{13}
&=
\frac14 G^2_{34} G^2_{42} G^2_{13} \widehat L^2_{12},
\\
\widehat G_{13}\widehat G_{42}\widehat G_{34}
&=
\frac14 G^2_{13} G^2_{42} G^2_{34} \widehat L^2_{12}
-G^2_{13} G^2_{42} G^3_{34} \widehat L^3_{12},
\\
\widehat G_{32}\widehat G_{14}\widehat G_{43}
&=
\frac14 G^2_{32} G^2_{14} G^2_{43} \widehat L^2_{12}
-G^2_{32} G^2_{14} G^3_{43} \widehat L^3_{12}.
\label{eq:product_GGG_final_5_appendix}
\end{align}

\subsection{Current vertex $I^\gamma_{12}(E)$}

\label{sec:subleading_terms_I}

We follow the approach of appendix~\ref{sec:subleading_terms_G} to evaluate the
subleading terms on the right-hand side of the RG equation for the current vertex, which
is obtained by replacing the first vertex in each of the terms in
Eq.~\eqref{eq:G_rg_eta_summation}, by a current vertex:
\begin{align}
&\widehat I^\gamma_{34}\widehat G_{12}\widehat G_{43},
&
&\widehat I^\gamma_{34}\widehat G_{13}\widehat G_{42},
&
&\widehat I^\gamma_{34}\widehat G_{42}\widehat G_{13},
\nonumber \\
&\widehat I^\gamma_{13}\widehat G_{42}\widehat G_{34},
&
&\widehat I^\gamma_{32}\widehat G_{14}\widehat G_{43}.
\label{eq:I_rg_subleading_terms}
\end{align}
Only the part $\sim\widehat L^1$ of the current vertex contributes to the
current kernel according to Eqs.~\eqref{eq:SigmaI_rg_V_spin_summation_lo}
and~\eqref{eq:SigmaI_rg_V_spin_summation_nlo}. The only way to obtain
contributions $\sim\widehat L^1$ from the products above is to consider the
part $\sim\widehat L^1$ of the current vertex, and the part $\sim\widehat L^2$
of both effective vertices:
\begin{multline}
\widehat I^\gamma_{n_1n_2} \widehat G_{n_3n_4} \widehat G_{n_5n_6}
\\
=
I^{\gamma1}_{n_1n_2} G^{2}_{n_3n_4} G^{2}_{n_5n_6}
\widehat L^{1}_{n_1n_2} \widehat L^{2}_{n_3n_4} \widehat L^{2}_{n_5n_6},
\end{multline}
where
\begin{multline}
\widehat L^{1}_{n_1n_2} \widehat L^{2}_{n_3n_4} \widehat
L^{2}_{n_5n_6}
\\
=
\sum_{i,j,k\in\{x,y,z\}}
L^{1}_{i} L^{2}_{j} L^{2}_{k}
\sigma^i_{n_1n_2} \sigma^j_{n_3n_4} \sigma^k_{n_5n_6},
\end{multline}
and we have to consider the products of Pauli matrix components from
Eqs.~(\ref{eq:product_GGG_PauliMatrixProducts_1}--
\ref{eq:product_GGG_PauliMatrixProducts_5}), except for the terms
$\sim\delta_{12}$, which do not contribute to the renormalization of
$I^{\gamma1}$. The products of superoperators which need to be evaluated
according to these Pauli matrix products are
\begin{align}
L^1_i L^2_i L^2_j \sigma^j_{12}&=0,
\\
L^1_i L^2_j L^2_j \sigma^i_{12}&=\frac 12 L^1_i L^a \sigma^i_{12}=\frac12
\widehat L^1_{12},
\\
L^1_i L^2_j L^2_i \sigma^j_{12}&=L^1_i L^2_i L^2_j \sigma^j_{12}
+L^1_i \left[L^2_j,L^2_i\right] \sigma^j_{12}
\nonumber \\
&=-\frac{i}{2}\epsilon_{jik} L^1_i L^2_k \sigma^j_{12}
\nonumber \\
&=-\frac12\left(i\underline{L}^1\times\underline{L}^2\right)\cdot\sigma_{12}
\nonumber \\
&=-\frac12\widehat L^1_{12}.
\end{align}
Finally, we get
\begin{align}
\widehat L^1_{34} \widehat L^2_{12} \widehat L^2_{43}
&=2 L^1_i L^2_j L^3_i \sigma^j_{12}
\nonumber \\
&=-\widehat L^1_{12},
\\
\widehat L^1_{34} \widehat L^2_{13} \widehat L^2_{42}
&=\widehat L^1_{34} \widehat L^2_{42} \widehat L^2_{13}
\nonumber \\
&= L^1_i L^2_i L^2_k \sigma^k_{12} + L^1_i L^2_j L^2_i \sigma^j_{12} - L^1_i
L^2_j L^2_j \sigma^i_{12}
\nonumber \\
&=-\widehat L^1_{12},
\\
\widehat L^1_{13} \widehat L^2_{42} \widehat L^2_{34}
&=\widehat L^1_{32} \widehat L^2_{14} \widehat L^2_{43}
\nonumber \\
&=L^1_i L^2_j L^2_j \sigma^i_{12} + L^1_i L^2_j L^2_i \sigma^j_{12} - L^1_i
L^2_i L^2_k \sigma^k_{12}
\nonumber \\
&=0.
\end{align}
The final result for the products in Eq.~\eqref{eq:I_rg_subleading_terms} is
\begin{align}
\widehat I^\gamma_{34}\widehat G_{12}\widehat G_{43}
&=
-I^{\gamma1}_{34} G^2_{12} G^2_{43} \widehat L^1_{12},
\label{eq:product_IGG_final_1_appendix}
\\
\widehat I^\gamma_{34}\widehat G_{13}\widehat G_{42}
&=
-I^{\gamma1}_{34} G^2_{13} G^2_{42} \widehat L^1_{12},
\\
\widehat I^\gamma_{34}\widehat G_{42}\widehat G_{13}
&=
-I^{\gamma1}_{34} G^2_{42} G^2_{13} \widehat L^1_{12},
\\
\widehat I^\gamma_{13}\widehat G_{42}\widehat G_{34}
&=0,
\\
\widehat I^\gamma_{32}\widehat G_{14}\widehat G_{43}
&=0.
\label{eq:product_IGG_final_5_appendix}
\end{align}

\end{appendix}


\end{document}